\documentclass{article}
\usepackage[utf8]{inputenc}

\usepackage[margin=1in]{geometry}
\usepackage{color}
\usepackage{graphicx}
\usepackage{epsfig}
\usepackage{epstopdf}
\usepackage{subfig}
\usepackage[font=small,skip=0pt]{caption}

\usepackage[dvipsnames]{xcolor}

\usepackage{amsmath}
\usepackage{amssymb}
\usepackage{amsthm}

\usepackage{rotating}
\usepackage{verbatim}
\usepackage{hyperref}
\usepackage{bm}
\usepackage[normalem]{ulem}
\usepackage[authoryear]{natbib}
\usepackage{bbm}
\usepackage{array}
\usepackage{enumerate}
\usepackage{float}

\usepackage[flushleft]{threeparttable}
\usepackage{tablefootnote}

\usepackage{ifthen}
\usepackage{url}
\usepackage{mathtools}

\usepackage{authblk}
\title{Powerful Knockoffs via Minimizing Reconstructability}
\author[]{Asher Spector}
\author[]{Lucas Janson}
\affil[]{Department of Statistics, Harvard University}
\date{}

\newtheorem{theorem}{Theorem}[section]
\newtheorem{proposition}{Proposition}[section]
\newtheorem{definition}{Definition}[section]

\newtheorem{corollary}{Corollary}[section] 

\newtheorem{lemma}[theorem]{Lemma}

\usepackage{algorithm}
\usepackage[noend]{algpseudocode}

\newcommand{\bX}{\mathbf{X}}
\newcommand{\by}{\mathbf{y}}

\newcommand{\cR}{\mathcal{R}}

\newcommand{\betan}{\beta^{(n)}}
\newcommand{\betaext}{\beta^{(\mathrm{ext})}}
\newcommand{\hatbetaext}{\hat{\beta}^{(\mathrm{ext})}}
\newcommand{\lassoext}{\hat{\beta}^{(\ell_1, \mathrm{ext})}}
\newcommand{\betanext}{\beta^{(n, \mathrm{ext})}}
\newcommand{\Sigman}{\Sigma^{(n)}}
\newcommand{\bjkn}{B_{j,k}}
\newcommand{\dkn}{D_k}
\newcommand{\bjknplus}{B_{j,k}^+}
\newcommand{\bjknminus}{B_{j,k}^-}
\newcommand{\bepsn}{B_{\epsilon_0}^{(n)}}
\newcommand{\package}{\texttt{knockpy}}

\newcommand{\mmaxent}{\mathrm{ME}} 
\newcommand{\tmaxent}{ME} 
\newcommand{\smaxent}{ME } 

\newcommand{\otherj}{t}

\newcommand\Perp{\protect\mathpalette{\protect\independenT}{\perp}}
\def\independenT#1#2{\mathrel{\rlap{$#1#2$}\mkern2mu{#1#2}}}
\newcommand{\iid}{\stackrel{\emph{i.i.d.}}{\sim}}
\DeclareMathOperator{\Var}{Var}
\DeclareMathOperator{\sorted}{sorted}
\DeclareMathOperator{\sign}{sign}
\DeclareMathOperator{\swap}{swap}
\DeclareMathOperator{\tv}{TV}
\DeclareMathOperator{\power}{Power}

\counterwithout*{footnote}{section}
\counterwithout*{footnote}{subsection}
\counterwithout*{footnote}{subsubsection}

\begin{document}

\maketitle

\begin{abstract}
Model-X knockoffs \citep{mxknockoffs2018} allows analysts to perform feature selection using almost any machine learning algorithm while still provably controlling the expected proportion of false discoveries. To apply model-X knockoffs, one must construct synthetic variables, called knockoffs, which effectively act as controls during feature selection. The gold standard for constructing knockoffs has been to minimize the mean absolute correlation (MAC) between features and their knockoffs, but, surprisingly, we prove this procedure can be powerless in extremely easy settings, including Gaussian linear models with correlated exchangeable features. The key problem is that minimizing the MAC creates strong joint dependencies between the features and knockoffs, which allow machine learning algorithms to partially or fully \textit{reconstruct} the effect of the features on the response using the knockoffs.  To improve the power of knockoffs, we propose generating knockoffs which \textit{minimize the reconstructability} (MRC) of the features, and we demonstrate our proposal for Gaussian features by showing it is computationally efficient, robust, and powerful. We also prove that certain MRC knockoffs minimize a natural definition of estimation error in Gaussian linear models. Furthermore, in an extensive set of simulations, we find many settings with correlated features in which MRC knockoffs dramatically outperform MAC-minimizing knockoffs and no settings in which MAC-minimizing knockoffs outperform MRC knockoffs by more than a very slight margin. We implement our methods and a host of others from the knockoffs literature in a new open source python package \package.\footnote{See \url{https://github.com/amspector100/knockpy}.}
\end{abstract}
\section{Introduction}

Model-X (MX) knockoffs \citep{mxknockoffs2018} has recently emerged as a powerful and flexible method to perform \textit{controlled variable selection}. Informally, given a set of features $(X_1, \dots, X_p)$ and an outcome of interest $Y$, knockoffs allows one to leverage almost any regression method to discover relationships between the features and the outcome. Notably, knockoffs exactly control the expected proportion of false positives in finite samples provided that the distribution of the features $X$ is known, while assuming nothing about the conditional distribution $Y \mid X$.

The knockoffs framework accomplishes this task by constructing synthetic variables, called knockoffs, which mimic the correlation structure of the original features. In principle, there are many possible ways to construct valid knockoff variables. However, the knockoffs literature has largely converged to a single measure of knockoff quality, namely that one should minimize the mean absolute correlation (MAC) between features and their knockoffs in order to maximize statistical power. Such knockoffs can be constructed via semidefinite programming (SDP) when the features are multivariate Gaussian, and almost the entire knockoffs literature has treated them as the ``gold standard'' of knockoff quality (see, e.g., \citet{fxknock, mxknockoffs2018, nodewiseknock, metro2019, fanok2020}).

\begin{figure}[h!]
    \centering
   \makebox[\textwidth]{\includegraphics[width=\textwidth]{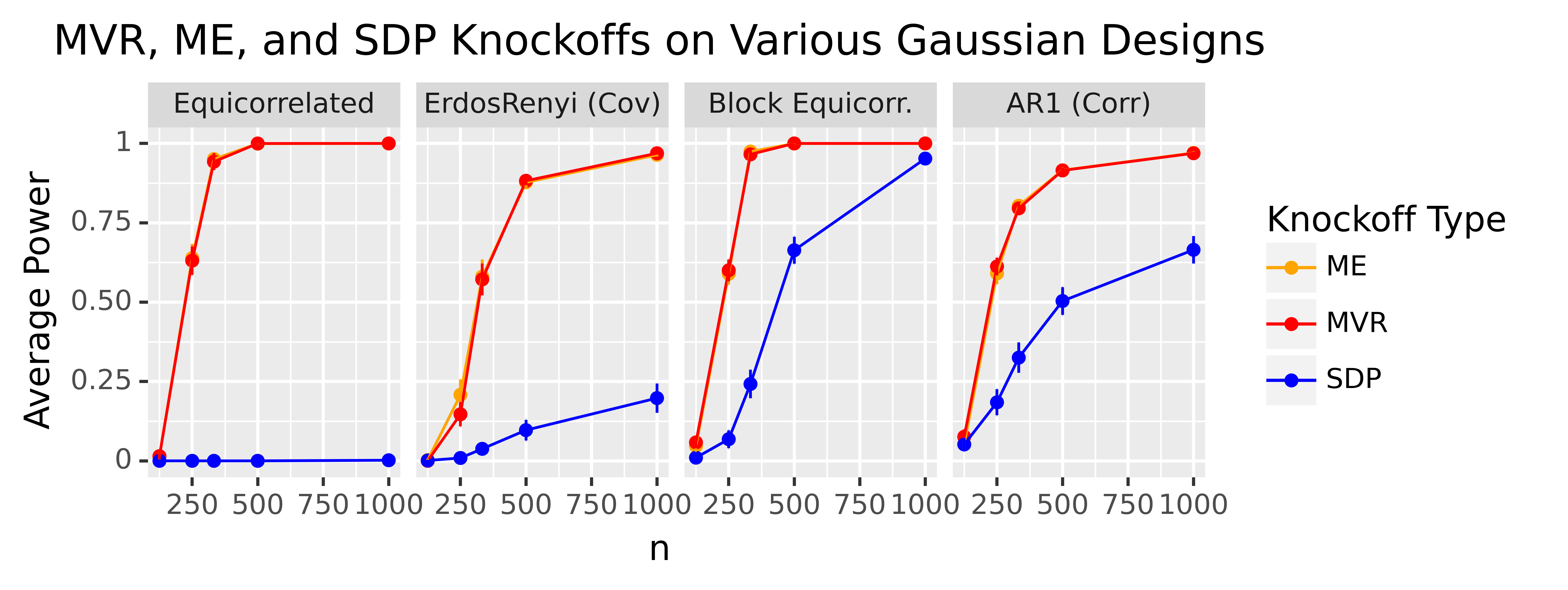}}
    \caption{This figure illustrates the main contributions of our work, namely that SDP knockoffs can have extraordinarily low power and the proposed MRC framework (MVR and \smaxent knockoffs) resolves this issue. In all cases, we sample $X \sim \mathcal{N}(0, \Sigma)$ with the different horizontal facets corresponding to different $\Sigma$, as defined in Section \ref{subsec::simgaussian}. We let $Y\mid X \sim \mathcal{N}(X \beta, 1)$ and use lasso coefficient differences as feature statistics with $p = 500$ and $50$ non-nulls. The non-nulls are sampled independently from $\mathrm{Unif}\left([-\delta, -\delta/2] \cup [\delta/2, \delta]\right)$ where $\delta = 2$ for the AR1 panel and $1$ for the others. We control the FDR at level $q = 0.1$.
    }
    \label{fig:contribution}
\end{figure}

\subsection{Contribution}

This paper describes an improved heuristic for generating powerful knockoffs. With this goal in mind, our work makes two key contributions.

\textbf{Identifying the reconstruction effect.} We prove that existing knockoff generators often create strong joint dependencies between the features and knockoffs. This allows predictive algorithms like the lasso to \textit{reconstruct} the effect of non-null features on the response using the knockoffs, which can substantially reduce statistical power. 
\begin{itemize}
    \item We prove that in a very simple setting---an exchangeable Gaussian design with correlation $\rho \ge 0.5$ and a Gaussian linear model for the response---almost every feature statistic has asymptotically zero power when used with knockoffs that minimize the MAC.
    \item We argue via both theory and simulations that minimizing the MAC frequently causes the reconstruction effect for correlated designs, and that this phenomenon will reduce power. We also identify several examples of the reconstruction effect in the previous knockoffs literature. 
\end{itemize}

\textbf{More powerful knockoffs via minimizing reconstructability (MRC).} We introduce a novel framework which generates powerful knockoffs by minimizing one's ability to reconstruct a feature using the other features and the knockoffs. We consider two concrete instantiations of this framework based on two measures of reconstructability: minimum variance-based reconstructability (MVR) knockoffs and maximum entropy (\tmaxent) knockoffs. These methods are well-defined for all design distributions, although they are particularly easy to analyze in the case when the features are Gaussian.
\begin{itemize}
    \item  We prove that when the features and response jointly follow a multivariate Gaussian distribution, MVR knockoffs exactly minimize the estimation error when using ordinary least squares (OLS) coefficients as feature importances. This means that even if MAC-minimizing knockoffs are perturbed to prevent the features from being exactly reconstructable, we still expect MVR knockoffs to have higher power.
    \item We demonstrate via simulations that MVR and \smaxent knockoffs often have dramatically higher power than MAC-minimizing knockoffs for correlated Gaussian features. As the features become less dependent, the performances of all three methods equalize, but crucially, we have not observed any examples where SDP knockoffs dramatically outperform either MVR or \smaxent knockoffs. We provide simulation results in both high and low dimensions for a wide variety of design distributions, response distributions, and feature statistics. Notably, this same conclusion holds even in our simulations with highly non-linear responses and feature statistics.
    \item We also apply the MRC framework in a variety of simulation settings beyond the Gaussian model-X case. We demonstrate MRC knockoffs can increase the power of fixed-X knockoffs \citep{fxknock}, second-order knockoffs \citep{mxknockoffs2018}, and the general Metropolized knockoff sampler for non-Gaussian features \citep{metro2019}.
\end{itemize}

Figure \ref{fig:contribution} encapsulates both of these contributions, namely that when the features are correlated, minimizing the MAC (via SDP when the features are Gaussian) often produces much-lower-power knockoffs than the MVR and \smaxent knockoffs considered in this paper.

\subsection{Notation}

Let $X = (X_1, \dots, X_p) \in \mathbb{R}^p$ be the set of $p$ features and let $Y \in \mathbb{R}$ be the response. We stack $n$ i.i.d. observations of the features and response into the design matrix $\bX \in \mathbb{R}^{n \times p}$ and the response vector $\by \in \mathbb{R}^n$. We use the non-bolded notation $(X, Y)$ to refer to an arbitrary single observation. For $p \in \mathbb{N}$, let $[p]$ denote the set $\{1, \dots, p\}$. For any subset $J \subset [p]$, we will denote $\bX_J$ as the matrix of columns of $\bX$ whose indices belong to $J$. $\bX_{\text{-}J}$ denotes all of the columns of $\bX$ whose indices do \textit{not} belong to $J$. $\bX_j$ denotes the $j$th column of $\bX$, and $\bX_{\text{-}j}$ denotes all of the columns of $\bX$ except column $j$. We let $I_p$ refer to the $p \times p$ identity. For a pair of square matrices $M_1, M_2\in \mathbb{R}^{d \times d}$, we say that $M_1 \succcurlyeq M_2$ if and only if $M_1 - M_2$ is positive semi-definite. We will let $\lambda_{\mathrm{min}}(M)$ and $\lambda_{\mathrm{max}}(M)$ denote the minimum and maximum eigenvalues of a square matrix $M$, respectively, and $\lambda_j(M)$ denotes the $j$th smallest eigenvalue of $M$. For two vectors $v \in \mathbb{R}^{k}, v' \in \mathbb{R}^{k'}$ with $k,k' \ge 1$, we let $(v, v') \in \mathbb{R}^{k + k'}$ denote their concatenation. For two matrices $M_1, M_2 \in \mathbb{R}^{n \times p}$, we will define $[M_1, M_2] \in \mathbb{R}^{n \times 2 p}$ to represent the column-wise concatenation of $M_1$ and $M_2$. For $j \in [p]$, we will let $[M_1, M_2]_{\swap(j)}$ denote the concatenation of $M_1$ and $M_2$ except that the $j$th column of $M_1$ has been swapped with the $j$th column of $M_2$. For $J \subset [p]$, $[M_1, M_2]_{\swap(J)}$ refers to the concatenation of $M_1$ and $M_2$ except with all columns $j \in J$ swapped. For any permutation $\sigma : [p] \to [p]$ and any matrix $M \in \mathbb{R}^{k \times p}$, let $\sigma(M)$ denote the same matrix as $M$ but with the columns permuted according to the permutation $\sigma$. E.g., if $\sigma(1) = 3$, then the first column of $M$ becomes the third column of $\sigma(M)$. For $J \subset [p]$, we let $- \mathbf{1}_J$ refer to the vector $v \in \mathbb{R}^p$ such that $v_j = -1$ for $j \in J$ and $v_j = 1$ when $j \not \in J$. For a scalar $k$, we let $\mathbf{1}_k$ represent the vector of all ones in $\mathbb{R}^k$. For a vector $v \in \mathbb{R}^p$, $\mathrm{diag}(v)$ represents the $p \times p$ diagonal matrix with diagonal $v$. If $M_1, \dots, M_k$ are square matrices, $\mathrm{blockdiag}(M_1, \dots, M_k)$ denotes the square block-diagonal matrix with blocks $M_1$ through $M_k$. 

\subsection{Review of model-X knockoffs}

 MX knockoffs aims to simultaneously test the hypotheses $H_j : X_j \Perp Y \mid X_{\text{-}j}$, for $j \in [p]$. If $\mathcal{H}_0 = \{j : H_j\}$ is the set of null hypotheses, the knockoffs procedure selects a set of features $\hat S$ and provably controls the false discovery rate (FDR) when the distribution of $X$ is known:
$$\text{FDR} \equiv \mathbb{E} \left(\frac{|\hat S \cap \mathcal{H}_0|}{|\hat S|} \right) \le q $$
for some prespecified $q \in [0,1]$. Note that when $\hat S = \emptyset$, we use the convention that $0/0 = 0$. The knockoffs procedure consists of three steps: constructing knockoffs, computing feature importances and feature statistics, and applying a data-dependent threshold.

\textit{Step 1: Constructing knockoffs}. First, given a feature vector $X \in \mathbb{R}^{p}$, we define knockoffs $\tilde{X} \in \mathbb{R}^{p}$ as random variables satisfying
\begin{equation} \label{eq::pairwiseexchange}[X, \tilde{X}]_{\swap(J)} \stackrel{d}{=} [X, \tilde{X}] \text{ and }   \tilde{X} \Perp Y \mid X
\end{equation}

for all $J \subset [p]$. 
This pairwise exchangeability condition implies that
\begin{equation}\label{eq::gdef}
\text{Cov}([X, \tilde{X}]) = G_S \equiv \begin{bmatrix} \Sigma & \Sigma - S \\ \Sigma - S & \Sigma \end{bmatrix} 
\end{equation}
for some diagonal matrix $S$ such that $G_S \succcurlyeq 0$, where $\Sigma = \text{Cov}(X)$. Note in the case where $X \sim \mathcal{N}(0, \Sigma)$, \citet{mxknockoffs2018} proved that when $[X, \tilde{X}] \sim \mathcal{N}(0, G_S)$, $\tilde{X}$ are valid knockoffs for $X$ for any $S$, as long as $G_S \succcurlyeq 0$. If one chooses to generate $\tilde{X}$ such that $[X, \tilde{X}]$ are multivariate Gaussian, then the choice of $S$ uniquely determines any valid knockoff-generation mechanism. 

\textit{Step 2: Feature importances and feature statistics.} The next step is to compute feature importances $Z \in \mathbb{R}^{2p}$ where for $j \in [p]$, $Z_j$ and $Z_{j+p}$ measure the importances of
$X_j$ and its knockoff $\tilde{X}_j$, respectively. We can use any function of the data $Z = z([\bX, \tilde{\bX}], \by) \in \mathbb{R}^{2p}$ to generate $Z$ under the restriction that swapping a feature with its knockoff also swaps the feature importances $Z_j, Z_{j+p}$. This allows one to use almost any feature importance measure to create $Z$, from cross-validated lasso coefficients to neural networks \citep{deeppink2018}.

We combine $Z_j, Z_{j+p}$ into a \textit{feature statistic} $W_j = f(Z_j, Z_{j+p})$, where $f$ is an antisymmetric function such that $f(x,y) = -f(y,x)$. For example, if $Z$ are absolute lasso coefficients, we might set $f(Z_j, Z_{j+p}) = Z_j - Z_{j+p}$. Here, $W_j$ represents the lasso (absolute) coefficient difference (LCD), where a high value of $W_j$ indicates that feature $X_j$ is more important than its knockoff in predicting $Y$. The antisymmetric property of $f$ implies that swapping $X_j$ with its knockoff flips the sign of $W_j$. We let $W = w([\bX, \tilde{\bX}], \by) \in \mathbb{R}^p$ be the vector of feature statistics. Note $Z, W$ are random variables, and $z, w$ are functions.

\textit{Step 3: The data-dependent threshold.} Finally, we define the data-dependent threshold
\begin{equation}\label{eq::ddthreshold}
    T = \min \left \{ t > 0 : \frac{\# \{j : W_j \le -t \} + 1}{\# \{ j : W_j \ge t \}} \le q \right\}.
\end{equation}
Formally, to ensure this minimum is well-defined, the minimum is only over $t \in \{|W_j| : j \in [p]\} \setminus \{0\}.$ By convention, $T = \infty$ if that set is empty. Selecting features $\hat S = \{ j : W_j \ge T \}$ guarantees FDR control at level $q$. 

\subsection{Related literature}

The MX knockoffs framework \citep{mxknockoffs2018} has received significant attention recently because it guarantees exact FDR control in feature selection even when the response $Y \mid X$ is nonlinear and the features $X$ are arbitrarily correlated. In particular, knockoffs has been applied successfully in genome-wide association studies (GWAS), with promising empirical performance (\citet{genehunting2017, knockoffzoom2019, popstruct2020}). Although the model-X framework does assume that the distribution of $X$ is known, prior work has shown through both theory \citep{robustness2020} and simulations (for example, \citet{genehunting2017}) that knockoffs is fairly robust to misspecification of the distribution of $X$. Additionally, a new line of research has relaxed the assumptions of knockoffs, such that knockoffs can control the false discovery rate as long as the distribution of $X$ is known up to a parametric model \citep{condknock2019}. Since a main advantage of knockoffs is that they control the false discovery rate under feature dependence, the main goal of our paper is to improve the \textit{power} of knockoffs under feature dependence. We pause to briefly compare our contribution in this respect to two relevant strands of the wider knockoffs literature.

First, our work draws heavily from the literature on sampling model-X knockoffs. Initially, \citet{fxknock} introduced the fixed-X knockoff framework and suggested creating knockoffs which minimize the MAC. \citet{mxknockoffs2018} built on \citet{fxknock} to introduce the model-X knockoffs filter, demonstrate how to sample knockoffs for Gaussian designs, and prove the existence of nontrivial knockoffs for general designs. Notably, \citet{mxknockoffs2018} also suggested generating knockoffs which minimize the MAC metric, although in their discussion they note the possibility of increased power from alternate knockoff constructions. Since then, several works have developed techniques to sample knockoffs for non-Gaussian designs. \citet{genehunting2017} developed a method to sample knockoffs for discrete Markov chains, and \citet{knockoffsmass2018} demonstrated how to sample knockoffs for some Bayesian networks. When the true model is unknown, \cite{deepknock2018, jordon2018knockoffgan} showed how to use deep generative networks and generative adversarial networks (GANs), respectively, to generate approximate knockoff constructions.
Most recently, \citet{metro2019} characterized all knockoff distributions and developed efficient algorithms to sample exact knockoffs in great generality, using tools from Markov Chain Monte Carlo (MCMC). However, very little of the existing literature has discussed \textit{which} knockoffs to generate, and most works have assumed implicitly that weaker feature-knockoff correlations improve power. Indeed, \citet{metro2019} did not even simulate the power of their general knockoff sampler---instead, they reported the MAC as a proxy for knockoff quality. The point of our paper is to demonstrate that the MAC heuristic can fail spectacularly and to propose a solution to this problem.

Second, four recent papers (\citet{altsign2017, multiknock2018, ciknockoff2019, ke2020}) have observed that knockoffs can sometimes lose power in correlated settings, and \citet{rank2017} prove the consistency of lasso-based knockoffs under certain conditions on the feature-knockoff distribution. Our work differs substantially from these, but we will be better able to explain why after we have formally introduced the reconstruction effect. 
See Section \ref{subsec::reconstlitcomp} for detailed comparisons to each of these works.

\subsection{Outline}

The outline of the rest of the paper is as follows. In Section \ref{sec::reconstruction}, we describe how minimizing the MAC causes the reconstruction effect for general Gaussian designs. We also demonstrate that reconstructability reduces the power of knockoffs for all designs. In Section \ref{sec::mvrknock}, we define two types of MRC knockoffs: MVR knockoffs and \smaxent knockoffs. We prove that MVR knockoffs are optimal in a sense for OLS coefficients in Gaussian linear models, and we further prove that MVR knockoffs are consistent in low dimensions, unlike SDP knockoffs. We also give intuition as to why MRC knockoffs are likely to improve power for nonlinear responses and discuss efficient algorithms for computing MRC knockoffs when the features are Gaussian. Finally, in Section \ref{sec::sim}, we present an extensive set of simulations comparing the power of MRC and MAC-minimizing knockoffs.

\section{The reconstruction effect reduces knockoffs' power}\label{sec::reconstruction}

The MAC heuristic applied in correlated settings can lead to significant power loss due to the reconstruction effect we identify in this section.

\subsection{Minimizing the MAC results in reconstructability for Gaussian designs}\label{subsec::rankdegen}

In this section, we demonstrate that minimizing the MAC often ensures that many $X_j$ can be reconstructed from $X_{\text{-}j}, \tilde{X}$ when $X$ is Gaussian and correlated. To begin with, note that when $[X, \tilde{X}] \sim \mathcal{N}(0, G_S)$, choosing a type of knockoffs to generate is equivalent to choosing the $S$-matrix from equation (\ref{eq::gdef}). When $X$ is Gaussian, we will refer to MAC-minimizing knockoffs as SDP knockoffs, as in the literature. When $X \sim \mathcal{N}(0,\Sigma)$, we will also assume $X$ is scaled such that $\Sigma_{jj} = 1$ for $j \in [p]$.
\begin{definition}[SDP Knockoffs]\label{def::sdp} Set $S_{\mathrm{SDP}} = \mathrm{diag}(s)$ where $s \in \mathbb{R}^p$ is the solution to the semidefinite program
\begin{align*}
    \mathrm{minimize} & \,\,\,  \sum_{j=1}^p |1 - s_j| \\
    \mathrm{s.t.} & \,\,\,  0 \preccurlyeq \mathrm{diag}(s) \preccurlyeq 2\Sigma.
\end{align*}
\end{definition}
A more computationally-efficient version of this procedure, called the equicorrelated method, minimizes the same objective under the constraint that $s$ is a constant vector. 

This SDP formulation will continue to increase the diagonal values of $S$ until it hits the boundary condition $\lambda_{\mathrm{min}}(2 \Sigma - S) = 0$ or attains the optimal $S = I_p$. Since the eigenvalues of $G_S$ are those of $S$ and $2 \Sigma - S$,  $G_{\mathrm{SDP}} \equiv G_{S_{\mathrm{SDP}}}$ will be low rank whenever $S_{\mathrm{SDP}} \ne I_p$.

\begin{lemma}\label{lem::generalrankdegen} Suppose $X \sim \mathcal{N}(0, \Sigma)$ and $\lambda_{\mathrm{min}}(\Sigma) \le 0.5$. Then $\mathrm{rank}(G_{\mathrm{SDP}}) < 2p$. Furthermore, if $\Sigma$ is block-diagonal with $b$ blocks, each with an eigenvalue below $0.5$, then $\mathrm{rank}(G_{\mathrm{SDP}}) \le 2p - b$.
\end{lemma}

As a running example throughout this section, we will often analyze the simple case where $\Sigma$ is ``equicorrelated'' to enable more explicit analysis of the power of SDP knockoffs. This means $X \sim \mathcal{N}(0, \Sigma)$ where 
$\Sigma_{jk} = \rho$ if and only if $j \ne k$ and $1$ otherwise, which implies $X$ is exchangeable.

\begin{lemma}\label{lem::rankdegen} In the equicorrelated case when $\rho \ge 0.5$, let $\tilde{X}$ be generated according to the SDP procedure. Then $G_{\mathrm{SDP}}$ has rank $p+1$, and $X_j + \tilde{X}_j = X_k + \tilde{X}_k$ for all $1 \le j, k \le p$. 
\end{lemma}

\subsection{Reconstructability and power}\label{subsec::reconstpower}

Why does the rank of $G_{\mathrm{SDP}}$ affect power? As an intuitive example, consider the equicorrelated case, where Lemma \ref{lem::rankdegen} tells us that one could reconstruct all information contained in the features $X$ simply by looking at $\tilde{X}$ and one other feature. This makes it difficult to assign feature importances to $X$ versus $\tilde{X}$. For example, suppose $Y$ follows a single-index model, meaning we can represent $Y = f(X \beta, U)$ for some deterministic function $f$ and $U \sim \mathrm{Unif}(0,1)$ independent of $X$. Then Lemma \ref{lem::rankdegen} implies that for any $J \subset [p]$, we can write 
\begin{equation}\label{eq::equireconstructability}
    X_J \beta_J = (X_1 + \tilde{X}_1) \left(\sum_{j \in J} \beta_j \right) - \tilde{X}_J \beta_J.
\end{equation}

To see why this is a problem, initially assume $\sum_{j \in J} \beta_j = 0$, which guarantees that $X_J \beta_J = - \tilde{X}_{J} \beta_J$. In this setting, the feature importances $Z = z([\bX, \tilde{\bX}], \by)$ will have difficulty distinguishing between two different models, the correct model where $Y = f([X_J, X_{\text{-}J}](\beta_J, \beta_{\text{-}J}), U)$ and an incorrect model where $Y = f([\tilde{X}_J, X_{\text{-}J}](- \beta_J, \beta_{\text{-}J}), U)$. Note that in the correct model, we would expect $\{Z_j\}_{j \in J}$ to be large and $\{Z_{j+p}\}_{j \in J}$ to be near zero, leading to large positive $\{W_j\}_{j \in J}$, but in the incorrect model, we would expect the opposite, leading to highly negative $\{W_j\}_{j \in J}$. Since both models fit the data equally well and the models differ only in some of the signs of their coefficients, $\{W_j\}_{j \in J}$ is equally likely to be positive or negative unless the feature statistic function $w$ incorporates prior information about the signs of $\beta_J$. For example, one could constrain the lasso to only assign nonnegative coefficient values to each feature. With enough data, the lasso would then correctly assign high feature importances to the set of features $\{j \in J : \beta_j > 0\}$. On the other hand, it would also ensure that the lasso would assign zero importance to features in the set $\{j \in J : \beta_j < 0\}$ and high importances to their knockoff counterparts, precluding the discovery of those features.

When $\sum_{j \in J} \beta_j \ne 0$, the incorrect model becomes $$Y = f\left(\left[X_1, \tilde{X}_1, \tilde{X}_J, X_{\text{-}J}\right]\left (\sum_{j \in J} \beta_j, \sum_{j \in J} \beta_j, - \beta_J, \beta_{\text{-}J}\right), U\right).$$
In this case, regularized feature statistics like the lasso, with enough data, may eventually identify the correct model, whose coefficients have smaller $\ell_1$ norm than the coefficients in the incorrect model. However, when the $\ell_1$ norms of both options are similar, it will take a very large amount of data for the lasso to identify the true model (see Theorem \ref{thm::avgequi}). This exemplifies a broader theme, which is that the power of common sparsity-inducing feature statistics will decline further when it is possible to reconstruct features using a \textit{sparse} subset of the other features and knockoffs. We refer to this phenomenon as \textit{sparse reconstructability}, and in the general Gaussian case, it will occur when many eigenvalues of $G_{\mathrm{SDP}}$ are close to zero.

Gaussian equicorrelated designs with a single-index response are only an example of the broader reconstruction effect, which occurs quite generally: whenever $X_j$ can be reconstructed from $X_{\text{-}j}, \tilde{X}$, feature importances such as those derived from a random forest are likely to confuse the contribution of a non-null $X_j$ with other features or knockoffs. In the worst case, however, Theorem \ref{thm::generalreconstruction} indicates that if $Y$ depends on $X_J$ through a statistic $g(X_J)$ which can be reconstructed using some function $g^*$ of $\tilde{X}_J$, no feature statistic will be able to use the data to distinguish between $X_J$ and $\tilde{X}_J$. This theorem does not assume $X$ to be Gaussian, nor does it assume $Y$ to follow a single-index model.

\begin{theorem}\label{thm::generalreconstruction}
Suppose we can represent $Y = f\left(g(X_J), X_{\text{-}J}, U \right)$ for some set $J \subset [p]$, functions $f$ and $g$, and independent noise $U \sim \mathrm{Unif}(0,1)$. Equivalently, this means $Y \Perp X_J \mid g(X_J), X_{\text{-}J}$. Suppose a function $g^*$ exists such that 
\begin{equation}\label{eq::genreconstructioncond}
g(X_J) = g^*(\tilde{X}_J)
\end{equation} holds almost surely. If $Y^* = f(g^*(X_J), X_{\text{-}J}, U)$, then
\begin{equation}\label{eq::unidentcause}
\left( [X, \tilde{X}], Y \right) \stackrel{d}{=} \left( [X, \tilde{X}]_{\swap(J)}, Y^* \right)
\, \, \, \, \, 
\text{ and }
\, \, \, \, \, 
\left( [X, \tilde{X}], Y^* \right) \stackrel{d}{=} \left( [X, \tilde{X}]_{\swap(J)}, Y\right).
\end{equation}
Furthermore, if we set $W = w([\bX, \tilde{\bX}], \by)$, and $W^* = w([\bX, \tilde{\bX}], \by^*)$, then for all $j \in J$,
\begin{equation}\label{eq::unidentconseq}
\mathbb{P}(W_j > 0) + \mathbb{P}(W_j^* > 0) \le 1.
\end{equation}
\end{theorem}

For instance, in the equicorrelated case when $Y = f(X\beta, U)$ is single-index, SDP knockoffs satisfy the assumption $g(X_J) = X_J \beta_J = - \tilde{X}_J \beta_J = g^*(\tilde{X}_J)$ when $\sum_{j \in J} \beta_j = 0$. In this setting, equation (\ref{eq::unidentconseq}) says that SDP knockoffs produce a no free lunch situation, where no feature statistic can have non-trivial power to select any $\{X_j\}_{j \in J}$ for both $Y$ and $Y^*$. Indeed, equation (\ref{eq::unidentcause}) tells us that even a limitless amount of data gives us no way to distinguish between the features and the knockoffs in $J$, since any statistic which consistently selects $X_J$ over $\tilde{X}_J$ when the response comes from $g$ faces \textit{exactly} the opposite choice when the response comes from $g^*$. Therefore, any feature statistic which has nontrivial power to select $X_J$ when the response follows $g$ will have ``negative'' power (less power than if all features in $J$ were null) when the response follows $g^*$. For example, any feature statistic which uses $X_J$ to predict $Y$ (using $g$) can do equally well using $\tilde{X}_J$ to predict $Y$ (using $g^*$). If we have no a priori bias towards $g$ or $g^*$, then we will be equally likely to select the features $X_J$ or their knockoffs $\tilde{X}_J$, making us powerless to detect any of $\{X_j\}_{j \in J}$, no matter the signal-to-noise ratio. Alternatively, a feature statistic which a priori biases towards (for example) $g$ over $g^*$ may gain power when the true model uses $g$, but as equation (\ref{eq::unidentconseq}) indicates, it will correspondingly lose power when the true model uses $g^*$. Furthermore, a preference for sparsity will not help in general. For example, when $g(X_J) = X_J \beta_J = - \tilde{X}_J \beta_J = g^*(\tilde{X}_J)$, both models have the same sparsity.

To gain intuition about when equation (\ref{eq::genreconstructioncond}) might hold, it may again be helpful to let $g(X_J) = X_J \beta_J$ for $\beta_J \in \mathbb{R}^{|J|}$. This includes (but is not limited to) the set of models where $Y \mid X$ is partially linear in $X_J$. In this case, equation (\ref{eq::genreconstructioncond}) would be satisfied if a $\beta_J^*$ exists such that $X_J \beta_J = \tilde{X}_J \beta_J^*$ holds almost surely. Although this condition may initially seem pathological, several remarks are in order here. First, even if this relation only holds approximately, reconstructability will still reduce the power of knockoffs in finite samples. For example, if one perturbed SDP knockoffs to prevent $G_{\mathrm{SDP}}$ from being \textit{exactly} low rank, we can still have $X_J \beta_J \approx \tilde{X}_J \beta_J^*$  as long as some of the eigenvalues of $G_{\mathrm{SDP}}$ are small. Second, as the approximate rank of $G_{\mathrm{SDP}}$ decreases, the approximate null space of $[X_J, \tilde{X}_J]$ will grow larger, which makes it more likely that for any $\beta_J$, there may be \textit{some} $\beta_J^*$ such that $X_J \beta_J \approx \tilde{X}_J \beta_J^*$. This corresponds to the sparse reconstructability phenomenon we identified earlier, where it is possible to reconstruct each feature $X_j$ using a small subset of the other features and knockoffs. Third, note that (\ref{eq::genreconstructioncond}) does not require $X_J$ to be reconstructable from $\tilde{X}_J$ alone---for example, in the equicorrelated case, $X_J$ is not reconstructable from $\tilde{X}_J$, but $X_J \beta_J = - \tilde{X}_J \beta_J$ anyway when $\sum_{j \in J} \beta_j = 0$. In general, certain classes of statistics $g(X_J)$ can be reconstructable from $\tilde{X}_J$ even when reconstructing $X_J$ requires joint information from $X_{\text{-}J}, \tilde{X}$. For this reason, it is important to prevent $X_J$ from being reconstructable from any of $X_{\text{-}J}, \tilde{X}$. Lastly, as $p$ grows larger, there are many more subsets $J$ for which (\ref{eq::genreconstructioncond}) can hold, especially when the distribution of $(X,Y)$ does not have any special structure. As a result, we expect this phenomenon to be more frequent in higher dimensions. 

To briefly summarize, so far, we have shown that the MAC heuristic often causes the diagonal entries of the $S$ matrix to become so large that many eigenvalues of the joint covariance matrix $G_S$ become quite small. Unfortunately, this creates strong joint dependencies in the distribution of $[X, \tilde{X}]$. In practice, this means that many features $X_j$ can largely be reconstructed from the knockoffs $\tilde{X}$ and the other features $X_{\text{-}j}$ in spite of low marginal correlations between $X_j$ and $\tilde{X}_j$. As a result, knockoff feature importances (such as lasso coefficients) may ignore a non-null feature $X_j$ and instead use $X_{\text{-}j}, \tilde{X}$ to reconstruct its effect on $Y$, dramatically reducing the power of knockoffs. In Section \ref{subsec::avgequi}, we show that the reconstruction effect can actually render knockoffs powerless in some settings, and as we will see in Section \ref{sec::sim}, this phenomenon occurs quite broadly, even when $X$ is not Gaussian.

\subsection{Main result for equicorrelated Gaussian designs}\label{subsec::avgequi}

Our main result for equicorrelated Gaussian designs and single-index response models largely follows from the intuition that as $p \to \infty$, there will often be a very large number of subsets $J$ such that $\sum_{j \in J} \beta_j \approx 0$, unless $\beta$ has very special structure. Theorem \ref{thm::generalreconstruction} tells us that when $\Sigma$ is equicorrelated, the presence of many such $J$ will dramatically reduce the power of knockoffs unless the feature statistic function $w$ encodes specific information about each of the signs of the non-null coefficients. To exclude this case, we impose a mild condition on $w$ which prevents it from treating any of the features of $X$ differently based on their position, and therefore prevents $w$ from incorporating different a priori information about features with positive and negative signs. In particular, we define a knockoff feature statistic function $w$ to be \textit{permutation invariant} if and only if for any permutation $\sigma : [p] \to [p]$, 
\begin{equation}\label{eq::defperminv}
\sigma(w([\bX, \tilde{\bX}], \by)) = w([\sigma(\bX), \sigma(\tilde{\bX})], \by).
\end{equation}
Since some feature statistics are randomized functions, our proofs also allow for equation (\ref{eq::defperminv}) to hold in distribution conditional on the data. Note almost all feature statistics in the knockoffs literature, including those derived from lasso coefficients and even neural networks, satisfy this property. To quickly define notation, we denote the average power by:
$$\power(w, \beta) = \frac{\mathbb{E}\left[\left| \hat S \cap \{j : \beta_j \ne 0\} \right|\right]}{\left| \{j : \beta_j \ne 0 \} \right|} $$
where the expectation above is over the data for a fixed feature statistic function $w$ and a fixed vector of coefficients $\beta$. As a reminder, $\hat S$ is the set of selected features. 

\begin{theorem}\label{thm::avgequi} Let $X$ be an equicorrelated Gaussian design with  correlation $\rho \ge 0.5$. Assume we can represent $Y = f(X\beta + \zeta, U)$ for $\zeta \sim \mathcal{N}(0, \sigma_0^2)$ and $U \sim \mathrm{Unif}(0, 1)$ independent of the data. Let $\mathcal{W}$ be the class of all permutation invariant feature statistic functions and suppose we aim to control the FDR at level $q \le 0.1$ using SDP knockoffs.

Sample $\betan \in \mathbb{R}^p$ uniformly from a $p$-dimensional hypercube centered at $0$ with a fixed side-length. Let $n$ be the number of data points and suppose there exists some $\epsilon > 0$ such that $n = o\left(p^{2-\epsilon}\right)$. Then as $n,p \to \infty$, 
$$ \sup_{w \in \mathcal{W}} \power(w, \betan) \stackrel{p}{\to} 0. $$
\end{theorem}

Theorem \ref{thm::avgequi} follows from a more general statement which applies to block-diagonal $\Sigma$ whose blocks are equicorrelated, which we prove in Appendix \ref{appendix::proofs}. Note that neither the precise distribution of $\beta$ nor the assumption that $q \le 0.1$ are essential, as we discuss in Appendix \ref{appendix::qconj}.

This theorem tells us that even in low-dimensional regimes, SDP knockoffs may have asymptotically zero average power, even when the feature statistic $w$ is chosen with oracle knowledge of the response model $Y \mid X$. This theorem applies to a variety of familiar single-index models, including Gaussian linear models and probit regression, which satisfy the assumption that $Y = f(X\beta + \zeta, U)$. Furthermore, since $\mathrm{Var}(\zeta) = \sigma_0^2$ can be arbitrarily small, we expect this conclusion to hold for most single-index and generalized linear models, which take the form $Y = f(X\beta, U)$.

    Of course, real datasets will rarely be exactly equicorrelated. The point of Theorem \ref{thm::avgequi} is to show that the reconstruction effect can prevent SDP knockoffs from discovering many non-null features, no matter the signal size or feature statistic. However, as we discuss at length in Section \ref{subsec::reconstpower}, the reconstruction effect can occur without equicorrelated features, and indeed, our empirical results in Section \ref{sec::sim} show that SDP knockoffs lose power in a variety of practical settings.

\subsection{SDP knockoffs infer signal magnitudes but not signs}\label{subsec::sdpopt}

To help understand why SDP knockoffs fail in the equicorrelated setting, we briefly analyze what SDP knockoffs do \textit{right}. In particular, SDP knockoffs can easily be used to detect that \textit{either} $X_j$ or $\tilde{X}_j$ has an effect on $Y$, but not to choose between them. This means that when using SDP knockoffs, non-null $W_j$ will often have the ``right'' magnitude but a negative sign.

\begin{proposition}\label{prop::sdpopt} Assume $X$ is equicorrelated and $Y \mid X \sim \mathcal{N}(X \beta, \sigma^2)$ with $n > 2p$ and correlation $\rho \ge 0.5$. Suppose we generate knockoffs such that the feature-knockoff correlation is constant and the MAC equals $\upsilon$. If $\hatbetaext \in \mathbb{R}^{2p}$ are OLS coefficients fit on $([\bX, \tilde{\bX}], \by)$, then the mean squared error of $\hatbetaext_{1:p} - \hatbetaext_{(p+1):2p}$ as an estimator of $\beta$ is increasing in $\upsilon$. 
\end{proposition}

This theorem proves that for OLS coefficients, the differences $\hatbetaext_j - \hatbetaext_{j+p}$ proxy $\beta_j$ better as the marginal correlations between $X_j$ and $\tilde{X}_j$ get smaller, apparently in support of the MAC heuristic. Unfortunately, setting $W_j = \hatbetaext_j - \hatbetaext_{j+p}$ is a poor feature statistic when $\beta_j$ may be negative, since $W_j$ will likely be negative when $\beta_j$ is negative, preventing the selection of features with negative coefficients. On the other hand, if we set $W_j = |\hatbetaext_j| - |\hatbetaext_{j+p}|$ to fix this problem, knockoffs will still have extremely low power as the MAC $\upsilon$ gets smaller because $\hatbetaext_j$ and $\hatbetaext_{j+p}$ individually will have high variance, making $W_j$ (approximately) equally likely to be positive or negative. 

Penalized regressions like lasso coefficients $\lassoext$ exhibit the same behavior. Consider how the lasso assigns coefficient values to a pair of features $\lassoext_j, \lassoext_k$ and their knockoffs $\lassoext_{j+p}, \lassoext_{k+p}$ when $\beta_j + \beta_k = 0$. As $n \to \infty$, the lasso will have to choose between at least two options:  setting $(\lassoext_j, \lassoext_{j+p}) \approx (\beta_j,0)$ and $\left(\lassoext_k,  \lassoext_{k+p}\right) \approx (\beta_k,0)$, or alternatively setting $\left(\lassoext_j, \lassoext_{j+p}\right) \approx (0, \beta_k)$ and $\left(\lassoext_k, \lassoext_{k+p}\right) \approx (0, \beta_{j})$. When using SDP knockoffs, both options deterministically have the same empirical mean-squared error as well as the same $\ell_1$ norm, so the lasso genuinely cannot choose between these two options. Moreover, this applies to \textit{every} $j,k$ such that $\beta_j + \beta_k = 0$. Note that since $W_j = \left|\lassoext_j\right| - \left| \lassoext_{j+p} \right|$ are symmetric but have large magnitudes, they actually make it more difficult to discover other features, since knockoffs can only make discoveries when the $W_j$ with large magnitudes have consistently positive signs (see equation (\ref{eq::ddthreshold}) or Figure \ref{fig:wstats} for an illustration of this).

\subsection{Relationship with the literature}\label{subsec::reconstlitcomp}

At this point, we pause to put our results on the reconstruction effect in context with the rest of the literature. To start, we note that the reconstruction effect appears to differ fundamentally from the ``alternating sign effect'' introduced by \citet{altsign2017}, who give heuristic evidence that lasso-path statistics may lose power when two features are positively correlated and their coefficients have opposite or ``alternating'' signs. For instance, the alternating sign effect only plagues certain feature statistics such as lasso-path statistics, and the initial version of \citet{altsign2017} shows that simple statistics like OLS statistics do not suffer from this effect. In contrast, Theorem \ref{thm::avgequi} shows that minimizing the MAC creates an \textit{unidentifiability} problem which makes nearly \textit{every} feature statistic asymptotically powerless in the equicorrelated case. Second, the reconstruction effect does not even rely on the presence of alternating signs in $\beta$---for example, we have shown in Theorem \ref{thm::generalreconstruction} that it can occur outside of single-index models, where no coefficient vector $\beta$ exists.

Furthermore, Theorem \ref{thm::avgequi} may seem to contradict Theorem $1$ of \citet{rank2017}, which proves that when $X$ is Gaussian with a Gaussian linear response $Y$, the power of lasso-based knockoffs approaches one asymptotically. However, that theorem assumes the regularity condition that $\lambda_{\mathrm{min}}(\Sigma)$ and $\lambda_{\mathrm{min}}(2S - S \Sigma^{-1} S)$ must be bounded away from zero. Yet Lemma \ref{lem::generalrankdegen} demonstrates that for SDP knockoffs, $G_{\mathrm{SDP}}$ and therefore its Schur complement $2S - S \Sigma^{-1} S$ will often be low rank, meaning the regularity conditions used by \citet{rank2017} will not hold. This suggests that we ought to use a different heuristic than minimizing the MAC, one that will automatically ensure $2S - S \Sigma^{-1} S$ is full rank whenever $\Sigma$ is. In the next section, we will define two heuristics which satisfy this property. Furthermore, we will prove that one of these heuristics is consistent under only the assumption that $\lambda_{\mathrm{min}}(\Sigma)$ is bounded away from zero. This result parallels Theorem $1$ of \cite{rank2017}, except we use different technical tools to avoid requiring the problematic assumption that $\lambda_{\mathrm{min}}(2S - S \Sigma^{-1} S)$ is bounded above zero. 

Like \citet{rank2017}, \citet{ciknockoff2019} observed that the consistency of knockoffs may depend on properties of the joint covariance matrix $G_S$. Although \citet{ciknockoff2019} did not analyze the power of SDP knockoffs, they proposed generating ``conditional independence'' (CI) knockoffs such that $X_j \Perp \tilde{X}_j \mid X_{\text{-}j}$, and they proved that some lasso-based feature statistics applied to CI knockoffs are consistent in Gaussian linear models. Unfortunately, the CI condition says little about whether $X_j$ is (approximately) reconstructable using \textit{joint} information from $X_{\text{-}j}, \tilde{X}$, and as a result, we show in Appendix  \ref{appendix::cicomp} that CI knockoffs actually can suffer from the same reconstructability problems that SDP knockoffs suffer from. This means that CI knockoffs may have low power in finite samples even if they are consistent asymptotically. Furthermore, CI knockoffs only exist for a restricted set of Gaussian designs, outside of which they violate the pairwise exchangeability condition (\ref{eq::pairwiseexchange}). While \citet{ke2020} recently suggested an extension of CI knockoffs to the general Gaussian case, their method involves computing $S_{\mathrm{CI}}$ without constraining $G_S \succcurlyeq 0$ and then performing a binary search to find the maximum $\gamma \in [0,1]$ such that $G_{\gamma \cdot S_{\mathrm{CI}}} \succcurlyeq 0$. Unfortunately, this new matrix $\gamma \cdot S_{\mathrm{CI}}$ lacks any conditional independence properties, and furthermore, $\gamma$ can be extremely small in practice, substantially reducing power. We discuss this more in Appendix \ref{appendix::cicomp}.

Lastly, Theorem $5.4$ of \citet{ke2020} proves that SDP (fixed-X) knockoffs match the power of an ``oracle'' procedure for a positive, block-equicorrelated correlation structure. However, \citet{ke2020} assume a fixed block-size of $\ell=2$ and vanishing sparsity, meaning that asymptotically, a vanishing proportion of the non-nulls lie in the same equicorrelated block as other non-nulls. Since the reconstruction effect for block-equicorrelated designs only occurs when multiple non-null features are correlated, it will asymptotically not occur in this regime.

\section{Minimum reconstructability knockoffs}\label{sec::mvrknock}

In Section \ref{sec::reconstruction}, we demonstrated that SDP knockoffs lack power in the equicorrelated case because they make it possible to ``reconstruct'' a feature $X_j$ using $X_{\text{-}j}, \tilde{X}$. To fix this problem, we suggest constructing $\tilde{X}$ in order to minimize the reconstructability (MRC) of each feature $X_j$ given the other features $X_{\text{-}j}$ and the knockoffs $\tilde{X}$. In this section, we describe two instantiations of this framework which can be efficiently computed when $X$ is Gaussian and perform well in a variety of settings. 

\subsection{Minimum variance-based reconstructability (MVR) knockoffs}\label{subsec::mvrdef}

One intuitive way to minimize ``reconstructability'' is to maximize the conditional variance $\Var(X_j | X_{\text{-}j}, \tilde{X})$ for each $j \in [p]$. This motivates the minimum variance-based reconstructability (MVR) knockoff construction.

\begin{definition}[Minimum Variance-Based Reconstructability (MVR) Knockoffs]\label{def::mvrdef} To sample MVR knockoffs, sample $\tilde{X} \mid X$ so as to minimize
\begin{equation}\label{eq::mvrdef}
L_{\mathrm{MVR}} = \sum_{j=1}^p \frac{1}{\mathbb{E}[\Var(X_j | X_{\text{-}j}, \tilde{X})]}.
\end{equation}
\end{definition}

MVR knockoffs minimize the \textit{inverse} expected conditional variances $\frac{1}{\mathbb{E}[\Var(X_j | X_{\text{-}j}, \tilde{X})]}$ in order to harshly penalize high levels of reconstructability and ensure $\mathbb{E}[\Var(X_j | X_{\text{-}j}, \tilde{X})] > 0$ for all $j$.
This is important because high levels of reconstructability often cause some feature statistics to have large magnitudes but negative signs, which dramatically reduces the power of knockoffs. For example, if $X_j$ is non-null and highly reconstructable, then feature statistics can easily reconstruct the effect of $X_j$ on $Y$ using $X_{\text{-}j}, \tilde{X}$, especially when the number of data points is small. When this happens, $X_j$ will likely be assigned a low feature importance and, more importantly, a knockoff variable such as $\tilde{X}_k$ or $\tilde{X}_j$ may be assigned a \textit{high} feature importance corresponding to the effect of $X_j$ on $Y$. This implies that the feature statistic $W_k$ (resp. $W_j$) may have a large magnitude but a negative sign. Ultimately, the knockoff filter can only make rejections if there exists some $t$ such that approximately $\frac{t}{1+q}$ of the $t$ feature statistics with the largest absolute values have positive signs. Thus, when even a few feature statistics have large magnitudes but negative signs, knockoffs may have very low power. We illustrate this effect in Figure \ref{fig:wstats} for equicorrelated designs. We discuss this phenomenon more in Appendix \ref{appendix::manywstats},
where we also show that the same phenomenon occurs for a diverse range of design matrices, including Gaussian Markov chains and Gaussian designs where the covariance matrix is $80\%$ sparse.

\begin{figure}[h!]
    \centering
    \makebox[\textwidth]{\includegraphics[width=\textwidth]{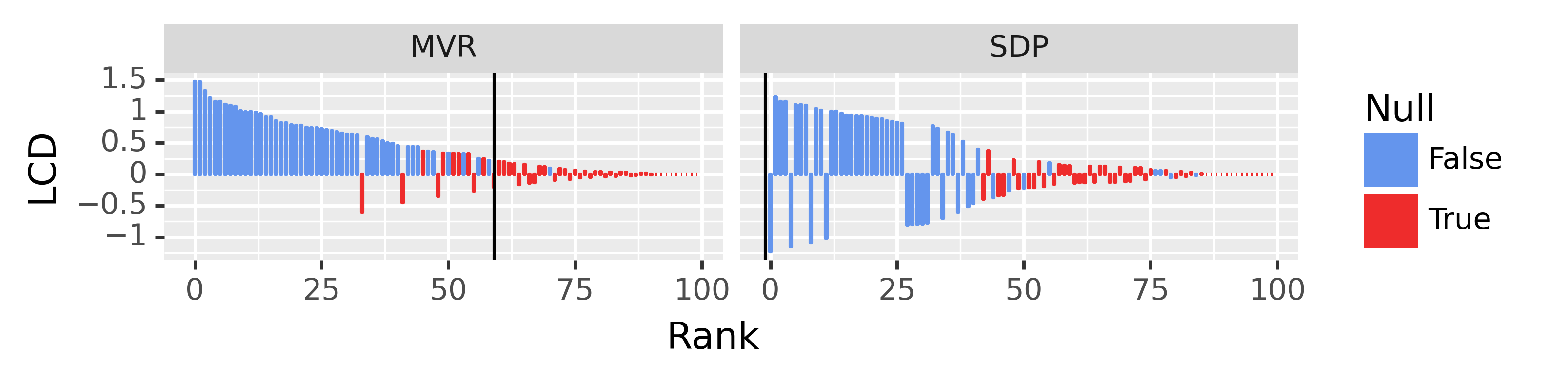}}
    \caption{$W$-statistic values for MVR vs. SDP knockoffs. We plot LCD feature statistics sorted in descending order of absolute value for an exchangeable Gaussian design with correlation $\rho = 0.6$, $n=190$, $p=100$, and $Y \mid X \sim \mathcal{N}(X \beta, 1)$ with $50$ non-null coefficients sampled as independent random signs $\pm 1$. The black line denotes the data dependent threshold $T$. SDP knockoffs have low power because many feature statistics have large magnitudes but negative signs.}    \label{fig:wstats}
\end{figure}

In the Gaussian case when $[X, \tilde{X}] \sim \mathcal{N}(0, G_S)$, we can write $L_{\mathrm{MVR}}$ as a simple and convex function of the $S$-matrix. In particular, $L_{\mathrm{MVR}}(S) = \frac{1}{2} \text{Tr}(G_S^{-1})$, as $1 / \Var(X_j | X_{\text{-}j}, \tilde{X}) = (G_S^{-1})_{j,j}$ for Gaussian features (see \citet{anderson2009}, Section $2.5$). See Appendix \ref{appendix::convexity} for a proof of convexity. This convenient formulation allows us to prove two appealing properties of MVR knockoffs when $X \sim \mathcal{N}(0, \Sigma)$, $Y \mid X \sim \mathcal{N}(X \beta, \sigma^2)$, and we use the absolute value of OLS coefficients as feature importances.

First, we prove a type of optimality result for MVR knockoffs. Although we would like to know which $S$-matrix maximizes power to detect non-nulls, the exact answer depends on the unknown coefficients $\beta$. As a proxy for this, however, we note that the power of knockoffs almost entirely depends on the accuracy of the OLS coefficients $\hatbetaext \in \mathbb{R}^{2p}$. As $n \to \infty$, we hope that $\hatbetaext_{1:p}$ will converge to $\beta$ and the knockoff feature importances $\hatbetaext_{(p+1):2p}$ will converge to $0$. In finite samples, it turns out that for \textit{any} $\Sigma$ and $\beta$, MVR knockoffs minimize the mean-squared error between $\hatbetaext$ and its target.

\begin{proposition}\label{prop::mvrolsopt} Suppose $X \sim \mathcal{N}(0, \Sigma)$ for any $\Sigma$ and $Y\mid X \sim \mathcal{N}(X \beta, \sigma^2)$. Let $\betaext \in \mathbb{R}^{2p}$ be the concatenation of $\beta \in \mathbb{R}^p$ with $p$ zeros. Suppose $\hatbetaext \in \mathbb{R}^{2p}$ are OLS coefficients fit on $([\bX, \tilde{\bX}], \by)$ and $n > 2p + 1$. Then
$$S_{\mathrm{MVR}} = \arg \min_S \mathbb{E}[||\hatbetaext-\betaext||_2^2].$$
\end{proposition}

Second, we prove that the MVR knockoff filter is consistent in low-dimensions for OLS feature statistics. This result would not be surprising except that we showed in Section \ref{sec::reconstruction} that in some low-dimensional settings, SDP knockoffs is inconsistent for \textit{every} feature statistic. 

\begin{theorem}\label{thm::olspower1} Suppose $X \sim \mathcal{N}(0, \Sigman)$, $Y\mid X \sim \mathcal{N}(X \betan, \sigma^2)$, and $\tilde{X}$ is generated using $S_{\mathrm{MVR}}$. Suppose $\hatbetaext \in \mathbb{R}^{2p}$ are OLS coefficients fit on $([\bX, \tilde{\bX}], \by)$, and set $w([\bX, \tilde{\bX}],\by) = |\hatbetaext_{1:p}| - |\hatbetaext_{(p+1):2p}|$.

Let $n, p \to \infty$ such that $p = o(n)$ and consider a sequence of covariance matrices $\Sigman \in \mathbb{R}^{p \times p}$ such that the minimum eigenvalue of $\Sigman$ is bounded above a fixed constant $\gamma \in \mathbb{R}^+$. Suppose we sample a sequence of random $\betan$ as follows. Let all but a uniformly drawn subset of $\lceil s_0 p \rceil$ entries of $\betan$ equal zero, for a fixed constant $s_0 \in (0, 1]$, and then sample the remaining (non-null) entries of $\betan$ from a $\lceil s_0 p \rceil$-dimensional hypercube centered at $0$ with any fixed side-length. Then 
$$\power(w, \betan) \stackrel{p}{\to} 1. $$
\end{theorem}

Note that the equicorrelated case discussed in Section \ref{sec::reconstruction} satisfies these regularity conditions on $\Sigman$, since its eigenvalues are bounded uniformly above $1 - \rho$. As noted in Section \ref{subsec::reconstlitcomp}, this theorem proves a similar consistency result to \cite{rank2017}, except our proof uses properties of MVR knockoffs to avoid assumptions about the minimum eigenvalue of $G_S$.

    This result should also make intuitive sense in the context of \citet{ciknockoff2019}, who show that knockoffs are consistent if and only if (informally) the diagonals of $\frac{1}{n} G_S^{-1}$ converge to $0$.     Technically, \cite{ciknockoff2019}'s theory does not directly apply to our setting, as it assumes a different asymptotic regime and relies on regularity conditions which the authors themselves admit are ``highly nontrivial" to verify unless $\Sigma^{(n)}$ is block-diagonal. We use different technical tools to avoid these assumptions. Conceptually, however, the key novelty of Theorem \ref{thm::olspower1} is that it shows for the first time that a concrete method---MVR knockoffs---can achieve \cite{ciknockoff2019}'s condition with minimal assumptions on $\Sigma^{(n)}$. To summarize, \cite{ciknockoff2019}'s theory established the importance of making the diagonals of $G_S^{-1}$ small, and MVR knockoffs explicitly follows this principle by minimizing $\text{Tr}(G_S^{-1})$, allowing us to prove its consistency in more generality than has been established for any other knockoff generation method.

Lastly, we note that the convexity of $L_{\mathrm{MVR}}$ in the Gaussian case allows us to develop an algorithm to compute $S_{\mathrm{MVR}}$ in $O(n_{\mathrm{iter}} p^3)$, which is the same time complexity or faster than the methods used to compute SDP knockoffs \cite{fanok2020}. Our method is inspired by \citet{fanok2020}, who introduced a coordinate-descent algorithm 
to compute $S_{\mathrm{SDP}}$ efficiently. The key idea is to use rank-one updates to maintain a running Cholesky decomposition of $2 \Sigma - S$. Our algorithm uses this same strategy, although extending their ideas to the MVR loss requires some nontrivial additional analysis. For brevity, we defer the details to Appendix \ref{appendix::computation}. Since the algorithms used to compute MVR and SDP knockoffs are extremely similar, we show in simulations in Appendix \ref{appendix::runtime} that not only do they have the same computational complexity, but they have very similar runtimes in practice. Additionally,
like the SDP formulation (\ref{def::sdp}), our algorithm can take advantage of block-diagonal approximations \citep{mxknockoffs2018} or low-rank factor structure in $\Sigma$ \citep{fanok2020} to dramatically speed up computations to be linear in $p$---see Appendix \ref{subsec::blockdiagapprx} 
for additional details. The overall point is that there is no computational reason to prefer either MVR or SDP knockoffs.

\subsection{Maximum entropy (\tmaxent) knockoffs}\label{subsec::maxentdef}

An information-theoretic alternative to MVR knockoffs is to minimize the mutual information between $X$ and $\tilde{X}$; as we shall see in a moment, this is equivalent to maximizing the entropy of $[X, \tilde{X}]$. \citet{multiknock2018} have previously considered \smaxent knockoffs, but only in the Gaussian case, and they introduce this method for a very different reason than we do. In particular, the authors motivate \smaxent knockoffs by observing that the SDP knockoff construction can induce sparsity in the diagonal of $S_{\mathrm{SDP}}$, meaning that some values of the diagonal of $S_{\mathrm{SDP}}$ become too small to distinguish between features and knockoffs. Although they demonstrate that \smaxent knockoffs will not suffer from this problem, they only compare the power of the two methods in a single case where the distributions of both $X$ and $Y \mid X$ are chosen adversarially against SDP knockoffs. In contrast, we advocate the use of \smaxent knockoffs to solve almost exactly the opposite problem: that the values of $S_{\mathrm{SDP}}$ are frequently \emph{too large}. Indeed, our most dramatic empirical results come in the equicorrelated case, where $S_{\mathrm{SDP}} = \gamma \cdot I_p$ for $\gamma$ as large as $1$. Clearly, in this case, the diagonal of $S_{\mathrm{SDP}}$ is not sparse at all. 

\begin{definition}[Maximum Entropy (\tmaxent) Knockoffs]\label{def::medef} Suppose $X$ is absolutely continuous on $\mathcal{X}$ with respect to a base measure $\mu$ with density $p(x)$. To sample \smaxent knockoffs, sample $\tilde{X} \mid X$ so as to minimize
\begin{equation}\label{eq::maxentdef}
L_{\mmaxent} = \int_{x \in \mathcal{X}} \int_{\tilde{x} \in \mathcal{X}} p(x, \tilde{x}) \log \left( p(x, \tilde{x}) \right) \text{d} \mu(\tilde{x}) \, \, \text{d} \mu(x), 
\end{equation}
where $\mathcal{X}$ is the support of $X$, and $p(x,\tilde{x})$ is the joint density of $[X, \tilde{X}]$. Note $L_{\mmaxent}$ corresponds to the negative entropy of $[X, \tilde{X}]$. If $[X, \tilde{X}]$ admits no joint density with respect to the product measure $\mu \times \mu$, we adopt the convention that $L_{\mmaxent} = \infty$.
\end{definition}

Given this definition, it is not immediately obvious that \smaxent knockoffs minimize any notion of reconstructability. Of course, the entropy of $[X, \tilde{X}]$ equals twice the entropy of $X$ minus the mutual information between $X$ and $\tilde{X}$, so \smaxent knockoffs also minimize the mutual information between $X$ and $\tilde{X}$. Since mutual information can account for joint dependencies between $X$ and $\tilde{X}$, we might expect it to perform better than the MAC metric, which only looks marginally at dependencies between $X_j$ and $\tilde{X}_j$. However, mutual information still may not necessarily capture the right notion of reconstructability. To perform feature selection with knockoffs, one must assign a feature importance to each individual feature and knockoff. To do this powerfully, we have argued that \emph{each} non-null feature $X_j$ must contain information that cannot be reconstructed using $X_{\text{-}j}$ and $\tilde{X}$. Although mutual information captures some notion of the aggregate dependencies between $X$ and $\tilde{X}$, it is not clear it coincides with this feature-level definition of reconstructability. 

To better connect ME knockoffs with reconstructability, we turn to the setting where $X$ is Gaussian, where $L_{\mmaxent}$ has a simple formulation. In particular, we can take $\tilde{X}$ to be jointly Gaussian with $X$, and then the entropy of $[X, \tilde{X}]$ equals $\log \det (G_S)$ up to a constant. Thus, maximizing the entropy of $[X, \tilde{X}]$ corresponds to minimizing $\log \det (G_S^{-1})$. This loss is actually quite similar to the MVR loss for Gaussian $X$, as both losses are convex, elementwise-decreasing functions of the eigenvalues of $G_S$. In particular,
\begin{equation}\label{eq::gaussianmvrmmicomp}
L_{\mathrm{MVR}}(S) \propto \text{Tr}(G_S^{-1}) = \sum_{j=1}^{2p} \frac{1}{\lambda_j(G_S)} \, \, \,  \text{ and }  \, \, \, L_{\mmaxent}(S) = \log \det(G_S^{-1}) = \sum_{j=1}^{2p} \log\left(\frac{1}{\lambda_j(G_S)}\right),
\end{equation}
where we express the $L_{\mmaxent}$ loss up to an additive constant. As a result, $S_{\mmaxent}$ and $S_{\mathrm{MVR}}$ are quite similar in the Gaussian case, suggesting that $L_{\mmaxent}$ captures a similar notion of reconstructability in this setting. This also indicates that \smaxent and MVR knockoffs will have very similar power in the Gaussian setting, which we confirm in Section \ref{sec::sim}. In Appendix \ref{appendix::computemaxent},
we modify the algorithm which computes $S_{\mathrm{MVR}}$ to compute $S_{\mmaxent}$ in the same time complexity.

Outside the Gaussian case, we cannot write $L_{\mmaxent}$ so simply, but we do prove in Lemma \ref{lem::mminondegen} that when $X$ has finite support, \smaxent knockoffs will not allow any feature to be perfectly reconstructable as long as any knockoff procedure can accomplish this. Intuitively, this result holds because if $X_j$ is nearly a deterministic function of $X_{\text{-}j}, \tilde{X}$, then almost all of the probability density of $[X, \tilde{X}]$ will lie along a $(2p-1)$ dimensional subset of $\mathcal{X} \times \mathcal{X}$. To ensure the density $p(x, \tilde{x})$ integrates to one over $\mathcal{X} \times \mathcal{X}$, the average log value of $p(x, \tilde{x})$ must become larger than is optimal. See Appendix \ref{appendix::discretemmi} 
for a more detailed discussion of the discrete case as well as a proof of Lemma \ref{lem::mminondegen}.

\begin{lemma}\label{lem::mminondegen} Let $X$ have finite support $\mathcal{X}$. For any $j \in [p]$ and any $x, \tilde{x} \in \mathcal{X} \times \mathcal{X}$, \smaxent knockoffs satisfy $\Var(X_j | X_{\text{-}j}=x_{\text{-}j}, \tilde{X}=\tilde{x}) > 0$, so long as this property does not contradict the definition of valid knockoffs.
\end{lemma}

At this point, one might wonder whether there are clear grounds to prefer MVR knockoffs over \smaxent knockoffs or vice versa. Our empirical results in Section \ref{sec::sim} suggest MVR and \smaxent knockoffs perform very similarly in the Gaussian case. There is more theoretical work to be done to better understand their relationship in general, but we can make two comments summarizing what we know (and do not know) so far. First, \smaxent knockoffs seem to minimize some notion of reconstructability in all examples we can tractably analyze, but it is not clear that our arguments generalize beyond the Gaussian case. We include \smaxent knockoffs in this section for the reader's consideration because they seem promising and they perform well empirically in Section \ref{sec::sim}. Second, MVR knockoffs are appealing because they enjoy exact optimality properties with OLS feature importances and are more readily proved to be consistent, as shown in Section \ref{subsec::mvrdef}. Note that since \smaxent knockoffs are not identical to MVR knockoffs, they do not enjoy this exact optimality, and it is difficult to prove the consistency of \smaxent knockoffs for general sequences of $\Sigma$.  That said, the solutions to MVR and \smaxent knockoffs are asymptotically identical in the case where $\Sigma$ is equicorrelated, so consistency and (approximate) optimality do hold for \smaxent knockoffs for exchangeable Gaussian designs---see Appendix \ref{appendix::equimvreqmmi}
for a precise statement. 

\section{Empirical results} \label{sec::sim}

In this section, we run an extensive set of simulations to demonstrate the power of MVR and \smaxent knockoffs. To do this, we developed a new open source python package \package, which implements a host of methods from the knockoffs literature, including a wide variety of feature statistics and knockoff sampling mechanisms for both fixed-X and model-X knockoffs. For example, \package\ includes a fully general Metropolized knockoff sampler \citep{metro2019} which can be multiple orders of magnitude faster than previous implementations. Most importantly, \package\ is written to be modular, so researchers can easily tweak its functionality or add other features on top of it.  We discuss \package\, further in Appendix \ref{appendix::knockpy}. 
All additional code for our simulations is available at \url{https://github.com/amspector100/mrcrep}.

We aim to demonstrate that the MRC framework offers real advantages over MAC-minimizing knockoffs in very practical settings, even when minimizing the MAC does not result in exact reconstructability. At the outset, we highlight some key conclusions.

\underline{Power for linear responses}: Both MVR and \smaxent knockoffs generally outperform SDP knockoffs in the setting where the design $X$ is Gaussian and $Y \mid X$ is linear. This result holds for a range of highly correlated covariance matrices and feature statistics, although as the features become less correlated, the performances of all three methods tend to equalize. We note further that MRC knockoffs frequently outperform their SDP counterparts by very large margins (as much as $100$ percentage points). Although there are examples where SDP knockoffs outperform MVR and \smaxent knockoffs, these are rare, and even in such cases, SDP knockoffs usually outperform MVR and \smaxent knockoffs by only a small margin. 

\underline{Power for nonlinear responses}: The preceding paragraph's conclusions \textit{also} hold when the conditional distribution $Y \mid X$ is highly nonlinear, even in cases where $Y$ does not follow a single-index model and linear feature statistics such as the lasso have zero power. Even very complex feature statistics like random forests with swap importances \citep{knockoffsmass2018} or DeepPINK \citep{deeppink2018} perform better with MVR and \smaxent knockoffs.

\underline{MVR vs. \smaxent knockoffs}: As our theory predicts, MVR and \smaxent knockoffs perform similarly in the Gaussian setting, although there are a few cases where \smaxent knockoffs slightly outperform MVR knockoffs. From now on, we use ``MRC knockoffs'' to refer to both MVR and \smaxent knockoffs at once.

\underline{Robustness}: Although we did not study robustness theoretically, we found empirically that MRC knockoffs can be both powerful and robust in the setting where $X$ is Gaussian but the covariance matrix $\Sigma$ is unknown and estimated using the data. In particular, in the high-dimensional setting where $\hat \Sigma$ is estimated with a shrinkage estimator, MRC knockoffs appear to violate FDR control less severely than SDP knockoffs, although we have no theoretical reason to believe that MRC knockoffs are in general more robust than SDP knockoffs.

\underline{Power for fixed-X (FX) knockoffs}: It is straightforward to define MRC knockoffs for the fixed-X knockoff filter of \citet{fxknock}. We demonstrate that MRC FX knockoffs are generally more powerful than their SDP counterparts.

\underline{Power for non-Gaussian designs}: MRC-inspired knockoffs even perform well in the setting where the features $X$ are not Gaussian, both for second-order knockoffs and as a guide for the Metropolized knockoff sampler. We will carefully define these generalizations in Section \ref{subsec::simnongaussian}.

In general, we will only plot average power, since knockoffs provably control the FDR. The exception to this, of course, is when we discuss the robustness of the knockoffs procedure. Unless otherwise specified, we use knockoffs to control the FDR at level $q = 0.1$. In all examples, we use Algorithms 1 and 2, detailed in Appendix \ref{appendix::computation},
to compute MVR and \smaxent knockoffs. In all instances, our plots include two standard deviation error bars, although in some cases the error bars are so small they are difficult to see. We also always plot both MVR and \smaxent knockoffs, although often the two methods have nearly identical performance, causing their power curves to entirely overlap.

\subsection{Simulations on equicorrelated designs}\label{subsec::simequi}

\begin{figure}[h]
    \centering
    \makebox[\textwidth]{\includegraphics[width=\textwidth]{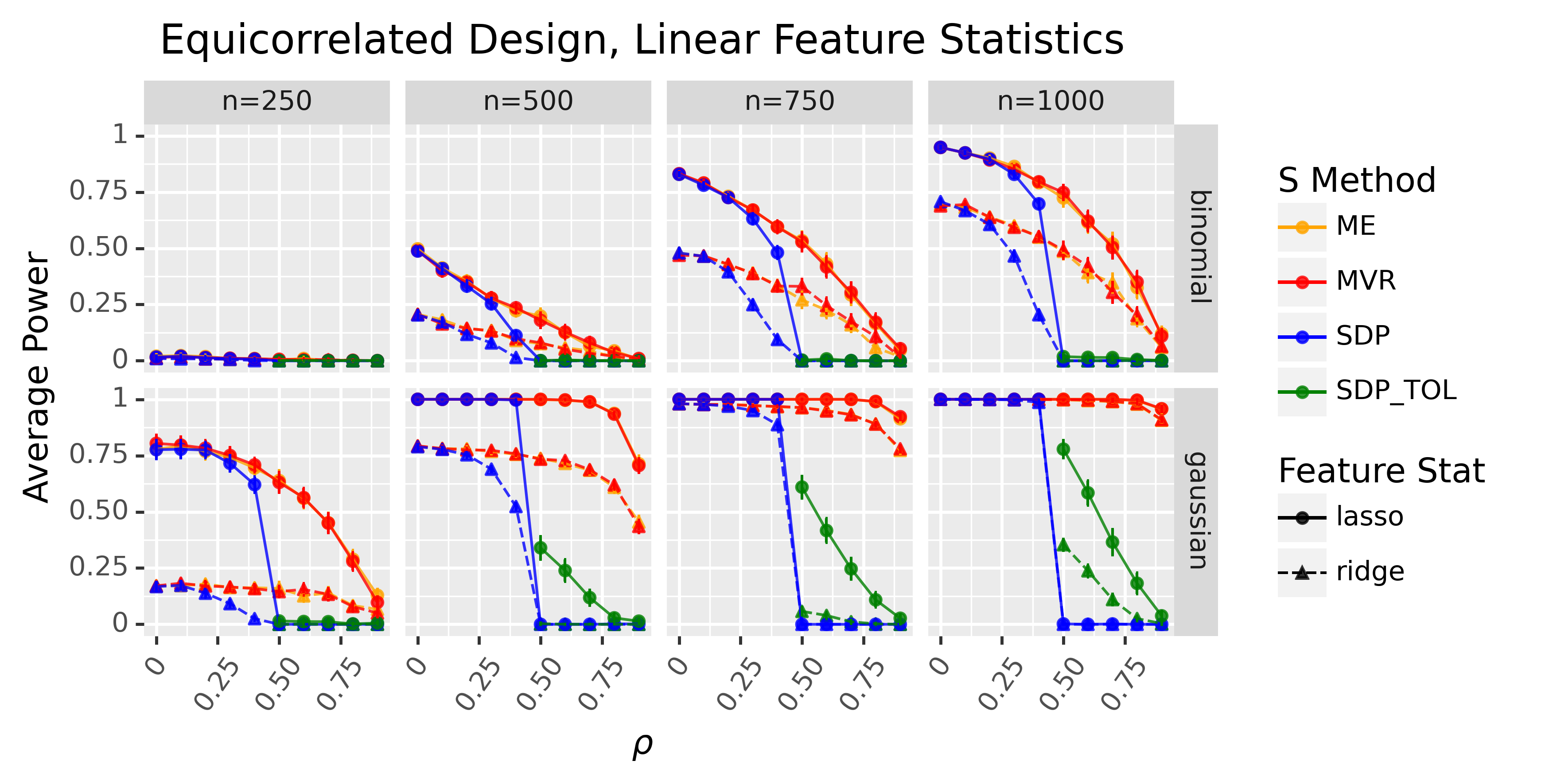}}
    \caption{Empirical powers when $X$ is an exchangeable Gaussian design. We set $p = 500$ and vary the correlation $\rho$ between $0$ and $0.9$ with the number of data points $n$ on the x-facets. In the lower panel, $Y\mid X \sim \mathcal{N}(X \beta, 1)$ and in the upper panel $Y\mid X \sim \text{Bin}(\pi(X \beta))$ where $\pi$ is the sigmoid function. In all cases the number of non-nulls is $50$ and the non-null values are sampled independently from $\mathrm{Unif}\left([-1, -1/2] \cup [1/2, 1] \right)$. We use lasso and ridge coefficient differences (and the logistic versions in the binomial case) and compare the performance of MVR, \tmaxent, and SDP knockoffs. Note SDP\_TOL knockoffs modify the SDP algorithm so that $\lambda_{\mathrm{min}}(G_S) \ge 0.01$---see equation (\ref{eq:perturbedsp}).}
    \label{fig:equiplot1}
\end{figure}

In this subsection, we investigate the performance of MRC and SDP knockoffs in the equicorrelated setting studied in Section \ref{sec::reconstruction}. In Figure \ref{fig:equiplot1}, we run simulations where $X \sim \mathcal{N}(0, \Sigma)$ with $\Sigma$ equicorrelated and $\rho$ varied between $0$ and $0.9$ for both a Gaussian and logistic linear response (see the caption for details). We compare four types of knockoffs: MVR, \tmaxent, SDP, and a perturbed version of SDP knockoffs where we set
\begin{equation}\label{eq:perturbedsp}
S_{\text{TOL}} = \gamma \cdot S_{\mathrm{SDP}}.
\end{equation}
We refer to this version as the ``sdp\_tol'' option because it ensures that the minimum eigenvalue of $G_{\text{TOL}}$ is above a tolerance $1 - \gamma$. When $\rho \ge 0.5$, we set $\gamma = 0.99$ to ensure that $\lambda_{\mathrm{min}}(G_{\text{TOL}}) \ge 0.01$. We only do this in the case where $\rho \ge 0.5$, since otherwise $G_{\mathrm{SDP}}$ will be full rank anyway. As our feature statistics, we use cross-validated lasso and ridge absolute coefficient differences of the form $W_j = |\hatbetaext_j| - |\hatbetaext_{j+p}|$, where $\hatbetaext_k$ refers to the $k$th estimated lasso or ridge coefficient, respectively, for $k \in [2p]$.

Figure \ref{fig:equiplot1} demonstrates that MRC knockoffs substantially outperform SDP knockoffs, even when $G_{\mathrm{SDP}}$ is not low rank. First, both MRC knockoff types uniformly outperform the perturbed SDP knockoffs, even though the perturbed SDP knockoffs outperform the exact SDP knockoffs. 
Moreover, \textit{even when $\rho < 0.5$} and $G_{\mathrm{SDP}}$ is full rank, MRC knockoffs can have much more power than SDP knockoffs: for example, when $\rho = 0.4$ and $n = 750$, logistic ridge has over twice the power when applied to MRC knockoffs versus SDP knockoffs. We note that this power gain in highly correlated settings does not come at the cost of power in settings with weaker correlations, since MRC knockoffs outperform or match the power of SDP knockoffs for every value of $\rho$.

Note that MVR and \smaxent knockoffs provably yield the same solution asymptotically in this setting (see Appendix  \ref{appendix::equimvreqmmi}), 
which is why their power curves almost entirely overlap in Figure \ref{fig:equiplot1}.

Since the perturbed SDP knockoffs outperform the SDP knockoffs, one might wonder how the choice of $\gamma$ affects power. To analyze this question, we perform a line search over all $S$-matrices which can be represented as a scaled identity matrix. In particular, we set 
\begin{equation}\label{eq:gammadef}
    S_{\gamma} = \gamma \cdot (2 - 2 \rho) \cdot I_p.
\end{equation}
Here, we can think of $\gamma$ as interpolating between the minimum possible $S$-matrix where $S = 0 \cdot I_p$ and the ``maximum'' $S$-matrix, where $S = 2 \lambda_{\mathrm{min}}(\Sigma) \cdot I_p = (2 - 2\rho) \cdot I_p$. The results indicate that up to Monte Carlo error, MVR and \smaxent knockoffs have the highest power among all such $S$-matrices in this setting for both lasso and ridge feature statistics. Notably, this result holds for all $\rho \in \{0.1, 0.3, 0.5, 0.7, 0.9\}$, whereas SDP knockoffs lose power compared to MVR knockoffs for all $\rho$ except $\rho = 0.1$, where all methods have nearly the same performance. For brevity, we present these results in Appendix \ref{appendix::equicorrelated}.

Lastly, we note that two previous papers \citep{xing2019, dai2020} have observed that in the equicorrelated case, knockoffs have surprisingly low power when compared to other feature selection methods. Since these papers used the SDP formulation, in  Appendix \ref{appendix:replications}, we exactly replicate these simulations but use MRC knockoffs instead. Our results show that MRC knockoffs have comparable or higher power than all of the other methods used in these papers.

\subsection{Simulations on Gaussian designs}\label{subsec::simgaussian}

In the previous section, we investigated the performance of MRC and SDP knockoffs on equicorrelated Gaussian designs with linear responses. In this section, we let $X \sim \mathcal{N}(0, \Sigma)$ but vary $\Sigma$ extensively. Furthermore, we allow the conditional distribution $Y \mid X$ to be highly nonlinear.

We defer a precise description of the covariance matrices $\Sigma$ in our simulations to Appendix 
\ref{appendix::simgaussiandescription}.
However, we will give a brief overview of the types of $\Sigma$ in order to emphasize that these $\Sigma$ differ substantially from the equicorrelated case. In the ``AR1'' setting, for example, $X$ is a standardized Gaussian Markov chain with correlations $\text{Cov}(X_j, X_{j+1})$ sampled from $\text{Beta}(3,1)$.  The ``AR1 (Corr)'' is the same setting except we cluster the non-nulls together along the chain, as we might expect in, e.g., genetic studies. Then, in the ``ER (Cov)'' and ``ER (Precision)'' settings, the covariance and precision matrices (respectively) are $80 \%$ sparse, where the nonzero entries are chosen uniformly at random, in accordance with an ErdosRenyi (ER) procedure. Lastly, we include simulations when $\Sigma$ is block-equicorrelated with a block size of $5$ and within-block correlations of $\rho = 0.5$.

\begin{figure}[h!]
    \centering
    \makebox[\textwidth]{\includegraphics[width=\textwidth]{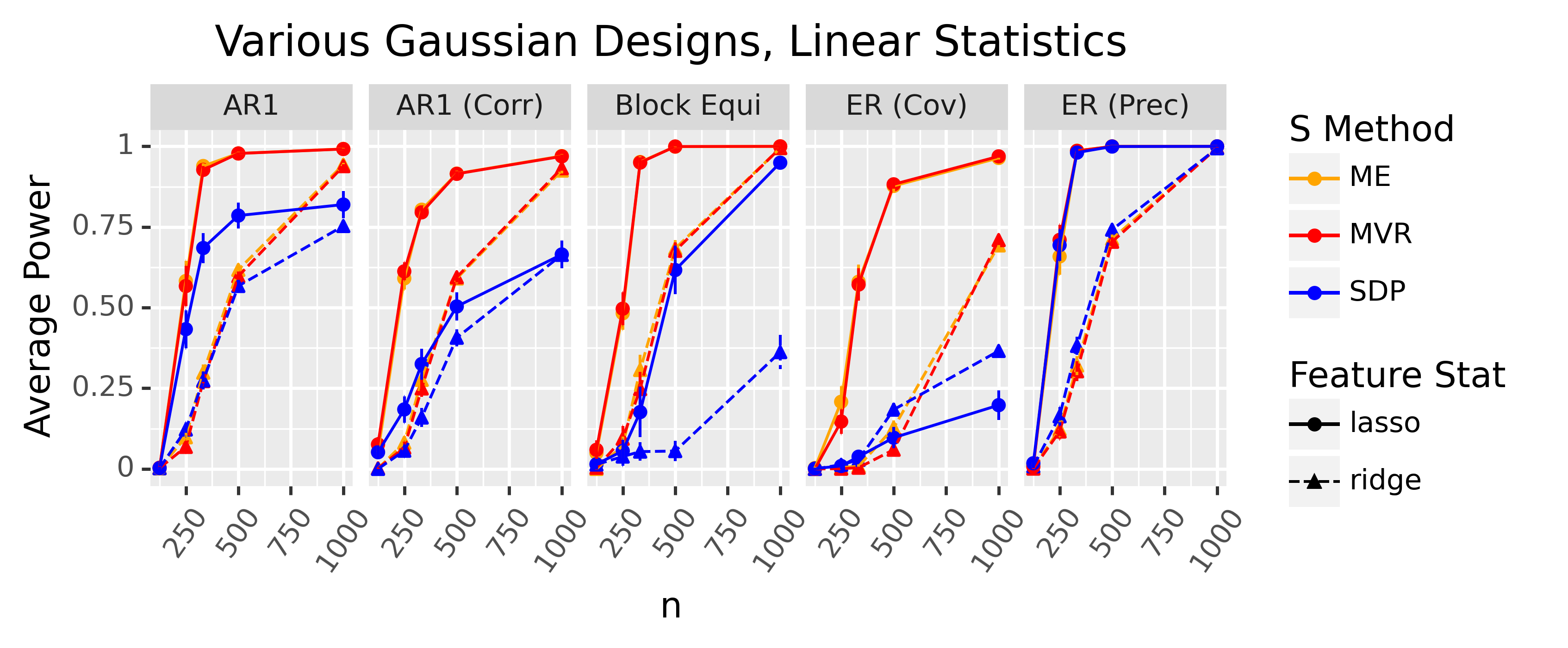}}
    \caption{Gaussian designs with a linear response. We set $X \sim \mathcal{N}(0, \Sigma)$ with $\Sigma$ as defined in Appendix \ref{appendix::simgaussiandescription}.
    and let $Y \mid X \sim \mathcal{N}(X \beta, 1)$. In all cases $p = 500$ and there are $50$ non-null with coefficients sampled independently from $\mathrm{Unif}(\left[-\delta, -\delta/2] \cup [\delta/2, \delta]\right)$, with $\delta = 2$ for the AR1 plots and $\delta = 1$ for the others.}
    \label{fig:gaussianlinear}
\end{figure}

First, in Figure \ref{fig:gaussianlinear}, we compare the power of MVR, \tmaxent, and SDP knockoffs when $Y \mid X \sim \mathcal{N}(X \beta, 1)$ for sparse $\beta$ (see the caption for details). We use cross-validated lasso and ridge coefficient differences as feature statistics, and once again, we find that both MVR and \smaxent knockoffs tend to substantially outperform SDP knockoffs. 

In Appendix \ref{appendix::ar1corr}, 
we also explore the effect of varying the between-feature correlations in the AR1 setting. As expected, higher correlations improve the performance of MRC knockoffs relative to SDP knockoffs, although interestingly, when the correlation $\text{Cov}(X_j, X_{j+1})$ is constant over all $j$, the performances of all methods are quite similar. This result is broadly consistent with our theory, although we defer discussion to Appendix \ref{appendix::ar1corr}.

Next, we consider the case where $Y \mid X \sim \mathcal{N}(\mu(X), 1)$ where $\mu$ is a nonlinear sparse model. We defer precise descriptions of these $\mu$ to Appendix \ref{appendix::simgaussiandescription}, 
but we emphasize that they are highly nonlinear and $Y \mid X$ never follows a single-index model. For example, in the ``pairint'' setting, $\mu(X) = \sum_{j,k \in [p]} \beta_{j,k} X_j X_k$ for a sparse $\beta$. Similarly, in the ``cos'' setting, $\mu(X) = \text{cos}(X) \beta$, and since $\cos$ is an even function, this ensures the features have no linear effect on $Y$. For this reason, linear feature statistics like the lasso frequently have zero power even when $n$ is as large as $15p$, as demonstrated in Appendix \ref{appendix::nonlinearstats}.

Since the conditional distribution $Y \mid X$ is very complicated, knockoffs will not have much power except in fairly low-dimensional settings. For example, when $Y$ is a linear response to pairwise interactions among the features, even a parametric feature statistic that searches explicitly for pairwise interactions must estimate $O(p^2)$ coefficients. In contrast, a single-index model only requires estimation of $O(p)$ parameters. As a result, we set $p = 200$ with $30$ non-null values and vary $n$ between $200$ and $3000$. 

We consider three feature statistics: a random forest feature statistic using the swap importance suggested by \citet{knockoffsmass2018}, the DeepPINK feature statistic from \citet{deeppink2018}, and, as a baseline, the lasso coefficient difference. We report results for the random forest statistic in Figure \ref{fig:gaussianrfplot}. Results for the DeepPINK feature statistic and the lasso feature statistic are in Appendix \ref{appendix::nonlinearstats}.
In all three cases, MVR and \smaxent knockoffs consistently outperform SDP knockoffs (although there are again a few exceptions to this, such as the truncated linear conditional mean on equicorrelated designs). 

\begin{figure}[h!]
    \centering
    \makebox[\textwidth]{\includegraphics[width=\textwidth]{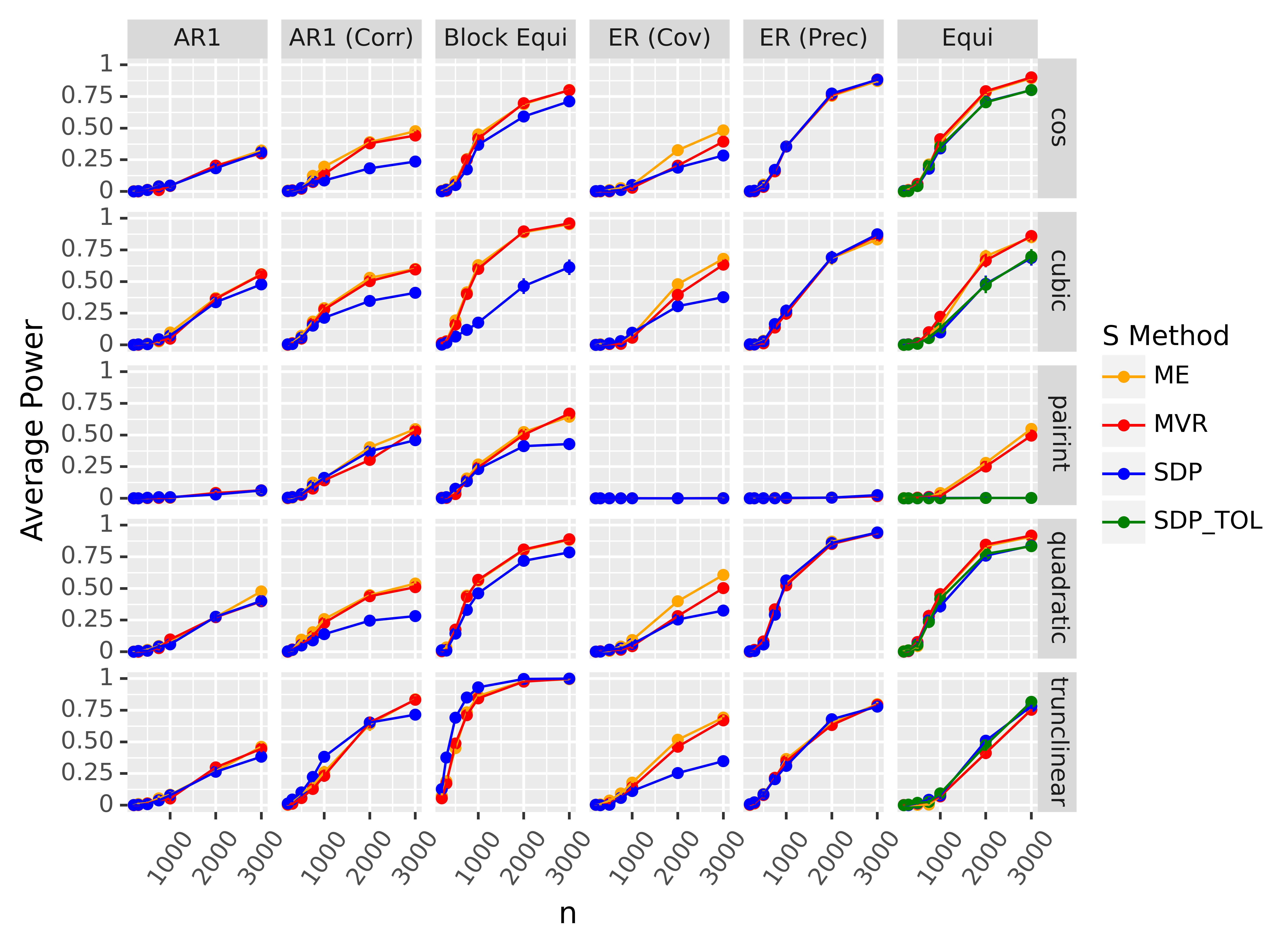}}
    \caption{Random forest statistic with swap importances on nonlinear responses: we let $X \sim \mathcal{N}(0, \Sigma)$ for various $\Sigma$ and $Y\mid X \sim \mathcal{N}(\mu(X), 1)$. We vary $\Sigma$ on the x-facets and $\mu$ on the y-facets. The precise definitions of the covariance matrices and conditional responses are presented in Appendix \ref{appendix::simgaussiandescription}. 
    We let $p = 200$ with $30$ non-nulls.}
    \label{fig:gaussianrfplot}
\end{figure}

\subsection{Robustness on Gaussian designs}\label{subsec::simrobustness}

\begin{figure}[h!]
    \centering
    \makebox[\textwidth]{\includegraphics[width=\textwidth]{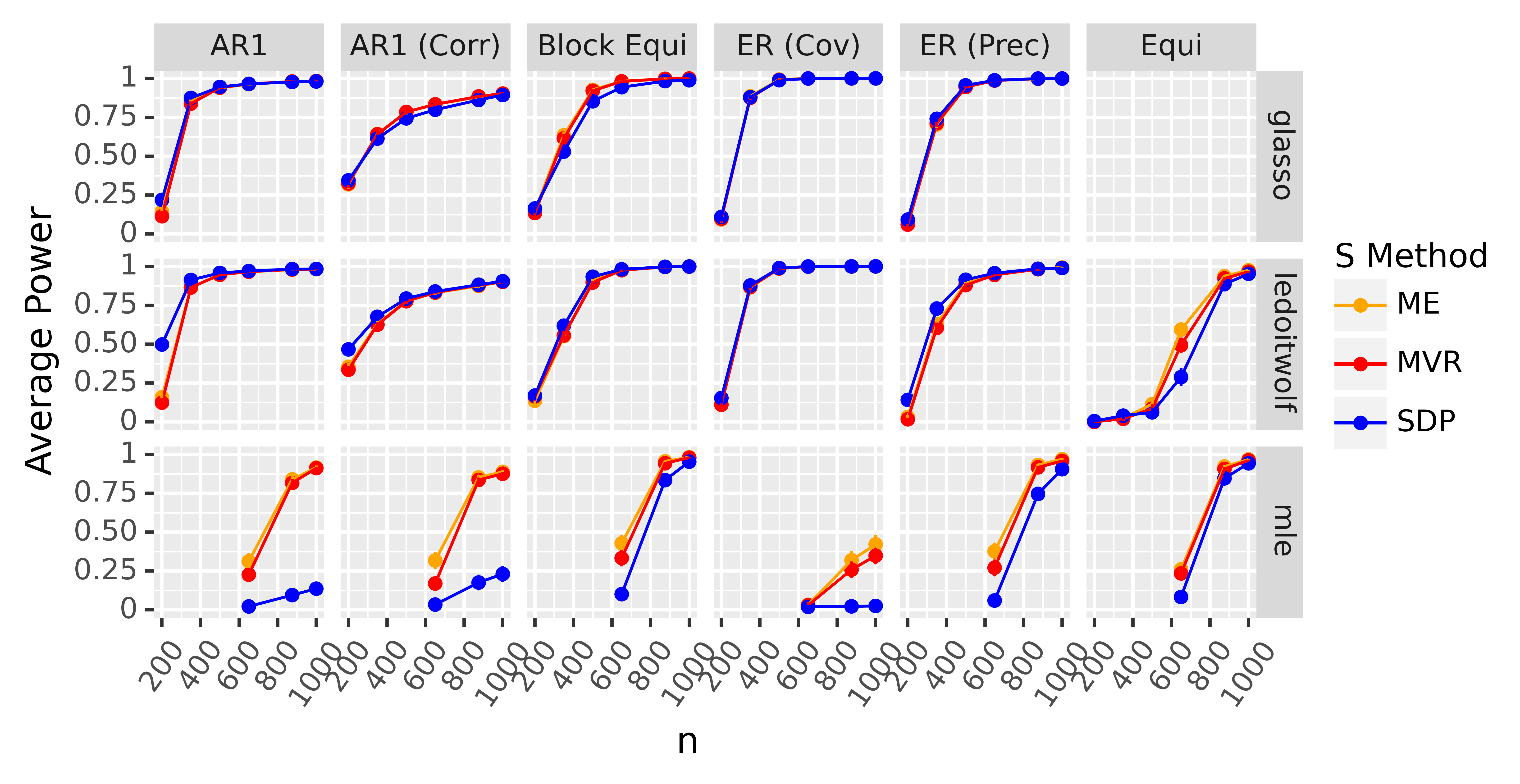}}
    \caption{Power of LCD statistics for estimated $\Sigma$: we let $X \sim \mathcal{N}(0, \Sigma)$ for various $\Sigma$ and $Y\mid X \sim \mathcal{N}(X\beta, 1)$. The definitions of $\Sigma$ follow Section \ref{subsec::simgaussian} and are formally defined in Appendix \ref{appendix::simgaussiandescription}.
    We let $p = 500$ with $50$ non-nulls with coefficients sampled independently from $\mathrm{Unif}(\left[-\delta, -\delta/2] \cup [\delta/2, \delta]\right)$, where $\delta = 1$ in ``ER (Cov)'' and ``ER (Prec)'' panels and $\delta = 0.5$ otherwise. Note that we do not apply the graphical lasso in the case where $\Sigma$ is equicorrelated, since in this case, the precision matrix is fully dense and the graphical lasso algorithm failed to converge.}
    \label{fig:robustpower}
\end{figure}

\begin{figure}[h!]
    \centering
    \makebox[\textwidth]{\includegraphics[width=\textwidth]{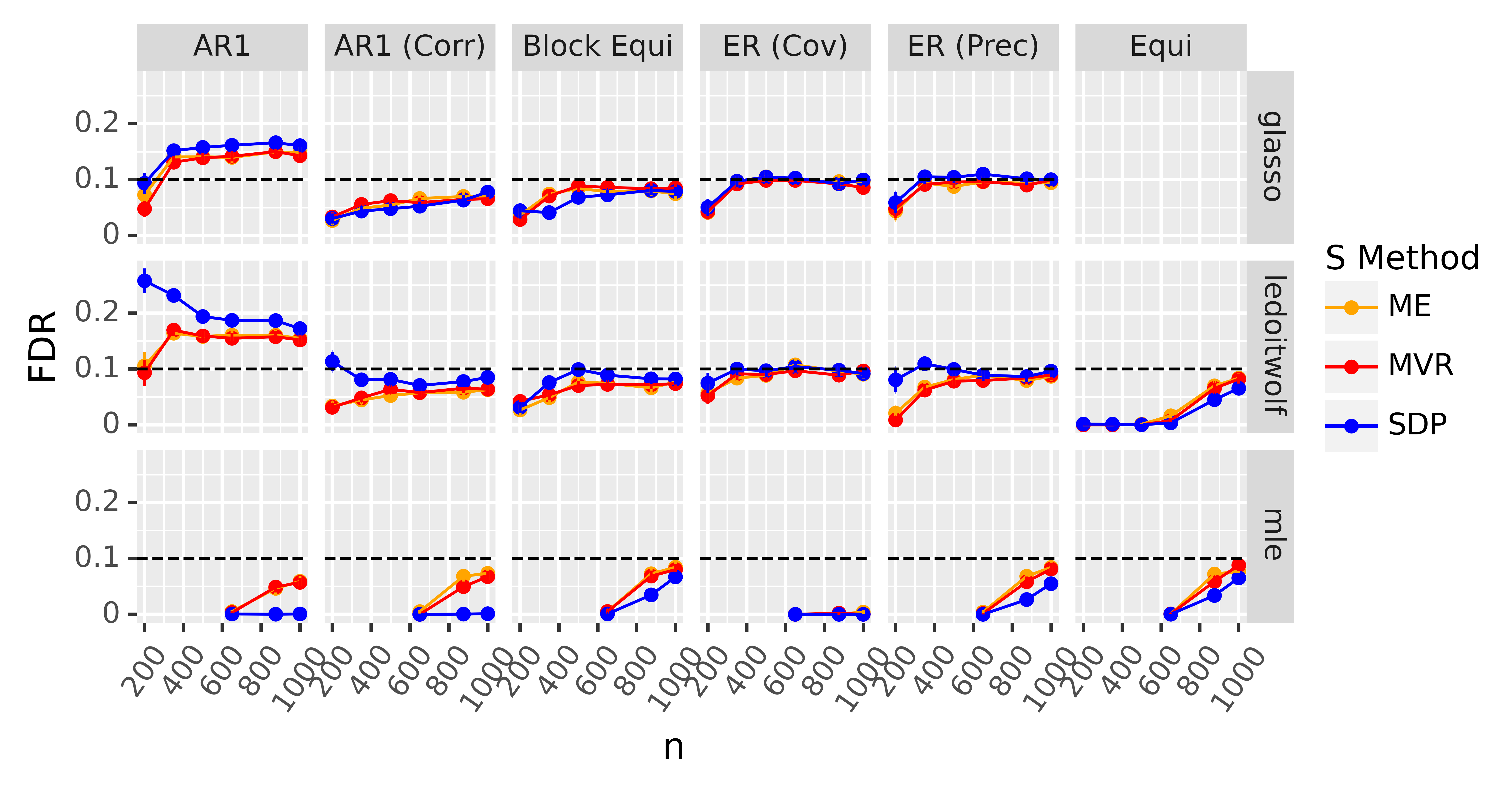}}
    \caption{The corresponding FDR plot for Figure \ref{fig:robustpower}.}
    \label{fig:robustfdr}
\end{figure}

In Sections \ref{sec::reconstruction} through \ref{subsec::simgaussian}, we have assumed that we know the true covariance matrix $\Sigma$. In this section, we analyze the robustness of MVR, \tmaxent, and SDP knockoffs when $\Sigma$ is not known and is estimated using the same data used to run knockoffs, using one of three methods: Ledoit--Wolf estimation \citep{ledoitwolf2004}, the graphical lasso algorithm \citep{glasso2008}, and the maximum likelihood estimate of $\Sigma$ when $n > p$.
 We let $Y\mid X \sim \mathcal{N}(X \beta, 1)$, and we use lasso coefficient difference statistics. Figures \ref{fig:robustpower} (power) and \ref{fig:robustfdr} (FDR) show that MVR and \smaxent knockoffs control the FDR better than SDP knockoffs, especially in high-dimensional settings where the covariance has been estimated using shrinkage methods. None of our theory predicted this, but it at least indicates that the power improvement of MVR/\smaxent knockoffs over SDP knockoffs does not come at the expense of robustness. See the figure caption for more simulation details.

\subsection{Application to fixed-X knockoffs}\label{subsec::simfx}

So far, we have worked with the MX knockoffs framework, where we assume we know the distribution of $X$ and we control the FDR in expectation over the distribution of $X$. However, the idea behind MVR and \smaxent knockoffs extends straightforwardly to the fixed-X (FX) knockoff filter, which treats the features $X$ as fixed and controls the FDR when $Y\mid X$ follows a Gaussian linear model. In the fixed-X setting, we assume the Gram matrix $\Sigma = \bX^{\top}\bX$ has diagonals equal to $1$ and construct knockoffs $\tilde{\bX}$ such that

\begin{equation}\label{eq::fxknockoffgen}
[\bX, \tilde{\bX}]^{\top} [\bX, \tilde{\bX}] = G_S \equiv \begin{bmatrix} \Sigma & \Sigma - S \\ \Sigma - S & \Sigma \end{bmatrix},
\end{equation}
where as previously, $S$ is a diagonal matrix such that $S \succcurlyeq 0$ and $2 \Sigma - S \succcurlyeq 0$. $\tilde{\bX}$ can be efficiently constructed when $n \ge 2p$ as outlined in \citet{fxknock}. From here on, the FX-knockoffs procedure is quite similar to the MX-knockoffs procedure, except that the feature statistics must obey a \textit{sufficiency} constraint that they only depend on the matrix $[\bX, \tilde{\bX}]^{\top} [\bX, \tilde{\bX}]$ and the empirical covariances $[\bX, \tilde{\bX}]^{\top} \by$. Then, when $Y \mid X \sim \mathcal{N}(X \beta, 1)$, the FX knockoff filter provably controls the FDR in finite samples.

Although the features do not have conditional variances or entropy in this setting since they are treated as fixed, the MVR and \smaxent losses in terms of $G_S$ naturally generalize to this setting. In particular, we set
\begin{equation}
S_{\mathrm{MVR}} = \arg \min_S \text{Tr}\left(G_S^{-1} \right) \,\, \, \text{ and } S_{\mmaxent} = \arg \min_S \log \det \left(G_S^{-1}\right).
\end{equation}
Proposition \ref{prop::mvrolsopt} naturally extends to the FX setting, where the $S_{\mathrm{MVR}}$ matrix minimizes the estimation error of OLS feature importances, except this time conditional on $\bX$. See the proof of Proposition \ref{prop::mvrolsopt} for details.

\begin{figure}[h!]
    \centering
    \makebox[\textwidth]{\includegraphics[width=\textwidth]{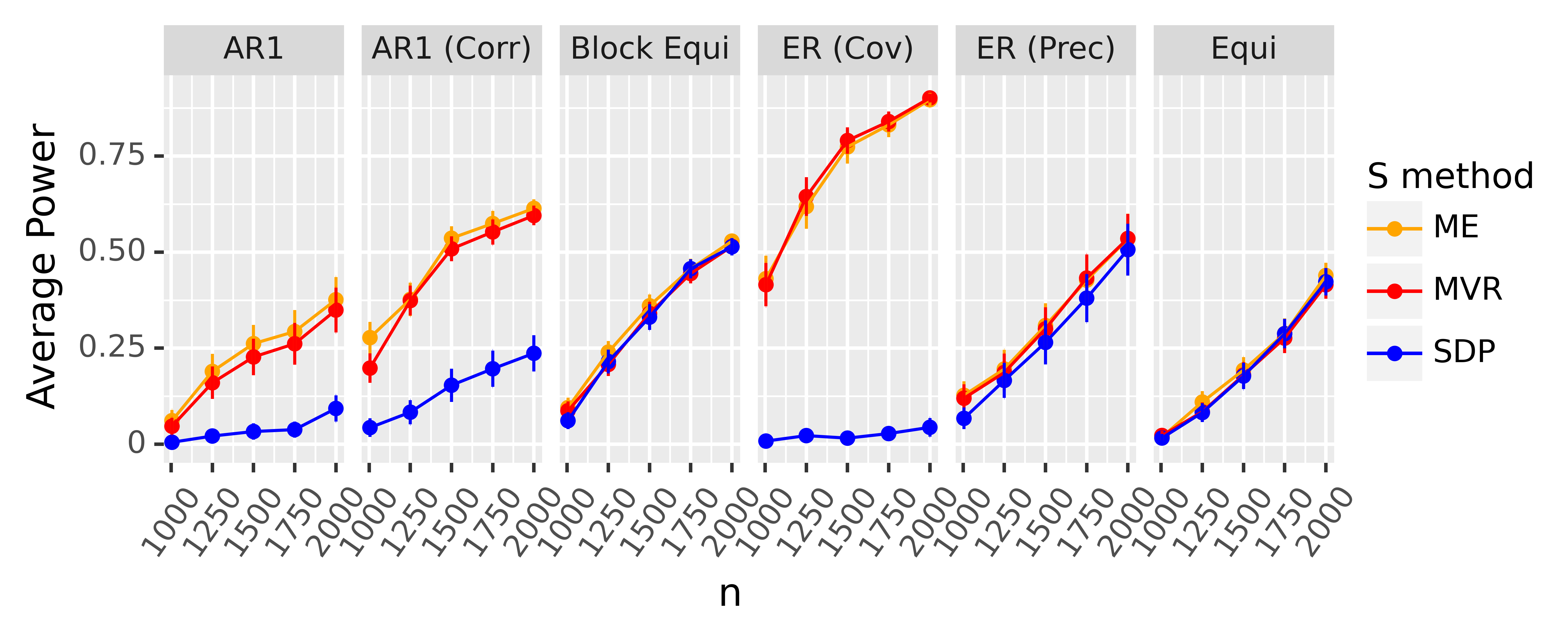}}
    \caption{Power for FX Knockoffs for Gaussian designs with lasso signed max statistics. We set $X \sim \mathcal{N}(0, \Sigma)$ with $\Sigma$ as defined in Appendix \ref{appendix::simgaussiandescription} 
    and let $Y\mid X \sim \mathcal{N}(X \beta, 1)$. In all cases $p = 500$ and there are $50$ non-nulls with coefficients sampled independently from $\mathrm{Unif}(\left[-\delta, -\delta/2] \cup [\delta/2, \delta]\right)$. For the AR1 plots, we let $\delta = 0.45$. For the Equi and Block-Equi plots, we let $\delta = 0.15$. For the ER (Prec) and ER (Cov) panels, we let $\delta = 0.25$ and $\delta = 0.5$, respectively.}
    \label{fig:fxpower}
\end{figure}

To study the power of MVR and \smaxent FX knockoffs, we simulate $X \sim \mathcal{N}(0, \Sigma)$ for the same covariance matrices as in Section \ref{subsec::simgaussian} and let $Y \mid X \sim \mathcal{N}(X \beta, 1)$ for sparse $\beta$ (see the caption of Figure \ref{fig:fxpower} for details). In order to obey the sufficiency property, FX feature statistics cannot use cross-validation, so we use the lasso signed max (LSM) statistic introduced in \citet{fxknock} instead of cross-validated lasso and ridge coefficient differences. Figure \ref{fig:fxpower} shows that MRC knockoffs generally outperform SDP knockoffs, although the power differential can be lower than in the MX case. 

\subsection{Application to group knockoffs}\label{subsec::groupknock} In this section, we show that the MRC framework can increase the power of group knockoffs \cite{daibarber2016}. In settings with highly correlated features, such as GWAS, some previous work has suggested partitioning the features into disjoint groups $G_1, \dots, G_m \subset [p]$. After clustering the features such that the between-group correlations are suitably low, \cite{daibarber2016} showed how to construct ``group knockoffs" which test the group-level hypotheses $H_{G_j} : Y \Perp X_{G_j} \mid X_{\text{-}G_j}$. Such group knockoffs need only satisfy a relaxed pairwise-exchangeability constraint, where the distribution of $[X, \tilde{X}]$ is invariant to swaps of entire groups, but not necessarily swaps of individual features. Although grouping the features can reduce correlations between the groups, it generally will not remove all dependence from the data  (see \cite{mxknockoffs2018, daibarber2016}), motivating the construction of MRC group knockoffs. One can construct MRC group knockoffs by minimizing the MRC loss functions $L_{\mathrm{MVR}}$ and $L_{\mathrm{ME}}$ subject to this relaxed constraint.

\begin{figure}[h!]
    \centering
    \makebox[\textwidth]{\includegraphics[width=\textwidth]{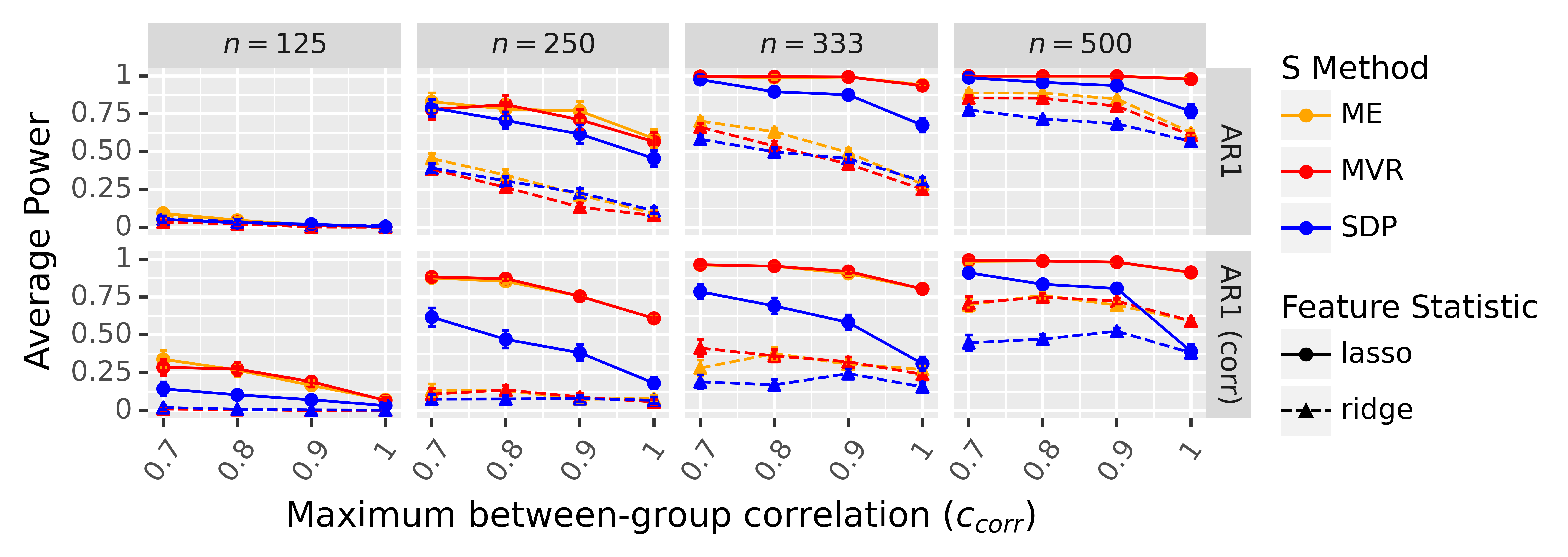}}
    \caption{Power for Group MX Knockoffs for Gaussian AR1 designs. We set $X \sim \mathcal{N}(0, \Sigma)$ with $\Sigma$ as defined in Appendix \ref{appendix::simgaussiandescription} 
    and let $Y\mid X \sim \mathcal{N}(X \beta, 1)$. In all cases $p = 500$ and there are $50$ non-nulls with coefficients sampled independently from $\mathrm{Unif}(\left[-\delta, -\delta/2] \cup [\delta/2, \delta]\right)$ with $\delta = 2$. To create groups, we hierarchically cluster the features using a single-linkage cutoff of $c_{\mathrm{corr}}$.}
    \label{fig::groupings}
\end{figure}

Figure \ref{fig::groupings} compares the powers of MRC and SDP group knockoffs in the AR1 settings studied in Section \ref{subsec::simgaussian}. To partition the features, we hierarchically cluster the features using correlations as a similarity measure and a single-linkage cutoff of $c_{\mathrm{corr}}$, where we vary $c_{\mathrm{corr}}$ between $1.0$, which recovers the ungrouped procedure, and $0.7$. The results show that MRC group knockoffs consistently outperform SDP group knockoffs.

\subsection{Application to non-Gaussian designs}\label{subsec::simnongaussian}

So far, we have focused on the case where $X \sim \mathcal{N}(0, \Sigma)$. Now, we consider the non-Gaussian case. To do this, we first review two general methods of constructing knockoffs for non-Gaussian features.

First, \citet{mxknockoffs2018} proposed an approximate second-order knockoff construction for the non-Gaussian case. In particular, if $X$ has mean zero and $\text{Cov}(X) = \Sigma$, the authors considered picking an $S$-matrix and sampling
\begin{equation}\label{eq::secondorderknockoffs}
    \tilde{X} \mid X \sim \mathcal{N}\left(X - X \Sigma^{-1} S, 2 S - S \Sigma^{-1} S \right).
\end{equation}
This guarantees that the first two moments of $X$ and $\tilde{X}$ match, and moreover that $\text{Cov}([X, \tilde{X}]) = G_S$ as in the Gaussian case. These are \textit{not} valid knockoffs and do not guarantee FDR control, but they may be fairly robust in practice. Furthermore, second-order knockoffs only require knowledge of the first two moments of $\tilde{X}$, as opposed to the joint density function. \citet{mxknockoffs2018} suggested setting $S = S_{\mathrm{SDP}}$ in equation (\ref{eq::secondorderknockoffs}). We will demonstrate below that $S_{\mathrm{MVR}}$ and $S_{\mmaxent}$ are more powerful alternatives.

Second, the Metropolized knockoff sampler introduced in \citet{metro2019} allows one to sample exact, valid knockoffs for arbitrary distributions of $X$ under the assumption that the unnormalized density $\Phi$ of $X$ is known. Given an ordering of the features $X_1, \dots, X_p$, the key idea is to sample $\tilde{X}_j$ by taking a step along a time-reversible Markov chain starting from $X_j$, such that $\mathcal{L}(X_j | X_{\text{-}j}, \tilde{X}_{1:(j-1)})$ is the stationary distribution of the chain. To accomplish this, \citet{metro2019} employ a Metropolis--Hastings style proposal and acceptance scheme to ensure the exact validity of the knockoffs. The authors suggested using Gaussian \textit{covariance-guided proposals}, where the Metropolis--Hastings proposals are sampled as second-order knockoffs according to equation (\ref{eq::secondorderknockoffs}). In the following simulations, we will generate second-order proposals $X^*$ using the MVR, \tmaxent, and SDP $S$-matrices and then compare their power.

We consider three types of non-Gaussian designs: a $t$-tailed Markov chain discussed in \citet{metro2019} with and without correlated signals, a ``block-equicorrelated'' design where each block is independently $t$-distributed with an equicorrelated covariance matrix, and a Gibbs measure on a $d \times d$ grid. In all cases, the features differ substantially from the Gaussian case: for example, in the first two settings, we let our $t$ distributions have $\nu = 3$ degrees of freedom. See Appendix \ref{appendix::metrodetails} 
for more details on the precise design distributions as well as our knockoff generation mechanism for the ``discrete grid'' model.

\begin{figure}[h!]
    \centering
    \makebox[\textwidth]{\includegraphics[width=\textwidth]{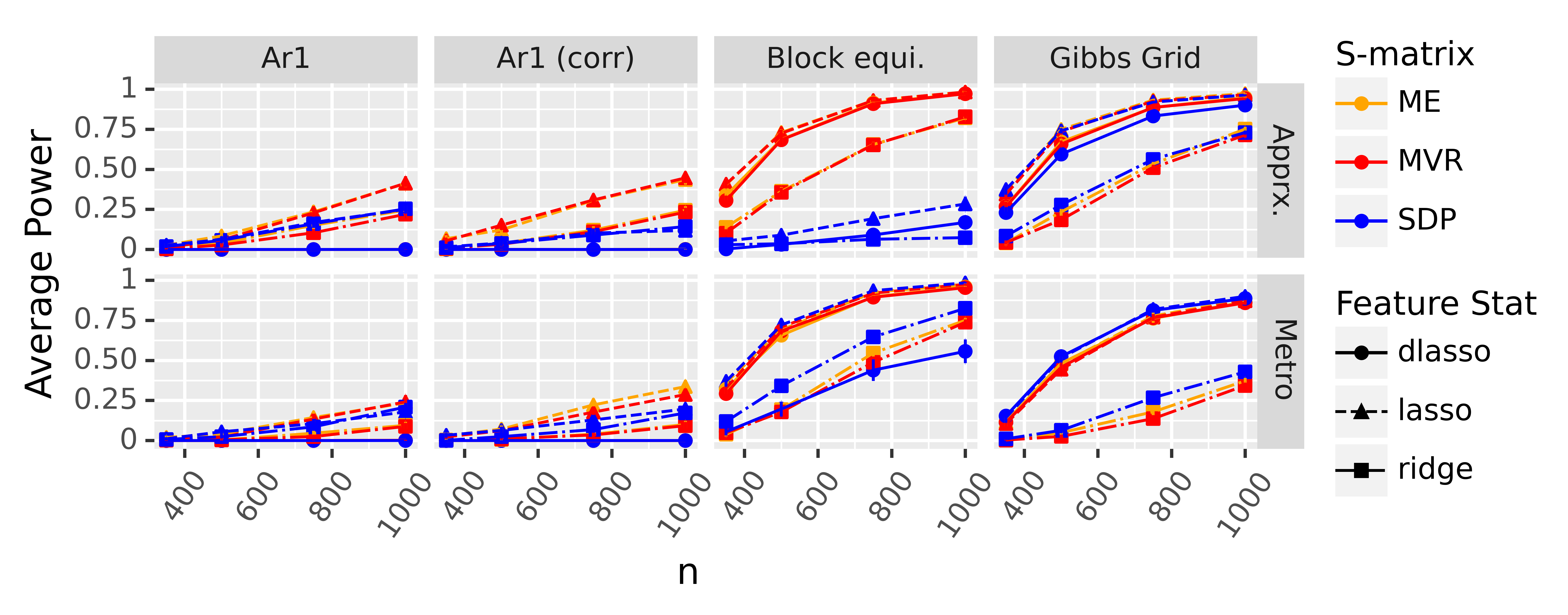}}
    \caption{Power for non-Gaussian designs with a linear response. The ``Apprx" panel refers to second-order knockoffs. The horizontal facets correspond to the designs described in Appendix \ref{appendix::metrodetails}. 
    We let $p = 500$ except for the Gibbs Grid, where $p = 625$. We let $Y \mid X \sim \mathcal{N}(X \beta, 1)$ with sparsity $10\%$ and non-null coefficients drawn from $\mathrm{Unif}(\left[-\delta, -\delta/2] \cup [\delta/2, \delta]\right)$ where $\delta = 0.3$ for the AR1 designs and $\delta = 0.4$ otherwise.}
    \label{fig::nongaussianlinearpower}
\end{figure}

In Figure \ref{fig::nongaussianlinearpower}, we let $Y \mid X \sim \mathcal{N}\left(X \beta, 1 \right)$ with similar $\beta$ as before (see the captions for details). We test three types of feature statistics: lasso coefficient differences, ridge coefficient differences, and debiased lasso (dlasso) coefficient differences \citep{javmont2014}.

We make two observations about these plots. First, the only difference between the second-order and Metropolized knockoff procedures is that the Metropolized sampling procedure takes the second-order knockoffs as proposals $X^*$ and sometimes rejects these knockoffs, setting $\tilde{X}_j = X_j$ in the case of a rejection. These rejections \textit{increase} the marginal correlations between $X_j$ and $\tilde{X}_j$, but outside of the Gibbs grid model, the SDP-guided Metropolized knockoffs have more power than their second-order counterparts. We interpret this as evidence that the reconstruction effect can still occur for non-Gaussian designs, and the Metropolis rejections (inadvertently) correct for this effect by increasing the marginal feature-knockoff correlations. Second, we observe that the MRC second-order knockoffs are substantially more powerful than their SDP counterparts. For the Metropolized knockoff sampler, the MVR and \smaxent proposals increase the power of debiased lasso statistics while decreasing the power of the ridge statistics. The lasso power is fairly similar between all three methods throughout. We present the corresponding FDR plot in Appendix  \ref{appendix::metrodetails}.

Lastly, a few other works have introduced methods to sample approximate knockoffs when the distribution of $X$ is unknown, using tools from the machine learning literature \cite{deepknock2018, jordon2018knockoffgan}. These constructions generally use pairwise dependency measures such as the MAC as part of their optimization criteria, but we expect it to be straightforward to ``plug in" the MRC objective criteria in place of the MAC. For example, one could repeatedly resample $X_j \mid X_{\text{-}j}, \tilde{X}$ to estimate $\mathrm{Var}(X_j \mid X_{\text{-}j}, \tilde{X})$, although we leave such details to future work.

\section{Discussion}\label{sec::discussion}

This paper identifies an important flaw in previous knockoff generation mechanisms, which reduces power by allowing regression test statistics like the lasso to reconstruct non-null features using the other features and knockoffs. To solve this problem, we introduced minimum reconstructability knockoffs, which substantially increase the power of knockoffs for correlated designs. However, our work leaves several questions open for future research.

One immediate question is how to construct exact MRC knockoffs for non-Gaussian designs. Our technique in Section \ref{subsec::simnongaussian} minimizes the reconstructability of the \textit{proposal} knockoffs $X^*$ in the Metropolized knockoff sampling framework, but the actual knockoffs $\tilde{X}$ may have different properties due to the complex acceptance scheme of the sampler. Additionally, \citet{condknock2019} introduced the idea of \textit{conditional knockoffs}, which enable FDR control when the distribution of $X$ is only known up to a parametric model. Constructing conditional MRC knockoffs may be practically useful, but it is not obvious even how to define conditional MRC knockoffs, with the exception of the Gaussian case, where conditional MRC knockoffs can be defined analagously to fixed-X MRC knockoffs (see Section \ref{subsec::simfx}).

Interestingly, for discrete designs with finite support, computing the distribution of \smaxent knockoffs corresponds to a maximum entropy problem with linear constraints (see Appendix \ref{appendix::discretemmi} 
for details). Such problems are convex and well studied  \citep{persson1986, boyd2004}, but the number of optimization variables and constraints grow exponentially with $p$, making the problem intractable. We discuss three potential ways around this in Appendix \ref{appendix::discretemmi}, including using conditional independence properties of $X$, restricting the class of feasible knockoff distributions, or finding approximate solutions, but further study is needed.

Another interesting future direction would be to better understand the notion of reconstructability. For example, we observed in Section \ref{sec::reconstruction} that the reconstruction effect gets worse when feature statistics can reconstruct non-null features using sparse subsets of the other features and knockoffs. This can occur when $G_S$ is particularly low rank, but it can also occur when the \textit{eigenvectors} of $G_S$ corresponding to small eigenvalues are (approximately) sparse. It may therefore be fruitful to incorporate the structure of the eigenvectors of $G_S$ into knockoff generation mechanisms. Alternatively, we have advocated minimizing two specific measures of reconstructability, but we have not thoroughly investigated other possibilities. Further analysis on this front may turn out to further improve power.

%
%

\section*{Acknowledgements}
The authors would like to thank Chenguang Dai, Buyu Lin, Jun Liu, Wenshuo Wang, and Xin Xing for valuable discussions and suggestions. The authors are also grateful to the anonymous referees for helpful comments. L. J. was partially supported by the William F. Milton Fund.

\bibliography{ref}
\bibliographystyle{apalike}
\appendix 

\section{Proofs for Section \ref{sec::reconstruction}}\label{appendix::proofs}

\subsection{Proofs for the general reconstruction effect}\label{appendix::generalreconstruction}

In this section, we prove some general facts about the reconstruction effect, which we will apply to the equicorrelated case in the following section. To start, we prove Lemma \ref{lem::generalrankdegen}, which shows that when $X$ is Gaussian, minimizing the MAC will often cause $G_{\mathrm{SDP}}$ to be low rank.

\begingroup
\def\thetheorem{\ref{lem::generalrankdegen}}
\begin{lemma} Suppose $X \sim \mathcal{N}(0, \Sigma)$ and $\lambda_{\mathrm{min}}(\Sigma) \le 0.5$. Then $\mathrm{rank}(G_{\mathrm{SDP}}) < 2p$. Furthermore, if $\Sigma$ is block-diagonal with $b$ blocks, each with an eigenvalue below $0.5$, then $\mathrm{rank}(G_{\mathrm{SDP}}) \le 2p - b$.
\begin{proof} The eigenvalues of $G_{\mathrm{SDP}}$ are the eigenvalues of $S_{\mathrm{SDP}}$ and $2 \Sigma - S_{\mathrm{SDP}}$. The case where $\lambda_{\mathrm{min}}(\Sigma) = 0.5$ is trivial, as we can set $S_{\mathrm{SDP}} = I_p$, which implies $2 \Sigma - S_{\mathrm{SDP}}$ will have at least one eigenvalue that equals $0$.

When $\lambda_{\mathrm{min}}(\Sigma) < 0.5$, assume for sake of contradiction that $G_{\mathrm{SDP}}$ is full rank, and thus $\lambda_{\mathrm{min}}(2 \Sigma - S_{\mathrm{SDP}}) = \gamma > 0$. This implies $2 \Sigma - S_{\mathrm{SDP}} - \gamma I_p \succcurlyeq 0$ as well. Represent $S_{\mathrm{SDP}} = \mathrm{diag}(s)$ and denote $s^* = \min(s + \gamma, 1)$, where the minimum is taken element-wise over the vector.

By the previous argument, $\mathrm{diag}(s^*)$ is a feasible $S$-matrix with lower mean absolute correlation than $S_{\mathrm{SDP}}$. Therefore, $S_{\mathrm{SDP}}$ cannot be the solution to the SDP. This is a contradiction and completes the proof of the non-block-diagonal statement. The block-diagonal statement follows because when $\Sigma$ is block-diagonal, the solution to the SDP is the SDP solution to each of the blocks \citep{fxknock}. 
\end{proof}
\end{lemma}
\endgroup

Next we prove Theorem \ref{thm::generalreconstruction}, which we very slightly restate to make it easier to apply in our later proofs. We use $\odot$ to denote elementwise multiplication.

\begingroup
\def\thetheorem{\ref{thm::generalreconstruction}}
\begin{theorem}
Suppose we can represent $Y = f\left(g(X_J), X_{\text{-}J}, U \right)$ for some set $J \subset [p]$, functions $f$ and $g$, and independent noise $U \sim \mathrm{Unif}(0,1)$. Equivalently, this means $Y \Perp X_J \mid g(X_J), X_{\text{-}J}$. Suppose a function $g^*$ exists such that $g(X_J) = g^*(\tilde{X}_J)$ holds almost surely. If $Y^* = f(g^*(X_J), X_{\text{-}J}, U)$, then
\begin{equation*}
\left( [X, \tilde{X}], Y \right) \stackrel{d}{=} \left( [X, \tilde{X}]_{\swap(J)}, Y^* \right)
\, \, \, \, \, 
\text{ and }
\, \, \, \, \, 
\left( [X, \tilde{X}], Y^* \right) \stackrel{d}{=} \left( [X, \tilde{X}]_{\swap(J)}, Y\right).
\end{equation*}
In particular, this implies
\begin{equation}\label{eq::negationreconst}
    w([\bX, \tilde{\bX}], \by) \stackrel{d}{=} -\mathbf{1}_J \odot \left(w([\bX, \tilde{\bX}],\by^*) \right),
\end{equation}
which ensures $\mathbb{P}\left(w([\bX, \tilde{\bX}], \by)_j > 0 \right) + \mathbb{P}\left(w([\bX, \tilde{\bX}], \by^*)_j > 0 \right) \le 1$.
\begin{proof} Note that since $[X_J, \tilde{X}_J] \stackrel{d}{=} [\tilde{X}_J, X_J]$, $g(X_J) = g^*(\tilde{X}_J)$ also implies $g(\tilde{X}_J) = g^*(X_J)$. Thus, we can rewrite
$$ Y^* = f(g^*(X_J), X_{\text{-}J}, U) = f(g(\tilde{X}_J), X_{\text{-}J}, U).$$
This means that $Y^* \mid [X, \tilde{X}]_{\swap(J)} \stackrel{d}{=} Y \mid [X, \tilde{X}]$. Since marginally $[X, \tilde{X}]_{\swap(J)} \stackrel{d}{=} [X, \tilde{X}]$, this implies
$$([X, \tilde{X}], Y) \stackrel{d}{=} ([X,\tilde{X}]_{\swap(J)}, Y^*). $$ The former equality holds for each i.i.d. row $(X, Y)$ and therefore holds for the matrix versions $\bX, \tilde{\bX}, \by$. Since $w$ is a function of $[\bX, \tilde{\bX}], \by$ and must obey the knockoff antisymmetry property, we obtain that
$$w([\bX, \tilde{\bX}], \by) \stackrel{d}{=} w([\bX,\tilde{\bX}]_{\swap(J)}, \by^*) = -\mathbf{1}_{J} \odot w([\bX, \tilde{\bX}], \by^*). $$
For each $j \in J$, this proves $\mathbb{P}\left(w([\bX, \tilde{\bX}], \by)_j > 0 \right) +$ $\mathbb{P}\left(w([\bX, \tilde{\bX}], \by^*)_j > 0 \right)$ $\le 1$.
\end{proof}
\end{theorem}
\addtocounter{theorem}{-1}
\endgroup

\subsection{Simple proofs for block-equicorrelated Gaussian designs}\label{appendix::simpleqpfs}

In this section, we will prove Lemma \ref{lem::rankdegen} and apply Theorem \ref{thm::generalreconstruction} to the equicorrelated case. These will be important tools for our later proof of Theorem \ref{thm::avgequi}. 
 
\begingroup
\def\thetheorem{\ref{lem::rankdegen}}
\begin{lemma} In the equicorrelated case when $\rho \ge 0.5$, let $\tilde{X}$ be generated according to the SDP procedure. Then $G_{\mathrm{SDP}}$ has rank $p+1$, and $X_j + \tilde{X}_j = X_k + \tilde{X}_k$ for all $1 \le j, k \le p$. 
\begin{proof} To find the solution for the SDP, we simply verify that $S_{\mathrm{SDP}} = (2 - 2\rho) \cdot I_p$ by checking the KKT conditions in Lemma \ref{lem::equicorrsoln}. Then, we have
$$[X, \tilde{X}] \sim \mathcal{N}(0, G_{\mathrm{SDP}}) \text{ for } G_{\mathrm{SDP}} = \begin{bmatrix} \Sigma & \Sigma - (2 - 2\rho) I_p \\ 
                       \Sigma - (2 - 2\rho) I_p & \Sigma \end{bmatrix} \in \mathbb{R}^{2p \times 2p}.  $$
First, we note that the eigenvalues of $G_{\mathrm{SDP}}$ are those of $(2-2 \rho) I_p$ and $2 \Sigma - (2 - 2\rho)I_p$. Note, however, that $\Sigma = (1-\rho) I_p + \rho \mathbf{1}_p \mathbf{1}_p^{\top}$, so $2 \Sigma - (2 - 2\rho) I_p = 2 \rho \mathbf{1}_{p} \mathbf{1}_{p}^{\top}$, which has a rank of $1$. Since $(2 - 2\rho) I_p$ has full rank, this implies that $G_{\mathrm{SDP}}$ has rank $p+1$.
             
To show that $X_j + \tilde{X}_j$ remains constant over all $1 \le j \le p$, denote the $j$th column of $G_{\mathrm{SDP}}$ as $G_j$. Simple arithmetic shows that $G_j + G_{j+p} = 2 \rho \cdot \mathbf{1}_{2p}$. This implies that if $\mu \in \mathbb{R}^{2p}$ is the vector of all zeros except $\mu_1, \mu_{p+1} = 1$ and $\mu_j, \mu_{j+p} = -1$, then $\mu^{\top} G_{\mathrm{SDP}} = 0 \in \mathbb{R}^{2p}$. If $\sqrt{G_{\mathrm{SDP}}}$ is a symmetric square root of $G_{\mathrm{SDP}}$, this implies $\mu^{\top} \sqrt{G_{\mathrm{SDP}}} = 0$.

Now, let $\epsilon \sim \mathcal{N}(0, I_{2p})$. We may represent 
$[X,\tilde{X}] \sim \sqrt{G_{\mathrm{SDP}}} \, \epsilon $. Using this representation, we conclude
$$X_1 + \tilde{X}_1 - X_j - \tilde{X}_j = \mu^{\top} [X, \tilde{X}] \sim \mu^{\top} \sqrt{G_{\mathrm{SDP}}} \, \epsilon = 0.$$
\end{proof}
\end{lemma}
\addtocounter{theorem}{-1}
\endgroup
\begin{corollary}\label{corr::blockrankdegen}
Suppose $\Sigma = \mathrm{blockdiag}(\Sigma_1, \dots, \Sigma_{\ell})$ and $\Sigma_k$ is equicorrelated with correlation at least $0.5$, for $1 \le k \le \ell$. Let $D_k$ be the set of indices corresponding to block $k$. Then for all $j_1,j_2 \in D_k$, $X_{j_1} + \tilde{X}_{j_1} = X_{j_2} + \tilde{X}_{j_2}$ if $\tilde{X}$ are SDP knockoffs.
\begin{proof} This follows directly from Lemma \ref{lem::rankdegen} since the solution to the SDP for a block-diagonal matrix is simply the diagonal matrix composed of the solutions to the blocks $\Sigma_1, \dots, \Sigma_{\ell}$.
\end{proof}
\end{corollary}

Corollary \ref{corr::blockrankdegen} allows us to apply Theorem \ref{thm::generalreconstruction} to the block-equicorrelated case. We define some notation before doing this.

\begin{definition}[Negation Notation] For coefficients $\beta \in \mathbb{R}^p$ and $J \subset [p]$, define $\beta_{\mathrm{neg}(J)} = - \mathbf{1}_J \odot \beta$. For the response variable $Y = f(X, U)$, define 
$Y_{\mathrm{neg}(J)} = f(- \mathbf{1}_{J} \odot X, U)$, 
where $\odot$ denotes elementwise multiplication.
\end{definition}

Note that in the follow Corollary, $Y_{\mathrm{neg}(J)}$ corresponds to $Y^*$ in the Theorem \ref{thm::generalreconstruction}.

\begin{corollary}\label{corr::eqreconstappendix} Let $X \sim \mathcal{N}(0, \Sigma)$ with $\Sigma = \mathrm{blockdiag}(\Sigma_1, \dots, \Sigma_{\ell})$, where each $\Sigma_k$ is equicorrelated with correlation $\rho_k \ge 0.5$. Let $Y = f(X\beta, U)$ and suppose we generate knockoffs $\tilde{X}$ according to the SDP procedure. Let $J \subset [p]$ such that $J$ is contained among a single equicorrelated block and $\sum_{j \in J} \beta_j = 0$. Then
$$w([\bX, \tilde{\bX}], \by) \stackrel{d}{=} -\mathbf{1}_J \odot \left(w([\bX, \tilde{\bX}],\by_{\mathrm{neg}(J)})\right), $$
which implies $\mathbb{P}\left(w([\bX, \tilde{\bX}], \by)_j > 0 \right) + \mathbb{P}\left(w([\bX, \tilde{\bX}], \by_{\mathrm{neg}(J)})_j > 0 \right) \le 1$.
\begin{proof} Without loss of generality assume $J = \{1, \dots, |J|\}$. Corollary \ref{corr::blockrankdegen} tells us that for $j \in J$, $X_j = X_1 + \tilde{X}_1 - \tilde{X}_j$. This implies that $X_J \beta_J = \sum_{j \in J} \beta_j (X_1 + \tilde{X}_1) - \tilde{X}_J \beta_J = - \tilde{X}_J \beta_J$, where the last equality follows because $\sum_{j \in J} \beta_J = 0$. This satisfies the assumptions of Theorem \ref{thm::generalreconstruction}, since $Y \Perp X_J \mid X_J \beta_J, X_{\text{-}J}$ and we can reconstruct $X_J \beta_J$ using $\tilde{X}_J$. Thus, if we set $Y_{\mathrm{neg}(J)} = f(-\mathbf{1}_J \odot X \beta, U) = f(X \beta_{\mathrm{neg}(J)}, U)$, Theorem \ref{thm::generalreconstruction} yields the result.
\end{proof}
\end{corollary}

\subsection{Proof of Theorem \ref{thm::avgequi}}

The proof for Theorem \ref{thm::avgequi} is presented below. Our first task is to strengthen the result of Corollary \ref{corr::eqreconstappendix} by applying the permutation invariance assumption and considering the case where $\sum_{j \in J} \beta_j \approx 0$. Then, we will prove our main technical theorem, Theorem \ref{thm::avgequiproof}. After that, we will prove Theorem \ref{thm::avgequi}. Throughout, we will defer overly technical details to Appendix \ref{appendix::avgequitechdetails}. In general, the value of universal constants $c_0, c_1, C_0, C_1, C_2$ may change by a constant from line to line.

The following proposition strengthens Corollary \ref{corr::eqreconstappendix}.

\begin{proposition}\label{prop::apprxeqreconst} Let $X \sim \mathcal{N}(0, \Sigma)$ for block-equicorrelated $\Sigma$ with within-block correlations at least $0.5$. Let $Y = f(X\beta + \zeta_0, U)$ for $\zeta_0 \sim \mathcal{N}(0, \sigma_0^2)$ and $U \sim \mathrm{Unif}(0,1)$, with $U, \zeta_0, X$ jointly independent. Suppose we generate knockoffs $\tilde{X}$ according to the SDP procedure and $W = w([\bX, \tilde{\bX}], \by)$ is a permutation invariant feature statistic. 
For any $j_1, j_2$ in the same equicorrelated block, assume $|\beta_{j_1} + \beta_{j_2}| < d \in \mathbb{R}$. Then if $J = \{j_1, j_2\}$, for a universal constant $c_0$ depending only on $\sigma_0^2$,
$$d_{\tv}([W_{j_2}, W_{j_1}, W_{\text{-}J}], [-W_{j_1}, -W_{j_2}, W_{\text{-}J}]) \le c_0 \sqrt{n}  d. $$
In particular, this implies
$$d_{\tv}([\sign(\sorted(W_J)), W_{\text{-}J}], [- \sign(\sorted(W_J)), W_{\text{-}J}]) \le c_0 \sqrt{n} d, $$
where $\sorted(W_J)$ refer to $W_J$ sorted in descending order of absolute value and $\sign(\sorted(W_J))$ are the signs of the sorted $W_J$.
\begin{proof} To begin with, we analyze the case where $d = 0$ and $\beta_{j_1} = - \beta_{j_2}$. Let $\sigma : [p] \to [p]$ be the permutation which swaps $j_1$ with $j_2$ but leaves all other indices constant. Then we observe $\beta_{\mathrm{neg}(J)} = \sigma(\beta)$, so 
$$Y_{\mathrm{neg}(J)} = f(X \sigma(\beta) + \zeta_0, U) = f(\sigma^{-1}(X) \beta + \zeta_0, U).$$
This implies that $Y_{\mathrm{neg}(J)} \mid [X,\tilde{X}] \stackrel{d}{=} Y \mid [\sigma(X), \sigma(\tilde{X})]$ (note that $\sigma = \sigma^{-1}$). Since $X$ and therefore $\tilde{X}$ are exchangeable within a block, this implies
$$(\bX, \tilde{\bX}, \by_{\mathrm{neg}(J)}) \stackrel{d}{=} (\sigma(\bX), \sigma(\tilde{\bX}), \by)  $$
which implies
$$w([\bX, \tilde{\bX}], \by_{\mathrm{neg}(J)}) \stackrel{d}{=} w([\sigma(\bX), \sigma(\tilde{\bX})], \by) \stackrel{d}{=} \sigma(w([\bX, \tilde{\bX}],  \by)), $$
where the right-hand side follows from the permutation invariance of $W$. We now apply Corollary \ref{corr::eqreconstappendix} to the term on the left-hand side to obtain
\begin{equation}\label{eq::permdegen}
-\mathbf{1}_J \odot w([\bX, \tilde{\bX}], \by) \stackrel{d}{=} w([\bX, \tilde{\bX}],  \by_{\mathrm{neg}(J)}) \stackrel{d}{=} \sigma(w([\bX, \tilde{\bX}], \by)).
\end{equation}
Now, we extend this analysis to the case where $|\beta_{j_1} + \beta_{j_2}| = d > 0$ using a total variation argument. Define the vector $\beta' \in \mathbb{R}^p$ where
$$\beta'_k = \begin{cases} \beta_k & k \ne j_1 \\ - \beta_{j_2} & k = {j_1} \\ \end{cases} $$
so in particular $\beta'_{j_1} = - \beta'_{j_2}$. Then let $Y' = f(X \beta' + \zeta_0, U)$ and let $\zeta \in \mathbb{R}^n$ denote the concatenation of all $\zeta_0$. Since $[\bX, \tilde{\bX}, \bX \beta + \zeta]$ are jointly multivariate Gaussian, it is simple to prove that
$$d_{\tv}([\bX, \tilde{\bX}, \bX \beta + \zeta], [\bX, \tilde{\bX},\bX \beta' + \zeta]) \le c_0 \sqrt{n} d $$
for some $c_0$ depending only on $\sigma_0^2$, which we show in Lemma \ref{lem::tvbound}. Since $\by$ is a function of independent noise and $[\bX, \tilde{\bX}, \bX \beta + \zeta]$, and respectively for $\mathbf{y'}$, this bound immediately implies
\begin{equation}\label{eq::sindextvbound}
d_{\tv}([\bX, \tilde{\bX}, \by], [\bX, \tilde{\bX},\mathbf{y'}]) \le c_0 \sqrt{n} d \, .
\end{equation}
Abbreviate $w([\bX, \tilde{\bX}], \by)$ as $W$ and $w([\bX, \tilde{\bX}], \mathbf{y'})$ as $W'$. Applying the triangle inequality, we obtain that
\begin{align*}
d_{\tv}([W_{j_1}, W_{j_2}, W_{\text{-}J}], [-W_{j_2}, -W_{j_1}, W_{\text{-}J}]) =& d_{\tv}(W, -\mathbf{1}_J \odot \sigma(W)) \\
\le&  d_{\tv}(W,W') + d_{\tv}(W', -\mathbf{1}_J \odot \sigma(W')) + d_{\tv}(-\mathbf{1}_J \odot \sigma(W'), -\mathbf{1}_J \odot \sigma(W)) \\
\le& c_0 \sqrt{n} d + 0 + c_0 \sqrt{n} d
\end{align*}
where the last line follows from applying (\ref{eq::permdegen}) to the middle term and (\ref{eq::sindextvbound}) to the other two terms. The proposition now follows if we reset $c_0$ to be twice its original value.
\end{proof}
\end{proposition}

The following corollary indicates that the conclusion of Proposition \ref{prop::apprxeqreconst} also holds conditionally with high probability. Since it is a straightforward property of joint total variation, we defer its proof to Appendix \ref{appendix::condapprxreconstproof}.

\begin{corollary}\label{corr::condapprxreconst} In the same setting as above, let $\alpha_1, \alpha_2$ be constants such that $\alpha_1 \cdot \alpha_2 > c_0 \sqrt{n} d$ and let $A$ be some event in the sigma-algebra generated by $[|\sorted(W_{J})|, W_{\text{-}J}]$. Then if $\mathbb{P}(A) > \alpha_2 $, we have that
\begin{equation}
d_{\tv}\left([\sorted(W_{J}),W_{\text{-}J}], [-\sorted(W_{J}), W_{\text{-}J}] \mid A \right) < \alpha_1. \label{eq::condtvbound}
\end{equation}
\end{corollary}

Below, we prove our main technical theorem, of which Theorem \ref{thm::avgequi} is a corollary.

\begin{theorem}\label{thm::avgequiproof} Let $X \sim \mathcal{N}(0, \Sigma)$ with $\Sigma = \mathrm{blockdiag}(\Sigma_1, \dots, \Sigma_{\ell})$ where for all $1 \le j \le \ell$, $\Sigma_j$ is equicorrelated with correlation $\rho_j \ge 0.5$. Suppose $Y = f(X \beta + \zeta_0, U)$ for $\zeta_0 \sim \mathcal{N}(0, \sigma_0^2)$ and $U \sim \mathrm{Unif}(0,1)$ with $U, \zeta_0, X$ jointly independent. Let $\tilde{X}$ be generated according to the SDP procedure and let $W = w([\bX, \tilde{\bX}], \by)$ be a permutation invariant feature statistic. Suppose we use knockoffs to control the FDR at level $q \le 0.1$.

Fix $b, m > 0$. Suppose we can group the features into $M$ disjoint pairs $G_1, \dots, G_M$ such that if $G_j = \{j_1, j_2\}$, then $X_{j_1}, X_{j_2}$ are in the same equicorrelated block and $|\beta_{j_1} + \beta_{j_2}| < \frac{b}{m}$. This leaves $M_{}' = p - 2 M$ ``singleton" features which are not in any pair.

Let $\alpha_1, \alpha_2$ be constants such that $\alpha_1 \cdot \alpha_2 > \frac{c_0 \sqrt{n} b}{m}$. Assume that $\alpha_1 < 0.001$ and $M_{}' \le \frac{p}{50}$. Then there exist universal constants $C_1, C_2, C_3$ such that if $\tau$ is the number of discoveries made by the procedure,

$$\frac{\mathbb{E}_{w, \beta}[\tau]}{p} \le \frac{C_1 M_{}' + C_2}{p} + C_3 \alpha_2.$$

\begin{proof} To start, relabel the features such that the ``singleton" statistics are labelled $W_1, \dots, W_{M_{}'}$. We will make the worst-case assumption that $W_1, \dots, W_{M_{}'} = + \infty$. This assumption is permissible since it only increases the number of discoveries made by the procedure, as formalized in Lemma \ref{lem::worstcaseavgequi}. Our proof now proceeds in three steps. 

\underline{Step 1}: First, we reformulate the problem in terms of $\eta = \sign(\sorted(W)) \in \{-1,1\}^p$, since the power of knockoffs depends only on the values of the random vector $\eta$. We introduce some notation for this purpose. For $\epsilon \in \{-1, 1\}^p$ and any $k \in \mathbb{N}$, let $V_k^+(\epsilon)$ be the number of positive $1$'s at or before the $k$th element of $\epsilon$. Formally,
$$V_k^+(\epsilon) = \# \{ j \le k : \epsilon_j = 1 \}. $$
Note for $k > p$, $V_k^+(\epsilon) = V_p^+(\epsilon)$ since $\epsilon$ only has $p$ coordinates. This means that  $V_k^+(\epsilon)$ is uniformly bounded by $p$. We also observe that $V_k^+(\epsilon)$ is nondecreasing in $k$. Define
\begin{equation}\label{eq::psidef}
\psi(\epsilon) = \max_{1 \le k \le \infty} \left\{ k : \frac{k - V_k^+(\epsilon) + 1}{V_k^+(\epsilon)} \le q \right\} = \max_{1 \le k \le \infty} \left\{ k : k \le (1+q) V_k^+(\epsilon) - 1 \right\} \le \lfloor (1+q) p - 1 \rfloor,
\end{equation}
where by convention $\psi(\epsilon) = 0$ if the set in the middle is empty. Intuitively, we can think of $\psi(\eta)$ as a reformulation of the data-dependent threshold in (\ref{eq::ddthreshold}), where the knockoffs procedure selects all features in $\sorted(W)$ which appear before position $\psi(\eta)$ and have positive signs. This implies that $\tau \le \psi(\eta)$, as we prove formally in Lemma \ref{lem::psibound}. Furthermore, note that we let $\psi$ take values greater than $p$ to ensure that the number of discoveries $\tau$ is a deterministic function of $\psi(\eta)$, which will be important in Step $3$---see Lemma \ref{lem::psibound} for details. By the prior analysis, it suffices to show that
$$\mathbb{E}_{w, \beta} [\psi(\eta)] \le C_1 M_{}' + C_2 + C_3 \alpha_2 p.$$ 
We may decompose
\begin{equation}\label{eq::indicatordecomp}
\psi(\eta) = \sum_{k=\lfloor (1+q) M_{}' - 1 \rfloor}^{\lfloor (1+q) p - 1 \rfloor} k \cdot \mathbb{I}(\psi(\eta) = k),
\end{equation}
where we begin the sum at $\lfloor (1+q) M_{}' - 1 \rfloor$ because we assume the first $M_{}'$ coordinates of $\eta$ correspond to the singleton features and are guaranteed to be $1$'s, so we must have that $\psi(\eta) \ge \lfloor (1+q) M_{}' - 1 \rfloor$. If we define
\begin{equation}\label{eq::deflambdak}
    \Lambda_k = \{\epsilon \in \{-1,1\}^p : \psi(\epsilon) = k \} \cap \{\epsilon : \mathbb{P}_{\beta, W}(\eta = \epsilon) \ne 0 \},
\end{equation}
then
\begin{equation}\label{eq::endstep1}
  \mathbb{E}_{w, \beta} \left[\psi(\eta) \right] = \sum_{k=\lfloor (1+q) M_{}' - 1 \rfloor}^{\lfloor (1+q) p - 1 \rfloor} k \cdot \mathbb{P}_{\beta, W}\left(\eta \in \Lambda_k \right).
\end{equation}

\underline{Step 2}: In this step, we will show that $\mathbb{P}\left(\eta = \epsilon \right)$ cannot be much larger than $2^{-M}$ for most $\epsilon \in \{-1,1\}^p$. The $2^{-M}$ comes because for each group $j \in [M]$, the conditional distribution of $\sorted(W_{G_j})$ given $(W_{-G_j}, |\sorted(W_{G_j})|)$ is approximately symmetric with high probability (see Corollary \ref{corr::condapprxreconst}). 

The main task of this section will be to introduce notation such that we can write the event $\eta = \epsilon$ in terms of the random variables $\sign(\sorted(W_{G_j}))$. Note for the rest of this proof and Lemma \ref{lem::swan} (stated later), we use $i$ and $\otherj$ to denote indices of $\sorted(W)$ and $j$ to denote indices of the unsorted feature statistics $W$ or the pairs $G_j$. In other words, $i, \otherj$ mean ``post-sorting," and $j$ means ``pre-sorting."

This step is the most conceptually challenging, so we will divide it into four sub-steps. Our notation will explicitly account for the singleton statistics, but we advise the reader to focus on the non-singleton statistics. 

\underline{Step 2a}: First, we will define a random vector $R = (R_1,\dots, R_p) \in \{0, \dots, M\}^p$. 
Intuitively, for $1 \le i \le p$, $R_i$ takes the (post-sorting) feature statistic $\sorted(W)_i$ and tells us which (pre-sorting) pair $G_j$ it came from. Formally, we define
$$R_i = \begin{cases} j & \sorted(W)_i \text{ came from pair } G_j \\ 
0 & \sorted(W)_i \text{ corresponds to a singleton statistic. }\end{cases}  $$
As an example, fix $p = 7$ and suppose the pairs are $G_1 = \{1, 2\}, G_2 = \{3, 4\}, G_3 = \{5, 6\}$, and the $7$th feature statistic is a singleton statistic (which we assume equals $+ \infty$). Then suppose we observe
\begin{equation}\label{eq::wstatexample}
W = (W_1, W_2, W_3, W_4, W_5, W_6, W_7) = (1.3, 5, 0, -1.5, -2,-2.3, +\infty).
\end{equation}
By the definition of our sorting function, this implies that 
\begin{equation}\label{eq::sortedwstatexample}
\sorted(W) = (+ \infty, 5, -2.3, -2, -1.5, 1.3, 0). 
\end{equation}
The first sorted coordinate $\sorted(W)_1 = + \infty$ corresponds to the unsorted feature statistic $W_7$, which is a singleton statistic. Thus, $R_1 = 0$. The second sorted coordinate $\sorted(W)_2 = 5$ corresponds to the unsorted feature statistic $W_2$ which belongs to pair $G_1$, so we set $R_2 = 1$. Continuing on, we find that
\begin{equation}\label{eq::Rvecexample}
    R = (0, 1, 3, 3, 2, 1, 2).
\end{equation}
In general, the first $M_{}'$ coordinates of $R$ will equal zero, since the $M_{}'$ singleton statistics equal $+\infty$ and make up the first $M_{}'$ coordinates of $\sorted(W)$. However, the important information contained by $R$ is captured by the last $p- M_{}'$ coordinates, which allow us to convert between $\sorted(W)_i$ and the groupings $W_{G_j}$. Crucially, $R$ does not contain any information about the within-pair rankings of feature statistics, and thus $R$ is in the sigma-algebra generated by $\{|\sorted(W_{G_j})|\}_{j=1}^{M}$. This is important, since it means that $\sorted(W_{G_j})$ is approximately symmetric conditional on $R$---recall that Corollary \ref{corr::condapprxreconst} tells us  $\sorted(W_{G_j})$ is approximately symmetric when we condition on $|\sorted(W_{G_j})|$ and $W_{-G_j}$, but not necessarily if we condition on $|W_{G_j}|$. 

\underline{Step 2b}: The random vector $R$ tells us for each $i$ that $\sorted(W)_i$ corresponds to one of the two elements of $\sorted(W_{G_{R_i}})$, since $G_{R_i}$ denotes the pair appearing at position $i$ in $\sorted(W)$, and $W_{G_{R_i}}$ denotes the feature statistics belonging to that pair. However, we do not know \textit{which} of the two elements in $\sorted(W_{G_{R_i}}) \in \mathbb{R}^2$ correspond to $\sorted(W)_i$. To fill this notational gap, we will introduce the random vector $N(R) = (N_1(R), \dots, N_p(R)) \in \{1, 2\}^p$ such that $\sorted(W)_i = \sorted(W_{G_{R_i}})_{N_i(R)}$. 

To do this, we note that both $\sorted(W)$ and $\sorted(W_{G_{R_i}})$ sort feature statistics in descending order of absolute value. For this reason, if $G_{j} = \{j_1, j_2\}$, then $W_{j_1}$ will appear before $W_{j_2}$ in $\sorted(W_{G_j})$ if and only if $W_{j_1}$ appears before $W_{j_2}$ in $\sorted(W)$. Therefore, we can define $N_i(R)$ as follows for $1 \le i \le p$:
$$N_i(R) = \begin{cases} 0 & \text{ sorted}(W)_i \text{ corresponds to a singleton statistic } \\
1 & \text{ sorted}(W)_i \text{ appears before the other member of its pair in } \sorted(W) \\
    2 & \text{ sorted}(W)_i \text{ appears after the other member of its pair in } \sorted(W). \end{cases} $$
This guarantees that $\sorted(W)_i = \sorted(W_{G_{R_i}})_{N_i(R)}$ for each $M_{}' + 1 \le i \le p$. More importantly, we have defined $N(R)$ such that it is a deterministic function of $R$, meaning that $W_{G_j}$ should be approximately symmetric even conditional on $N(R)$.

As a concrete example, recall equations (\ref{eq::wstatexample}), (\ref{eq::sortedwstatexample}), and (\ref{eq::Rvecexample}). To find $N_3(R)$, we note that $R_3 = 3$, meaning that $\sorted(W)_3$ comes from $G_3$. Since no previous coordinate of $R$ equals $3$, we conclude $N_3(R) = 1$. On the other hand, to find $N_4(R)$, we note $R_4 = 3$ as well, meaning $\sorted(W)_4$ also comes from $G_3$. Since $\sorted(W)_4$ is not the first coordinate of $\sorted(W)$ to come from pair $G_3$, we set $N_4(R) = 2$. Continuing, we get that
\begin{equation}\label{eq::Nvecexample}
N(R) = (0, 1, 1, 2, 1, 2, 2).
\end{equation}
As always, the $0$ in $N(R)$ signals the presence of a (nonrandom) singleton statistic.

\underline{Step 2c}: We are now ready to write the event $\eta = \epsilon$ in terms of $\{\sign(\sorted(W_{G_j}))\}_{j=1}^M$ and $R$. Let $\epsilon \in \{-1, 1\}^p$ such that $\epsilon_{\otherj} = 1$ for $1 \le {\otherj} \le M_{}'$. Recall that our notation guarantees that $\sorted(W)_i = \sorted(W_{G_{R_i}})_{N_i(R)}$. Thus, the following events are equivalent:
\begin{align*}
\sign(\sorted(W)) = \epsilon & \Leftrightarrow
 \left(\bigcap_{i=M_{}'+1}^{p} \sign(\sorted(W_{G_{R_i}}))_{N_i(R)} = \epsilon_{i}\right) .
\end{align*}
 This motivates the following decomposition by the chain-rule of conditional probability. Define probabilities
$$p(r) = \mathbb{P}\left(R = r\right), $$
\begin{equation}\label{eq::defprobepsi}
p(\epsilon_i | \epsilon_{(M_{}'+1):(i-1)}, r) = 
    \mathbb{P}\left( \sign(\sorted(W_{G_{r_i}}))_{N_i(r)} = \epsilon_i 
    \, \Bigg| \, 
    \bigcap_{\otherj=M_{}'+1}^{i-1} \sign(\sorted(W_{G_{r_{\otherj}}}))_{N_{\otherj}(r)} = \epsilon_\otherj \, ,\,  R = r \right), 
\end{equation}
where we also adopt the convention that for all $i$, $p(\epsilon_i | \epsilon_{(M_{}'+1):(i-1)}, r) = 0$ unless $\epsilon_{\otherj} = 1$ for $1 \le {\otherj} \le M_{}'$, since the first $M_{}'$ statistics in $\sorted(W)$ correspond to the (positive) singleton statistics. Note also that in the previous statement, we define $i:\otherj$ to be the empty set when $i > \otherj$. Now, by the law of total probability and the chain-rule of conditional probability, 

\begin{equation}\label{eq::grouprankcond}
    \mathbb{P}\left(\eta = \epsilon \right)
    = 
    \sum_{r : p(r) > 0} p(r) \prod_{i=M_{}'+1}^p p(\epsilon_i | \epsilon_{(M_{}'+1):(i-1)},r),
\end{equation}
where again, the product begins at $M_{}'+1$ because we assume the first $M_{}'$ values correspond to the singleton statistics.

Intuitively, we should expect (\ref{eq::grouprankcond}) to have a value which does not substantially exceed $2^{-M}$.  To see this, imagine iterating through $p(\epsilon_i | \epsilon_{(M_{}'+1):(i-1)}, r)$ as we increment $i$ from $M_{}'+1$ to $p$. For any $i$, if $\sorted(W)_i$ corresponds to a group we have not seen before---meaning $N_i(r) = 1$---then we should expect $p(\epsilon_i | \epsilon_{(M_{}'+1):(i-1)}, r) \approx 1/2$ since $\sorted(W_{G_{R_i}})$ is approximately symmetric. Slightly more formally, this holds because when $N_i(r) = 1$, we are only conditioning on $R$ and the signs of the other groups, both of which are contained in the sigma-algebra generated by $\{|\sorted(W_{G_{r_i}})|, W_{-{G_{r_i}}}\}$, in accordance with Corollary \ref{corr::condapprxreconst}. On the other hand, when we consider an $i$ such that $N_i(r) = 2$, we can only trivially upper bound $p(\epsilon_i | \epsilon_{(M_{}'+1):(i-1)}, r)$ by $1$, since we are conditioning on the sign of $\sorted(W_{G_{r_i}})_1$, which violates the assumption of Corollary \ref{corr::condapprxreconst}.

To formalize this idea, we need one final piece of notation, which we define in Step 2d.

\underline{Step 2d}: The vector $R$ allows us to convert from sorted coordinates $i$ into unsorted groups $G_j$. The random vector $C(R) \in [p]^{M}$ will do (approximately) the inverse. We define $C(R)$ elementwise, such that for $j \in [M]$, $C_j(R)$ equals the coordinate of the first element of pair $G_j$ in $\sorted(W)$. In the example from equations (\ref{eq::wstatexample}), (\ref{eq::sortedwstatexample}), (\ref{eq::Rvecexample}), we would have that
$$C(R) = (2, 5, 3). $$
For example, $C_1(R) = 2$ because pair $j=1$ appears for the first time in $\sorted(W)$ at coordinate $i=2$. In particular, $C_j(R) = i$ guarantees that $\sorted(W)_i$ corresponds to pair $G_j$ and $N_i(R) = 1$. With this in mind, we can state our main Lemma for Step $2$.

\begin{lemma}\label{lem::swan} Let $\alpha_1, \alpha_2$ be constants such that $\alpha_1 \alpha_2 > \frac{c_0 \sqrt{n} b}{m}$. For any $1 \le j \le M$ and for any $\Lambda \subset \{-1,1\}^p$, 
\begin{equation}\label{eq::swan}
\alpha_2 > 
\sum_{\epsilon \in \Lambda} 
\sum_{r : p(r) > 0} p(r) 
 \mathbb{I} \left(p(\epsilon_{C_j(r)} | \epsilon_{(M_{}'+1):(C_j(r)-1)}, r) > 0.5 + \alpha_1 \right) \prod_{i \ne C_j(r), i \ge M_{}' + 1} p(\epsilon_i | \epsilon_{(M_{}'+1):(i-1)}, r).
\end{equation}
\begin{proof}
We defer the proof to Appendix \ref{appendix::avgequicomputations} for brevity, but note that this follows directly from Corollary \ref{corr::condapprxreconst}. See the preceding logic for intuition.
\end{proof}
\end{lemma}

Intuitively, $C_j(r)$ singles out the first term in the conditional decomposition of $\mathbb{P}(\eta = \epsilon)$ corresponding to $G_j$, and (\ref{eq::swan}) indicates that this term $p(\epsilon_{C_j(r)} | \epsilon_{(M_{}'+1):(C_j(r)-1)}, r) \approx 1/2$ with high probability. Slightly more formally, we could capture this idea by stating that $$\mathbb{P}\left( (\eta, R) \in \{ (\epsilon, r) : p(\epsilon_{C_j(r)} | \epsilon_{(M_{}'+1):(C_j(r)-1)}, r) > 0.5 + \alpha_1\}\right)$$ is quite small.

\underline{Step 3}: In this step, we combine Steps $1$ and $2$ to prove the final result. Combining equation (\ref{eq::endstep1}) and equation (\ref{eq::grouprankcond}), we obtain
$$ \mathbb{E}_{w, \beta}[\psi(\eta)] = \sum_{k=\lfloor (1+q) M_{}' - 1 \rfloor}^{\lfloor (1+q) p - 1 \rfloor} k \sum_{r : p(r) > 0} \sum_{\epsilon \in \Lambda_k} p(r) \prod_{i=M_{}'+1}^{p} p(\epsilon_i | \epsilon_{(M_{}'+1):(i-1)},r).$$
For convenience of notation, we define
\begin{equation}\label{eq::condepsr}
p(\epsilon|r) = \prod_{i=M_{}'+1}^{p} p(\epsilon_i | \epsilon_{(M_{}'+1):(i-1)},r) = \mathbb{P}\left(\eta = \epsilon \mid R = r \right)
\end{equation}
and 
$$n(\epsilon,r) = \sum_{j=1}^{M} \mathbb{I}\left(p(\epsilon_{C_j(r)} | \epsilon_{(M_{}'+1):(C_j(r)-1)}, r) \ge 0.5 + \alpha_1\right).  $$
Intuitively, $n(\epsilon,r)$ counts the number of pairs $G_j$ which violate the (unconditional) total variation bound from Proposition \ref{prop::apprxeqreconst} by more than $\alpha_1$ in each conditional decomposition.

Note that Lemma \ref{lem::swan} guarantees that because the sets $\{\Lambda_k\}$ are disjoint, for each pair $1 \le j \le M$, we have
\begin{equation}\label{eq::preswanconstraint}
\sum_{k=\lfloor (1+q) M_{}' - 1 \rfloor}^{\lfloor (1+q) p - 1 \rfloor} 
\sum_{\epsilon \in \Lambda_k} 
\sum_{r : p(r) > 0} \
p(r)  p(\epsilon|r)  \frac{\mathbb{I}\left(p(\epsilon_{C_j(r)} | \epsilon_{(M_{}'+1):(C_j(r)-1)}, r) \ge 0.5 + \alpha_1\right)}{p(\epsilon_{C_j(r)} | \epsilon_{(M_{}'+1):(C_j(r)-1)}, r) }
\le \alpha_2.  
\end{equation}
The definition of $\Lambda_k$ explicitly guarantees the denominator in the above expression is nonzero. Moreover, since $p(\epsilon_{C_j(r)} | \epsilon_{(M_{}'+1):(C_j(r)-1)}, r) \le 1$, this implies
$$ \sum_{k=\lfloor (1+q) M_{}' - 1 \rfloor}^{\lfloor (1+q) p - 1 \rfloor} 
\sum_{\epsilon \in \Lambda_k} 
\sum_{r : p(r) > 0} \
p(r)  p(\epsilon|r) \mathbb{I}\left(p(\epsilon_{C_j(r)} | \epsilon_{(M_{}'+1):(C_j(r)-1)}, r) \ge 0.5 + \alpha_1\right)
\le \alpha_2, $$
and since this holds for all $1 \le j \le M$, we sum over $j$ to obtain
\begin{align}
    M  \cdot \alpha_2 
    \ge & \sum_{k=\lfloor (1+q) M_{}' - 1 \rfloor}^{\lfloor (1+q) p - 1 \rfloor} 
    \sum_{\epsilon \in \Lambda_k} 
    \sum_{r : p(r) > 0} p(r)  p(\epsilon|r)  \sum_{j=1}^{M} 
    \mathbb{I}\left(p(\epsilon_{C_j(r)} | \epsilon_{(M_{}'+1):(C_j(r)-1)}, r) \ge 0.5 + \alpha_1\right)  
    \nonumber \\
    = & \sum_{k=\lfloor (1+q) M_{}' - 1 \rfloor}^{\lfloor (1+q) p - 1 \rfloor} 
    \sum_{\epsilon \in \Lambda_k} 
    \sum_{r : p(r) > 0} p(r)  p(\epsilon|r)  n(\epsilon,r) \label{eq::swanconstraint} 
\end{align}
by the definition of $n(\epsilon ,r)$. Using this notation, we want to bound
\begin{equation}\label{eq::avgequiobjective}
    \mathbb{E}_{w, \beta}[\psi(\eta)] = \sum_{k=\lfloor (1+q) M_{}' - 1 \rfloor}^{\lfloor (1+q) p - 1 \rfloor} k \sum_{r : p(r) > 0} \sum_{\epsilon \in \Lambda_k} p(r)  p(\epsilon|r).
\end{equation}

Fix $\ell_0 \in \mathbb{N}$ such that $\ell_0 > 200$. The main idea will be to split the sum in (\ref{eq::avgequiobjective}) into the two parts defined below:
$$U_k(r) = \left \{\epsilon \in \Lambda_k : n(\epsilon,r) \ge  \frac{\min(k,p)}{\ell_0} \right \}  \text{ and } U_k^c(r) = \left \{\epsilon \in \Lambda_k : n(\epsilon,r) <  \frac{\min(k,p)}{\ell_0} \right \}.  $$
Intuitively, $U_k(r)$ denotes the set of $\epsilon$ such that in the conditional decomposition of $\mathbb{P}\left(\eta = \epsilon | R = r\right)$ as defined by (\ref{eq::condepsr}), many of the conditional probabilities in the product deviate substantially from $0.5$. However, Lemma \ref{lem::swan} tells us that the overall probability assigned to this set must be fairly low. In contrast, $U_k^c(r)$ contains a very large number of $\epsilon$, but the probability that $\eta$ equals any $\epsilon \in U_k^c(r)$ is approximately $2^{-M}$.

Applying this decomposition, we have
\begin{equation}\label{eq::splitavgequi}
   \mathbb{E}_{w, \beta}[\psi(\eta)] = \sum_{k=\lfloor (1+q) M_{}' - 1 \rfloor}^{\lfloor (1+q) p - 1 \rfloor} k \sum_{r : p(r) > 0} p(r) \left[ \sum_{\epsilon \in U_k(r)} p(\epsilon|r) +  \sum_{\epsilon \in U_k^c(r)} p(\epsilon|r) \right].
\end{equation}
We will bound the two terms in (\ref{eq::splitavgequi}) separately. First, we consider the set $U_k(r)$. In this set, $n(\epsilon, r)$ is fairly large, which intuitively indicates that many of the $\sign(\sorted(W_{G_j}))$ differ in conditional total variation from $- \sign(\sorted(W_{G_j}))$ conditional on the values of the other groups. We can bound this term fairly quickly using (\ref{eq::swanconstraint}). In particular, we have guaranteed that $(1+q) \cdot \ell_0 \cdot n(\epsilon, r) \ge (1+q) \min(k,p) \ge k$ for $\epsilon \in U_k(r)$. Applying (\ref{eq::swanconstraint}), this implies,
\begin{align*}
 (1+q) \ell_0 M \alpha_2  &\ge  (1 + q) \ell_0 \sum_{k=\lfloor (1+q) M_{}' - 1 \rfloor}^{\lfloor (1+q) p - 1 \rfloor} 
    \sum_{r : p(r) > 0}  
    \sum_{\epsilon \in \Lambda_k} 
    p(r) p(\epsilon|r)  n(\epsilon,r) \\
    &\ge \sum_{k=\lfloor (1+q) M_{}' - 1 \rfloor}^{\lfloor (1+q) p - 1 \rfloor} 
    \sum_{r : p(r) > 0} 
     \sum_{\epsilon \in U_k(r)} k \cdot p(r)  p(\epsilon|r).
\end{align*}
Together with (\ref{eq::splitavgequi}), this implies
\begin{equation}\label{eq::splitavgequi2}
   \mathbb{E}_{w, \beta}[\psi(\eta)] \le (1+q) \ell_0 M \alpha_2  + \sum_{k=\lfloor (1+q) M_{}' - 1 \rfloor}^{\lfloor (1+q) p - 1 \rfloor} k \sum_{r : p(r) > 0} p(r) \sum_{\epsilon \in U_k^c(r)} p(\epsilon|r).
\end{equation}
To bound the term including $U_k^c(r)$, we note $n(\epsilon, r)$ is fairly small for each $\epsilon \in U_k^c(r)$. This means that most of the conditional distributions of $\sign(\sorted(W_{G_j}))$ do not differ too much from those of $- \sign(\sorted(W_{-G_j}))$. With some effort, we will be able to compare this sum to a geometric series to show it is bounded by a linear function of $M_{}'$. 

The key intuition is as follows. For any group ranking $r : p(r) > 0$, the first $k$ coordinates of $\sorted(W)$ must contain elements from at least $(k-M_{}')/2$ unique pairs. In particular, this means it contains at least $(k-M_{}')/2$ feature statistics which appear ``first" in their pair, i.e., have a larger absolute value than the other feature statistic in their pair. For any $\epsilon \in U_k^c(r)$, we know that in the conditional decomposition of $p(\epsilon | r)$, all but at most $\frac{\min(k,p)}{\ell_0}$ of the groups $G_j$ are mostly symmetric conditional on $|\sorted(W_{G_j})|$ and the values of $W_{-G_j}$. This means that in the conditional decomposition of $p(\epsilon | r)$, the first $k$ terms in the product should yield a value less than $(0.5 + \alpha_1)^{(k-M_{}')/2 - \frac{\min(k,p)}{\ell_0}}$. Then, using the definition of $\Lambda_k$, we will bound the number of distinct values that $\epsilon_{1:k}$ can take.

To formalize this intuition, for $k \le p$, define $L_k^c(r)$ as the possible values $\epsilon_{1:k}$ can take for $\epsilon \in U_k^c(r)$,  i.e.,
$$L_k^c(r) = \{\epsilon_{1:k} : \epsilon \in U_k^c(r) \} \subset \{-1, 1\}^k .$$
For $k > p$, we define $L_k^c(r) = U_k^c(r)$, since in this case $\epsilon$ has only $p < k$ coordinates. Observe by definition of $p(\epsilon | r)$, when $k \le p$,
\begin{align}
    \sum_{\epsilon \in U_k^c(r)} p(\epsilon|r) 
=& 
    \sum_{\epsilon \in U_k^c(r)} \mathbb{P}\left(\eta = \epsilon \mid R = r \right) \nonumber \\
\le&
    \sum_{\epsilon_{1:k} \in L_k^c(r)} \mathbb{P}\left(\eta_{1:k} = \epsilon_{1:k} \mid R = r \right) \nonumber \\
=& 
    \sum_{\epsilon_{1:k} \in L_k^c(r)}
    \prod_{i=M_{}'+1}^{k} p(\epsilon_i | \epsilon_{(M_{}'+1):(i-1)},r). \nonumber
\end{align}
Of course, the same logic applies when $k > p$ as long as we are careful to remember that $\eta$ and $\epsilon$ only have $p$ coordinates. This implies that for any $k$,
\begin{equation}
\sum_{\epsilon \in U_k^c(r)} p(\epsilon|r)
\le 
\sum_{\epsilon \in L_k^c(r)} \, \, 
\prod_{i=M_{}'+1}^{\min(k,p)} p(\epsilon_i | \epsilon_{(M_{}'+1):(i-1)},r) \label{eq::avgequilowerprod}
\end{equation}
where we are abusing notation slightly in the last equation, since any $\epsilon \in L_k^c(r)$ is an element of $\{-1,1\}^{\min(k,p)}$. Now, consider the product in (\ref{eq::avgequilowerprod}). When $k > M_{}'$, there must be $\min(k, p) - M_{}'$ terms corresponding to non-singleton feature statistics, and at least $(\min(k,p) - M_{}')/2$ of these correspond to the first sign from one of the groups. However, by the definition of $U_k^c(r)$, at most $\min(k, p)/\ell_0$ of these terms can be greater than $0.5 + \alpha_1$. This implies
\begin{equation*}
    \sum_{\epsilon \in U_k^c(r)} p(\epsilon|r) 
    \le 
    |L_k^c(r)| (0.5 + \alpha_1)^{(\min(k,p) - M_{}')/2 - \min(k,p)/\ell_0}.
\end{equation*}
Note that for every $\epsilon \in U_k^c(r)$, however, we must have that $\psi(\epsilon) = k$ since $U_k^c(r) \subset \Lambda_k$. By the definition of $\psi$, this implies that $V_k^+(\epsilon) = \left \lceil \frac{k+1}{1+q}\right \rceil$. Briefly, this must hold because if $V_k^+(\epsilon)$ were any larger, then $\psi(\epsilon) \ge k+1$, but if $V_k^+(\epsilon)$ were any smaller, then $\frac{k - V_k^+(\epsilon) + 1}{V^+_k(\epsilon)} \ge q$, contradicting the definition of $\psi$. We prove this formally in Lemma \ref{lem::psibound}. Combining this constraint with the fact that the first $M_{}'$ coordinates of $\epsilon$ must equal one, this means that for $k > (1+q)M_{}'$, we have
$$|L_k^c(r)| \le \# \{ \epsilon_{1:k} : \epsilon \in \{-1,1\}^p \text{ and }  \psi(\epsilon) = k \} 
 \le \binom{\min(k, p) - M_{}'}{\lceil \frac{k+1}{1+q} \rceil-M_{}'},$$
where the binomial coefficient results from the fact that there are $\min(k,p) - M_{}'$ free coordinates of $\epsilon$, of which $\left \lceil \frac{k+1}{1+q} \right \rceil - M_{}'$ must equal one. 

For an arbitrary constant $C_0 > (1 + q)$, the previous logic yields the following bound:
 
 \begin{align}
    \sum_{k=\lfloor (1+q) M_{}' - 1 \rfloor}^{\lfloor (1+q) p - 1 \rfloor} k \sum_{r : p(r) > 0} p(r) \sum_{\epsilon \in U_k^c(r)} p(\epsilon|r) 
\le & 
    C_0 M_{}' + \sum_{k = C_0 M_{}'}^{\lfloor (1+q) p - 1 \rfloor} k \sum_{r : p(r) > 0} p(r) \sum_{\epsilon \in U_k^c(r)} p(\epsilon|r) \nonumber \\
\le &
    C_0 M_{}' + \sum_{k= C_0 M_{}'}^{\lfloor (1+q) p - 1 \rfloor} k \sum_{r : p(r) > 0} p(r) \binom{\min(k, p) - M_{}'}{\lceil \frac{k+1}{1+q} \rceil - M_{}'} (0.5 + \alpha_1)^{\frac{\min(k,p) - M_{}'}{2} - \frac{\min(k,p)}{\ell_0} } \nonumber \\
\le & C_0 M_{}' + 
    \sum_{k=C_0 M_{}'}^{\lfloor (1+q) p - 1 \rfloor} k \cdot \binom{\min(k, p) - M_{}'}{\lceil \frac{k+1}{1+q} \rceil - M_{}'} \cdot (0.5 + \alpha_1)^{(\min(k,p) - M_{}')/2 - \min(k,p)/\ell_0 }. \label{eq::lastnontecheq}
 \end{align}
 
The main idea is now to approximate the binomial term as an exponential term that can be dominated by the exponential term. This allows us to simplify (\ref{eq::lastnontecheq}) and bound it by a linear function of $M_{}'$. 

 \begin{lemma}\label{lem::binomial2geometric} For sufficiently large $C_0$ and some universal constants $C_1, C_2$, 
 $$\sum_{k=C_0 M_{}'}^{\lfloor (1+q) p - 1 \rfloor} k \cdot \binom{\min(k, p) - M_{}'}{\lceil \frac{k+1}{1+q} \rceil - M_{}'} \cdot (0.5 + \alpha_1)^{(\min(k,p) - M_{}')/2 - \min(k,p)/\ell_0 } \le C_1 M_{}' + C_2.$$
Note we use the assumptions that $M_{}' \le \frac{p}{50}$, $\ell_0 > 200$, $\alpha_1 < 0.001$, and $q \le 0.1$.
\begin{proof} Although this argument is fairly simple, it is quite tedious, so we prove it in Appendix \ref{appendix::avgequicomputations}.
\end{proof}
\end{lemma}

Combining this with (\ref{eq::splitavgequi2}), we conclude that
\begin{align}
   \mathbb{E}_{w, \beta}[\psi(\eta)] \le &  
        (1+q) \ell_0 M \alpha_2 
        + \sum_{k=\lfloor (1+q) M_{}' - 1 \rfloor}^{\lfloor (1+q) p - 1 \rfloor} k \sum_{r : p(r) > 0} p(r) \sum_{\epsilon \in U_k^c(r)} p(\epsilon|r) \nonumber \\
& \le
    (1+q) \ell_0 M \alpha_2 + C_1 M_{}' + C_2 \nonumber \\
& \le 
    (1+q) \ell_0 p \alpha_2 + C_1 M_{}' + C_2,
\end{align}
where the last line follows because $M$ is the number of groups, so $M \le p$. This implies that
$$ \power(w, \beta) = \frac{\mathbb{E}_{w, \beta}[\tau]}{p} \le \frac{C_1 M_{}'}{p} + (1+q)\ell_0 \alpha_2 + \frac{C_2}{p}$$
as desired.
\end{proof}
\end{theorem}

Now we can prove Theorem \ref{thm::avgequi}.

\begin{corollary}[Theorem \ref{thm::avgequi}] Let $X$ be a block-equicorrelated Gaussian design with $\ell$ blocks of equal size and within-block correlations of at least $0.5$. Assume $Y \mid X$ follows a noisy single-index model and let $\mathcal{W}$ be the class of all permutation invariant feature statistic functions. Suppose we aim to control the FDR at level $q \le 0.1$ using SDP knockoffs.

Sample $\betan$ uniformly from $\mathcal{C}_{p,b}$, the $p$-dimensional hypercube centered at $0$ with fixed side-length $b$. Let $n$ be the number of data points and suppose there exists some $\epsilon > 0$ such that $n = o\left(\left(\frac{p}{\ell}\right)^{2-2\epsilon}\right)$. Further assume $\ell = o\left(\frac{p}{\log^{1/\epsilon} p}\right)$. Then as $n,p \to \infty$, for any $\delta > 0$,
$$\lim_{n \to \infty} \mathbb{P}_{\betan \sim \mathcal{C}_{p,b}} \left( \sup_{w \in \mathcal{W}} \power(\betan,W) \ge \delta \right) = 0.$$
\begin{proof} Note each coefficient $\betan_j$ is sampled independently and uniformly from $(-b, b)$. To construct pairs of approximately alternating signs, as in Theorem \ref{thm::avgequiproof}, fix an integer $m \in \mathbb{N}$ and let $\dkn$ be the set of features in the $k$th equicorrelated block of $\Sigman$. Define buckets $\{ \bjkn \}$ for $1 \le j \le m$, $1 \le k \le \ell$ as follows:
\begin{equation}\label{eq::bucketdef}
    \bjkn = \left\{i \in \dkn  : \frac{(j-1) \cdot b}{m} \le |\betan_i| \le \frac{j \cdot b}{m}  \right\}.
\end{equation}

Furthermore, define $\bjknplus = \left \{i \in \bjkn : \betan_i > 0\right\}$ and $\bjknminus = \left \{i \in \bjkn : \betan_i < 0\right\}$. For each bucket $\bjkn$, we can create $\min\left(\left|\bjknplus\right|, \left|\bjknminus\right|\right)$ pairs of coefficients such that the sum of each pair is less than $\frac{b}{m}$.  Using this bucketing, we have $M_{}'$ singleton features where
$$M' = p - 2 \sum_{k=1}^\ell \sum_{j=1}^{m} \min(|\bjknplus|, |\bjknminus|).$$
Now, we are ready to show that $\mathbb{P}_{\betan \sim \mathcal{C}_{p,b}}(\sup_{w \in \mathcal{W}} \power(\betan, W) > \delta)$ is bounded by the probability that $M_{}'$ is larger than $\delta_{n,m,p} \cdot p$, for $\delta_{n,m,p}$ defined in a moment. Note that for the rest of this proof, we use $\mathbb{P}$ as a shorthand for $\mathbb{P}_{\betan \sim \mathcal{C}_{p,b}}$. Using the notation of Theorem \ref{thm::avgequiproof}, if we set $$\alpha_2 = \frac{2000 c_0 \sqrt{n} b}{m} $$
and we set $\alpha_1 = 0.0005 < 0.001$, then there are constants $C_1, C_2, C_3$ such that
$$ \sup_{w \in \mathcal{W}} \power(w, \betan) \le \frac{C_1 M_{}' + C_2}{p} + C_3 \cdot \frac{b \sqrt{n}}{m}.$$
Note that we obtain the ``sup" in the preceding equation because Theorem \ref{thm::avgequiproof} holds uniformly over all $w \in \mathcal{W}$. Now, the event $\sup_{w \in \mathcal{W}} \power(w, \betan) \ge \delta$ implies that
$$\delta \le \sup_{w \in \mathcal{W}} \power(w, \betan) \le \frac{C_1 M_{}' + C_2}{p} + C_3 \cdot \frac{b \sqrt{n}}{m},$$
which implies
$$  M_{}'
\ge 
    \frac{p}{C_1} \left(\delta - C_3 \cdot \frac{b \sqrt{n}}{m}\right) - \frac{C_2}{C_1}.
$$
Define $\delta_{p,m,n}$ to be this quantity divided by $2p$, i.e.,
$$\delta_{p,m,n} = \frac{1}{2 C_1} \left(\delta - C_3 \cdot \frac{b \sqrt{n}}{m} - \frac{C_2}{p} \right). $$
For now, assume that $\delta_{p,m,n} > 0$: we will show that this holds for a carefully chosen $m$ and sufficiently large $p$ later. By the preceding argument, we have that
\begin{equation}\label{eq::powerintermsofT}
\mathbb{P} \left(\sup_{w \in \mathcal{W}} \power(w, \betan) \ge \delta \right) \le \mathbb{P} \left(M_{}' \ge 2 \delta_{p,m,n} p \right).
\end{equation}
In Appendix \ref{subsec::randomcoef}, we apply concentration inequalities to prove that for any $\delta > 0$ and for $K$ such that $m K \le \frac{p}{4 \ell}$,
\begin{equation}\label{eq::finaltresult}
\mathbb{P} \left(M_{}' \ge 2 \delta p \right) \le 2 \ell m \exp(-\delta^2 K) + \ell m \exp \left( - \frac{p}{5\ell m} \right).
\end{equation}
As a result,
\begin{equation}\label{eq::powerintermsofT2}
    \mathbb{P} \left(\sup_{w \in \mathcal{W}} \power(w, \betan) \ge \delta \right)
    \le 
      2 \ell m \exp(-\delta_{m,n,p}^2 K) + \ell m \exp \left( - \frac{p}{5\ell m} \right).
\end{equation}
For sufficiently large $p$, we may choose $m$ such that $\delta_{p,m,n} \ge \delta/(4C_1)$. This corresponds to choosing
$$m = \left \lceil \left(\frac{\delta}{2} - \frac{C_2}{p}\right)^{-1} C_3 b \sqrt{n} \right \rceil. $$
Note when $p > (4C_2)/\delta$, we have that $\delta/2 - C_2/p \ge \delta/4$, so   $$
m \le \frac{4 C_3 b \sqrt{n}}{\delta}.$$
These two computations yield
$$ \mathbb{P} \left(\sup_{w \in \mathcal{W}} \power(w, \betan) \ge \delta \right) \le \frac{8 b C_3 \cdot \ell \sqrt{n}}{\delta} \exp \left(- \frac{\delta^2}{16C_1^2}  K \right) + \frac{4 b C_3 \cdot \ell \sqrt{n}}{\delta} \exp \left( - \frac{\delta p}{4 C_3 b \cdot \ell \sqrt{n}} \right) $$
$$ \le O \left( \ell \sqrt{n}\right) \exp \left(- \frac{\delta^2}{16C_1^2}  K \right) + O \left( \ell \sqrt{n}\right) \exp \left( - \frac{\delta p}{4 C_3 b \cdot \ell \sqrt{n}} \right). $$

At this point, we use the assumption that for some $0 < \epsilon < 1$, $n = o\left(\left(\frac{p}{\ell}\right)^{2 -2 \epsilon}\right)$ which implies $\sqrt{n} = o\left(\left(\frac{p}{\ell}\right)^{1-\epsilon}\right)$.  Let $K = \left(\frac{p}{\ell}\right)^{\epsilon}$, so $K m = O\left(\sqrt{n} \left(\frac{p}{\ell}\right)^{\epsilon}\right) =  o\left(\frac{p}{\ell}\right)$. This satisfies the constraint $Km < \frac{p}{4 \ell}$ for sufficiently large $p$, which allows us to substitute $\left(\frac{p}{\ell}\right)^{\epsilon}$ for $K$ and rewrite the constant terms to conclude
$$\mathbb{P} \left(\sup_{w \in \mathcal{W}} \power(w, \betan) \ge \delta \right) \le O \left( \ell \sqrt{n}\right) \exp\left(-c_0 \left(\frac{p}{\ell}\right)^{\epsilon} \right) + O \left( \ell \sqrt{n}\right) \exp \left( - c_1 \frac{p}{\ell \sqrt{n}} \right),$$
for positive $c_0$ and $c_1$. We can use the condition on $n$ to bound this by
\begin{equation}\label{eq::finaleq}
    \mathbb{P} \left(\sup_{w \in \mathcal{W}} \power(w, \betan) \ge \delta \right) \le o \left(\ell^{\epsilon} p^{1-\epsilon} \right) \exp\left(-c_0 \left(\frac{p}{\ell}\right)^{\epsilon}\right) + o \left(\ell^{\epsilon} p^{1-\epsilon} \right) \exp \left( - c_1 \left(\frac{p}{\ell} \right)^{\epsilon} \right).
\end{equation}
To show that (\ref{eq::finaleq}) vanishes, we use the assumption that $\ell = o\left(\frac{p}{ \log^{1/\epsilon}(p)}\right)$, which implies $\ell = o\left(\frac{p}{ \log^{1/\epsilon}(p^d)}\right)$ for any fixed $d > 0$. Now, to analyze the right-hand term, note that $\frac{p}{\ell} = \Omega\left(\log^{1/\epsilon}(p^d) \right)$, so $\exp\left(- c_1 \left(\frac{p}{\ell} \right)^{\epsilon} \right) = o(p^{- c_1 d})$. For large enough $d$,  $p^{-c_1 d}$ dominates the $\ell^{\epsilon} p^{1-\epsilon}$ term, since $\ell^{\epsilon} p^{1-\epsilon} \le p^2$. This means that the right-hand term in (\ref{eq::finaleq}) vanishes. The left-hand term vanishes by the same logic if we replace $c_1$ with $c_0$. Thus, the bound in (\ref{eq::finaleq}) vanishes, proving the theorem.
\end{proof}
\end{corollary}

\subsection{Proof of Proposition \ref{prop::sdpopt}}\label{appendix::sdpopt}

Below, we restate Proposition \ref{prop::sdpopt} and prove a slightly more specific result, from which the original statement follows.

\begingroup
\def\theproposition{\ref{prop::sdpopt}}
\begin{proposition} Suppose $X \sim \mathcal{N}(0, \Sigma)$ for equicorrelated $\Sigma$ with correlation $\rho \ge 0.5$, and let $Y\mid X \sim \mathcal{N}(X \beta, \sigma^2)$ with $n > 2p$ \textit{and} $\gamma < 2 - 2 \rho$. Let $S = \gamma I_p$, and let $\hatbetaext \in \mathbb{R}^{2p}$ be OLS coefficients fit on $[\bX, \tilde{\bX}], \by$. Then
$$ 
\mathbb{E}\left(\sum_{j=1}^p (\hatbetaext_j - \hatbetaext_{j+p} - \beta_j)^2\right) \propto \frac{2p}{\gamma} = \frac{2p}{1-\upsilon},
$$
where $\upsilon = 1 - \gamma$ is the MAC. Since $\gamma, \upsilon \in (0,1)$, this is increasing in $\upsilon$.
\begin{proof}
Note that because $\Sigma = \rho \mathbf{1}_{p} \mathbf{1}_{p}^{\top} + (1-\rho) I_p$, we can represent 
$$G_S = \begin{bmatrix} 
\rho \mathbf{1}_{p} \mathbf{1}_{p}^{\top} + (1-\rho) I_p &
\rho \mathbf{1}_{p} \mathbf{1}_{p}^{\top} + (1-\rho - \gamma) I_p \\
\rho \mathbf{1}_{p} \mathbf{1}_{p}^{\top} + (1-\rho - \gamma) I_p & 
\rho \mathbf{1}_{p} \mathbf{1}_{p}^{\top} + (1-\rho) I_p
\end{bmatrix} = \rho \mathbf{1}_{2p} \mathbf{1}_{2p}^{\top} + A, $$
where
$$A = \begin{bmatrix} (1 - \rho) I_p & (1 - \gamma - \rho) I_p \\ (1 - \gamma - \rho) I_p & (1 - \rho) I_p \end{bmatrix}. $$
It is straightforward to check that if we set $c_1 = \frac{1}{2} \left(\frac{1}{2 - 2\rho - \gamma} + \frac{1}{\gamma} \right)$ and $c_{2} = \frac{1}{2}\left(\frac{1}{2 - 2\rho - \gamma} - \frac{1}{\gamma} \right)$, then $A^{-1} =  \begin{bmatrix} c_{1} I_p & c_2 I_p \\ c_2 I_p & c_1 I_p \end{bmatrix} $. Using the Sherman--Morrison inversion formula, we set
\begin{equation}\label{eq::sdpoptgformula}
G_S^{-1} = A^{-1} - \frac{\rho \cdot A^{-1} \mathbf{1}_{2p} \mathbf{1}_{2p}^{\top} A^{-1}}{1 + \rho \cdot \mathbf{1}_{2p} A^{-1} \mathbf{1}_{2p}^{\top}}
= A^{-1} - c_3 \mathbf{1}_{2p} \mathbf{1}_{2p}^{\top} \text{ for some } c_3 \in \mathbb{R}. 
\end{equation}
Now we are ready to analyze the estimation error in the theorem statement. We concatenate $\beta$ and $\mathbf{0} \in \mathbb{R}^p$ to obtain $\betaext = (\beta, \mathbf{0})$. Since $\hatbetaext$ are OLS coefficients, we see that conditional on $[\bX, \tilde{\bX}]$, 
$$\hatbetaext - \betaext \mid [\bX, \tilde{\bX}] \sim \mathcal{N}\left(0, \sigma^2 ([\bX, \tilde{\bX}]^{\top}[\bX, \tilde{\bX}])^{-1}\right). $$
Since $[X, \tilde{X}]$ is Gaussian, we know that $\mathbb{E}\left(([\bX, \tilde{\bX}]^{\top}[\bX, \tilde{\bX}])^{-1}\right) \propto G_S^{-1}$ and therefore for all $1 \le j \le p$, $$\mathbb{E}[(\hatbetaext_j - \hatbetaext_{j+p} - \beta_j)^2] \propto (G_S^{-1})_{j,j} + (G_S^{-1})_{j+p,j+p} - 2 (G_S^{-1})_{j,j+p}.$$ Combining this with equation (\ref{eq::sdpoptgformula}), we note the $c_3$ constants cancel to yield that
$$\mathbb{E}\left((\hatbetaext_j - \hatbetaext_{j+p} - \beta_j)^2\right) \propto 2 c_1 - 2 c_2 = \frac{2}{\gamma}. $$
Summing over $1 \le j \le p$ proves the theorem.
\end{proof}
\end{proposition}
\addtocounter{theorem}{-1}
\endgroup

\subsection{Discussion of assumptions on $q$ and the coefficient distribution}\label{appendix::qconj}

Recall from Section \ref{subsec::sdpopt} and Proposition \ref{prop::apprxeqreconst} that for equicorrelated designs with $\rho \ge 0.5$, if $\beta_j + \beta_k \approx 0$, a feature statistic like the lasso will be (approximately) equally likely to set $W_j, W_k \approx \beta_j, \beta_k$ or alternatively $W_j, W_k \approx - \beta_k, - \beta_j$. This observation indicates that we probably do not need the number of positive and negative signs in $\beta$ to balance exactly, nor do we need $q \le 0.1$. To see why, for simplicity, consider the case where $\beta$ has coefficients with a constant absolute value with $p^+$ positive coefficients, $p^-$ negative coefficients, and no null features. Assume without loss of generality that $p^+ > p^-$.
    
In this setting, the lasso will have $p^-$ opportunities to accidentally flip the sign of a pair of coefficients with alternating signs with probability $\frac{1}{2}$. Heuristically, this means that we should expect $\frac{p^-}{p^+ + p^-}$ of the signs of the non-null $W$ to be negative, even though all of the feature statistics should have roughly the same absolute value (see Proposition \ref{prop::apprxeqreconst}). In the knockoffs procedure, however, we can only reject any features at all if there is some set of features with high absolute values such that a proportion greater than $\frac{1}{1+q}$ of them have positive signs, where $q$ is the targeted level of FDR control. Heuristically, this suggests that the lasso will only have power if $p_{\mathrm{pos}} = \frac{p^+}{p^+ + p^-} > \frac{1}{1+q}$.

\begin{figure}
    \centering
    \includegraphics{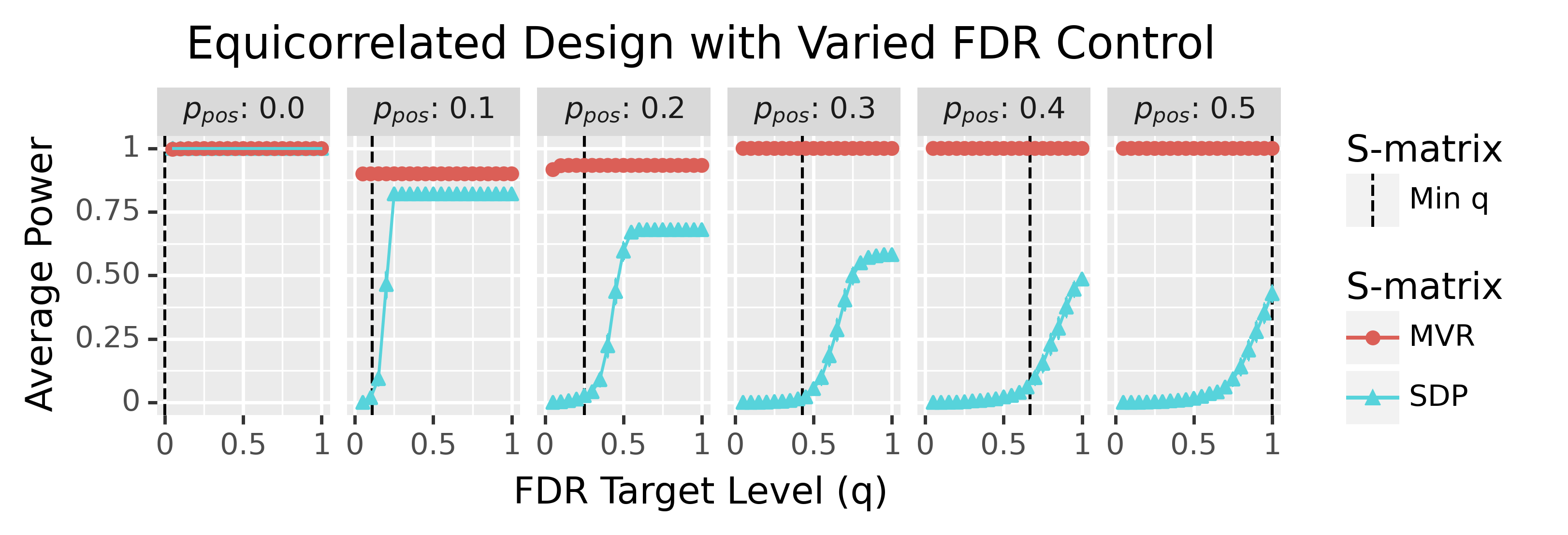}
    \caption{This figure empirical validates the informal conjecture from Appendix \ref{appendix::qconj}. We let the design $X$ be correlated with correlation $\rho = 0.6$, $Y \sim \mathcal{N}(X \beta, 1)$, with $n = 2000$ and $p = 500$. $\beta$ has $250$ non-null values with coefficient magnitudes sampled from $\mathrm{Unif}(2.5,5)$ and signs sampled from $\text{Bern}(p_{\mathrm{pos}})$. We use lasso coefficient differences as our feature statistics. The dotted black lines are the conjectured minimum $q$-values. }
    \label{fig::qconjecture}
\end{figure}

We confirm this informal conjecture in Figure \ref{fig::qconjecture}, and we see that knockoffs have nearly zero power when $p_{\mathrm{pos}} \le \frac{1}{1+q}$ even though the non-null coefficients do not have a constant absolute value. The exception is that when $p_{\mathrm{pos}} = 0.5$ and $q \ge 0.75$, where knockoffs have up to $50\%$ power. This behavior is not unexpected, however, since our analysis is asymptotic and $p$ and $n$ may need to be large for our conjecture to hold when $q$ is close to $1$. For example, when $q = 0.9$, the procedure will select \textit{every} feature with a positive feature statistic as long as $\frac{1}{1+q} \approx 52.6\%$ of the feature statistics are positive. This will happen fairly frequently unless $n$ and $p$ are quite large, even under the global null, where the signs of the feature statistics are independent and perfectly symmetric. Despite this, note that in all cases, the power of SDP knockoffs when $p_{\mathrm{pos}} = 0.5$ is much less than $q$, meaning that SDP knockoffs still always have trivial power.

\section{Technical proofs for Theorem \ref{thm::avgequi}}\label{appendix::avgequitechdetails}

In this section, we prove some of the technical lemmas used in Appendix \ref{appendix::proofs}.

\subsection{Verifying the SDP solution for equicorrelated Gaussian designs}

\begin{lemma}\label{lem::equicorrsoln}
 Suppose $X \sim \mathcal{N}(0, \Sigma)$ where $\Sigma$ is an equicorrelated correlation matrix with correlation $\rho \ge 0.5$. Then $S_{\mathrm{SDP}} = (2 - 2 \rho) \cdot I_p$. 
\begin{proof}
Recall that
$$S_{\mathrm{SDP}} = \mathrm{diag}(s_{\mathrm{SDP}}),$$
where $s_{\mathrm{SDP}} \in \mathbb{R}^p$ is the solution to the semidefinite program:
$$s_{\mathrm{SDP}} = \arg \min_{s \in \mathbb{R}^p} ||1 - s||_1 \text{ s.t. } s_j \ge 0 \text{ for each $j \in [p]$ and } \mathrm{diag}(s) \preceq 2 \Sigma.$$
We can rewrite this in a standardized dual form, as below:
$$s = \arg \max \sum_{j=1}^p s_j \text{ s.t. } \begin{bmatrix} 2 \Sigma & 0 & 0 \\ 0 & 0 & 0 \\ 0 & 0 & I_p  \end{bmatrix} - \sum_{j=1}^p s_j A_j \succeq 0, $$
where $A_j \in \mathbb{R}^{3p \times 3p}$ is zero everywhere except $A_{j,j} = A_{2p+j,2p+j} = 1$ and $A_{p+j,p+j} = -1$. This means that
$$s = \arg \max \sum_{j=1}^p s_j \text{ s.t. } \begin{bmatrix} 2 \Sigma & 0 & 0 \\ 0 & 0 & 0 \\ 0 & 0 & I_p  \end{bmatrix} - \begin{bmatrix} \mathrm{diag}(s) & 0 & 0 \\ 0 & - \mathrm{diag}(s) & 0 \\ 0 & 0 & \mathrm{diag}(s)  \end{bmatrix} \succeq 0. $$
Since a block matrix is positive semidefinite if and only if its blocks are, this constraint guarantees that $2\Sigma - \mathrm{diag}(s) \succeq 0$ and also that $0 \preceq \mathrm{diag}(s) \preceq I_p$.

This is the dual form of the corresponding primal SDP below:
\begin{align*}
\mathrm{minimize} & \,\,\,\,\, \begin{bmatrix} 2 \Sigma & 0 & 0 \\ 0 & 0 & 0 \\ 0 & 0 & I_p  \end{bmatrix} \bullet K  \\
\text{s.t.} & \,\,\,\,\,  A_j \bullet K = 1 \,\,\,\,\, \text{ for all } 1 \le j \le p \\
& \,\,\,\,\, K \succeq 0, \\
\end{align*}
where $K \in \mathbb{R}^{3p \times 3p}$ and for any matrix $D$, $D \bullet K \equiv \sum_{ij} D_{ij} K_{ij} = \text{tr}(DK)$. At this point, for $\rho \ge 0.5$,  we define
$$s = (2 - 2 \rho) \cdot \mathbf{1}_p, $$
$$M = - \frac{1}{p-1} \cdot \mathbf{1}_p \mathbf{1}_p^{\top} + \frac{p}{p-1} I_p \in \mathbb{R}^{p \times p}, $$
$$K = \begin{bmatrix} M & 0 & 0 \\ 0 & 0 & 0 \\ 0 & 0 & 0 \\ \end{bmatrix},  $$
which guarantees that $K$ is positive semi-definite. Note $s$ is a feasible solution to the dual problem because (i) each $s_j = 2 - 2 \rho \le 1$ when $\rho \ge 0.5$ and (ii) $\lambda_{\mathrm{min}}(\Sigma) = 1 - \rho$ (see Lemma \ref{lem::rankdegen}), so $0 \preccurlyeq \mathrm{diag}(s) \preccurlyeq 2 \Sigma$. Similarly, $A_j \bullet K = 1$ by construction of $K$ and definition of $A_j$ for each $j$, so $K$ is a feasible solution as well. Finally, we note that the duality gap is zero, as
$$\sum_{j=1}^p s_j = (2 - 2\rho) \cdot p$$
and similarly
$$\begin{bmatrix} 2 \Sigma & 0 & 0 \\ 0 & 0 & 0 \\ 0 & 0 & I_p  \end{bmatrix} \bullet K = 2 \sum_{1 \le i,j \le p}  \Sigma_{ij} M_{ij} = 2p - \frac{1}{p-1} \cdot p \cdot (p-1) \cdot (2 \rho) = (2 - 2\rho) \cdot p, $$
where the $2p$ term comes from the diagonals of $M$ and the second term comes from the $p \cdot (p-1)$ off-diagonal elements of $M$. Since the dual gap is zero, the optimal solution for $s$ is $(2 - 2\rho) \cdot \mathbf{1}_p$, which proves that $S_{\mathrm{SDP}} = (2 - 2\rho) \cdot I_p$.
\end{proof}
\end{lemma}

\subsection{Proof of Corollary \ref{corr::condapprxreconst}}\label{appendix::condapprxreconstproof}

In Proposition \ref{prop::apprxeqreconst}, we proved that for block-equicorrelated Gaussian designs with correlations greater than $0.5$ and a single-index response model, if $J = \{j_1, j_2\} \subset [p]$ lie in the same equicorrelated block and $|\beta_{j_1} + \beta_{j_2}| < d$, then if we apply a permutation invariant $w$ to SDP knockoffs,
$$d_{\tv}([\sign(\sorted(W_J)), W_{\text{-}J}], [- \sign(\sorted(W_J)), W_{\text{-}J}]) \le c_0 \sqrt{n} d.$$
Now, we prove Corollary \ref{corr::condapprxreconst}, which extends this result to apply conditionally.

\begin{theorem}[Corollary \ref{corr::condapprxreconst}] In the same setting as Proposition \ref{prop::apprxeqreconst}, let $\alpha_1, \alpha_2$ be constants such that $\alpha_1 \cdot \alpha_2 > c_0 \sqrt{n} d$ and let $A$ be some event in the sigma-algebra generated by $[|\sorted(W_{J})|, W_{\text{-}J}]$. Then if $\mathbb{P}(A) > \alpha_2 $, we have that
\begin{equation}\label{eq::condapprxreconsteq}
d_{\tv}\left([\sorted(W_{J}),W_{\text{-}J}], [-\sorted(W_{J}), W_{\text{-}J}] \mid A \right) < \alpha_1.
\end{equation}
\begin{proof}
Suppose for the sake of contradiction that equation (\ref{eq::condapprxreconsteq}) fails to hold and that $\mathbb{P}(A) > \alpha_2$.  Then there must be some set $B_1$ such that if we define 
$$I^+(W) = \{[\sorted(W_{J}), W_{\text{-}J}] \in B_1\} \text{ and } I^{-}(W) = \{[-\sorted(W_{J}), W_{\text{-}J}] \in B_1\},$$
then we have
$$\mathbb{P}(I^{+}(W) \mid A) - \mathbb{P}(I^{-}(W) \mid A) \ge \alpha_1 .$$
By the definition of conditional probability, this means that
$$\mathbb{P}(I^+(W) \cap A) - \mathbb{P}(I^-(W) \cap A) \ge \alpha_1 \cdot \mathbb{P}(A) > \alpha_1 \cdot \alpha_2 > c_0 \sqrt{n} d  .$$
 Since $A$ is in the sigma-algebra generated by $[|\sorted(W_{J})|, W_{\text{-}J}]$, there is a measurable set $B_2$ such that
$$\{[\sorted(W_{J}), W_{\text{-}J}] \in B_2\} = \{[-\sorted(W_{J}), W_{\text{-}J}] \in B_2\} = A.$$
Applying this equivalence and the definitions of $I^+(W), I^-(W)$, we obtain a violation of the marginal total variation bound in Proposition \ref{prop::apprxeqreconst}:
$$\mathbb{P}([\sorted(W_{J}), W_{\text{-}J}] \in B_1 \cap B_2) - \mathbb{P}([-\sorted(W_{J}), W_{\text{-}J}] \in B_1 \cap B_2) > c_0 \sqrt{n} d.$$
\end{proof}
\end{theorem}

\subsection{Basic properties of the knockoff filter}

In this section we prove two simple lemmas about $\psi$ and the knockoff filter. As a quick reminder, recall that for any $\epsilon \in \{-1, 1\}^p$ and any $k \in \mathbb{N}$, we define $V_k^+(\epsilon)$ to count the number of ones in the first $k$ coordinates of $\epsilon$. More precisely,
$$V_k^+(\epsilon) = \# \{j \le k : \epsilon_j = 1 \}.$$
Furthermore, as in (\ref{eq::psidef}), we define $\psi(\epsilon)$ as follows:
\begin{equation*}
\psi(\epsilon) = \max_{1 \le k \le \infty} \left\{ k : \frac{k - V_k^+(\epsilon) + 1}{V_k^+(\epsilon)} \le q \right\}.
\end{equation*}
As discussed in the proof of Theorem \ref{thm::avgequi}, intuitively $\psi$ is a reformulation of the knockoff data-dependent threshold $T$, such that the threshold depends only on the signs of the sorted feature statistics $W$. Indeed, the next lemma tells us that the number of discoveries made by the knockoffs procedure is a deterministic, increasing function of $\psi(\sign(\sorted(W)))$.

\begin{lemma}\label{lem::psibound} Let $W$ be feature statistics generated by any knockoff procedure in any setting, and let $\eta = \sign(\sorted(W))$ be the signs of the sorted feature statistics sorted by absolute value in descending order. 
Let $\tau$ be the number of discoveries of this procedure. Then when $\tau > 0$:
\begin{equation}\label{eq:taupsieq}
\tau = \left \lceil\frac{\psi(\eta)+1}{1+q}  \right \rceil \le \psi(\eta).
\end{equation}
Moreover, let $\epsilon \in \{-1, 1\}^p$ such that $\psi(\epsilon) = k > 0$. Then
\begin{equation}\label{eq:vkcondition}
V_{k}^+(\epsilon) = \left \lceil \frac{k+1}{1+q}  \right \rceil.
\end{equation}
\begin{proof} First we prove (\ref{eq:vkcondition}). If $\psi(\epsilon) = k$, then
$$ \frac{k - V_k^+(\epsilon) + 1}{V_k^+(\epsilon)} \le q \implies V_k^+(\epsilon) \ge \frac{k+1}{1+q}$$
by the definition of $\psi$. As a result, to show (\ref{eq:vkcondition}), it suffices to show that $V_k^+(\epsilon) < \frac{k+2}{1+q}$. Suppose for the sake of contradiction that $V_k^+(\epsilon) \ge \frac{k+2}{1+q}$. This implies that
$$\frac{k - V_k^+(\epsilon)+2}{V_k^+(\epsilon)} \le q.$$
Since $V_{k+1}^+(\epsilon) \ge V_k^+(\epsilon)$, this implies that 
$$\frac{k+1 - V_{k+1}^+(\epsilon) + 1}{V_{k+1}^+(\epsilon)} \le \frac{k + V_k^+(\epsilon) - 2}{V_k^+(\epsilon)} \le q,$$
where we are using the fact that $(k-V_k^+(\epsilon)+2)/V_k^+(\epsilon)$ is decreasing in $V_k^+(\epsilon)$. However, by definition of $\psi$, this implies that $\psi(\epsilon) \ge k+1$. This is a contradiction, since we assumed $\psi(\epsilon) = k$. Thus (\ref{eq:vkcondition}) must hold.

Now we prove (\ref{eq:taupsieq}). In the knockoffs procedure, to control the FDR at level $q$, we define the data-dependent threshold
$$T = \min_{1 \le i \le p} \left \{|W_i| : \frac{\# \{ j : W_j \le - |W_i| \} + 1}{\# \{j : W_j \ge |W_i| \}} \le q \right \} \setminus \{0\} $$
and then we reject the null for features $\{j : W_j \ge T\}$. Without loss of generality assume $W$ is sorted by absolute value, e.g. $|W_1| \ge |W_2| \ge \dots |W_p|$, since neither $\psi(\eta)$ nor $T$ depends on the initial order of the feature statistics. Then note that because $\eta = \sign(W)$,
$$\{ j : W_j \ge T \} = \{j \le \psi(\eta) : \eta_j = 1 \}.  $$
This implies that $\tau = V_{\psi(\eta)}^+(\eta)$, since both procedures accept all of the positive feature statistics between $1$ and $\psi(\eta)$. By (\ref{eq:vkcondition}), we conclude that
$$\tau = V_{\psi(\eta)}^+(\eta) = \left \lceil\frac{\psi(\eta)+1}{1+q}  \right \rceil \le \psi(\eta). $$
\end{proof}  
\end{lemma}

\begin{lemma}\label{lem::worstcaseavgequi} Consider an arbitrary set of feature statistics $W = W_1, \dots, W_t, \dots, W_p$. Define $W^* = W_1^*, \dots, W_p^*$ such that for a fixed $t \in [p]$,
$$W_j^* = 
\begin{cases} 
    + \infty & j \in \{1, \dots, t \} \\ 
    W_j & \text{ else } \\
\end{cases} $$
Let $\tau(W)$ and $\tau(W^*)$ be the number of discoveries made according to the knockoff procedure when applied to $W$ and $W^*$.  Then $\tau(W) \le \tau(W^*)$.
\begin{proof} Let $\eta = \sign(\sorted(W))$ and $\eta^* = \sign(\sorted(W^*))$. In Lemma \ref{lem::psibound}, we proved that the number of discoveries is fully determined by and increasing in $\psi(\eta)$. Therefore it suffices to show  $\psi(\eta) \le \psi(\eta^*) $. By definition of $\psi$, we can show this by showing that for all $k$, $V_k^+(\eta^*) \ge V_k^+(\eta)$. To see this, note by definition of the sort function that since the absolute values of $W_1^*, \dots, W_t^*$ are infinite and all of their signs are positive, we must have that $\eta^* = [\mathbf{1}_t, \sign(\sorted(W_{t+1:p})]$. This implies that switching from $W$ to $W^*$ can only add positive signs to the first $k$ elements of the path, for any $k$, since any feature statistic it swaps out is replaced with a positive sign. Therefore $V_k(\eta^*) \ge V_k^+(\eta)$ for all $k$.
\end{proof}
\end{lemma}

\subsection{Computations from Theorem \ref{thm::avgequiproof}}\label{appendix::avgequicomputations}

In this section, we prove Lemmas \ref{lem::swan} and \ref{lem::binomial2geometric}. Before doing so, it may be helpful to recall some context from Theorem \ref{thm::avgequiproof}. In Theorem \ref{thm::avgequiproof}, we grouped the coordinates of $W$ into $M_{}'$ ``singleton statistics" and $M$ pairs (so $M_{}' + 2M = p$). We denote the pairs $G_1, \dots, G_M$, and we constructed them such that if $j_1, j_2 \in G_j$, then $|\beta_{j_1} - \beta_{j_2}| < \frac{b}{m}$, where $b$ is the maximum absolute value of any coefficient in the single-index model and $m > 0$ is an arbitrary natural number. We showed earlier in Corollary \ref{corr::condapprxreconst} that $\sorted(W_{G_j})$ has an approximately symmetric distribution, even conditional on $|\sorted(W_{G_j})|$ and $W_{-G_j}$. The purpose of Lemma \ref{lem::swan} is to convert this result into a result that will help us show that $\mathbb{P}\left(\sign(\sorted(W)) = \epsilon\right)$ cannot stray above $2^{-M}$ for too many $\epsilon \in \{-1,1\}^p$, where the $2^{-M}$ comes because there are $M$ approximately symmetric pairs of feature statistics.

Lastly, for Lemma \ref{lem::swan}, it will be necessary to recall the notation introduced in step $2$ of Theorem \ref{thm::avgequi}. As a brief but incomplete refresher, we let $\eta = \sign(\sorted(W))$, and we defined $R$ to be the random vector which takes a coordinate $\eta_i$ of the sorted feature statistics and returns the index of the group $G_j$ which $\eta_i$ corresponds to. We used $R$ to carefully decompose
$$\mathbb{P}\left(\eta = \epsilon \right) = \sum_{r : p(r) > 0} p(r) p(\epsilon | r) = \sum_{r : p(r) > 0} p(r) \prod_{i=M_{}'+1}^{p} p(\epsilon_i | \epsilon_{(M_{}'+1):(i-1)}, r), $$
where the product begins at $M_{}'+1$ because we make the worst-case assumption that the $M_{}'$ singleton statistics are equal to $+ \infty$, meaning that the first $M_{}'$ coordinates of $\eta$ will always equal $1$. Crucially, we noted that the conditional probabilities $p(\epsilon_i | \epsilon_{(M_{}'+1):(i-1)}, r)$ can be written purely in terms of the features $\{\sorted(W_j)\}_{j=1}^M$ and $R$, allowing us to apply Corollary \ref{corr::condapprxreconst}. In Lemma \ref{lem::swan}, we use the notation $C_j(r)$ to denote the first coordinate of $\eta$ that corresponds to group $G_j$---for this reason, one can intuitively interpret the statement of Lemma \ref{lem::swan} to mean that $\sorted(W_{G_j}) \mid R, \sorted(W)_{1:C_j(r)-1}$ is symmetric with very high probability.

For more details, see step $2$ of Theorem \ref{thm::avgequiproof}.

\begingroup
\def\thetheorem{\ref{lem::swan}}
\begin{lemma} Let $\alpha_1, \alpha_2$ be constants such that $\alpha_1 \alpha_2 > \frac{c_0 \sqrt{n} b}{m}$. For any $1 \le j \le M$ and for any $\Lambda \subset \{-1,1\}^p$, 
\begin{equation}\label{eq::swanrepeated}
\alpha_2 > 
\sum_{\epsilon \in \Lambda} 
\sum_{r : p(r) > 0} p(r) 
 \mathbb{I} \left(p(\epsilon_{C_j(r)} | \epsilon_{(M_{}'+1):(C_j(r)-1)}, r) > 0.5 + \alpha_1 \right) \prod_{i \ne C_j(r), i \ge M_{}' + 1} p(\epsilon_i | \epsilon_{(M_{}'+1):(i-1)}, r). 
\end{equation}
\begin{proof} 
It suffices to prove (\ref{eq::swanrepeated}) for $\Lambda = \{-1, 1\}^p$ since every element in the above sum is nonnegative. We will employ a proof by contradiction, so at this point we assume that (\ref{eq::swanrepeated}) does not hold. The proof proceeds in two steps. First, we simplify the sum in (\ref{eq::swanrepeated}). Second, we will use (\ref{eq::swanrepeated}) to construct a violation of Corollary \ref{corr::condapprxreconst}. In particular, we will define a set $A$ such that $\mathbb{P}(A) \ge \alpha_2 / 2$ but conditional on $A$, the total variation distance between $\sorted(W_{G_j})$ and $-\sorted(W_{G_j})$ is at least $2 \alpha_1$.

\underline{Step 1: Simplification}. Note that for any $r$, if we fix the first $C_j(r)$ coordinates of $\epsilon$, we are summing over every possible value of the last $p-C_j(r)$ coordinates of $\epsilon$. Since the last $p - C_j(r)$ coordinates of $\epsilon$ must take \textit{some} value, all conditional probabilities depending on these last $p - C_j(r)$ coordinates should sum to one. The effect of this is that in the above product, we can drop all terms in the product where $i > C_j(r)$ and only sum over the first $C_j(r)$ coordinates of $\epsilon$, as below.\footnote{Note that this logic does not apply to all $p$ coordinates of $\epsilon$ since the indicator variable in (\ref{eq::swanrepeated}) is not a conditional probability. However, it only depends on the first $C_j(r)$ coordinates of $\epsilon$.} 

\begin{align*}
    \alpha_2 \le & 
    \sum_{r : p(r) > 0} 
    \sum_{\epsilon \in \{-1,1\}^p}
    p(r) 
    \mathbb{I} \left(p(\epsilon_{C_j(r)} | \epsilon_{(M_{}'+1):(C_j(r)-1)}, r) > 0.5 + \alpha_1\right) 
    \prod_{i \ne C_j(r), i \ge M_{}'+1} p(\epsilon_i | \epsilon_{(M_{}'+1):(i-1)}, r)  \nonumber \\
    & \nonumber \\
    = & 
    \sum_{r : p(r) > 0} \sum_{\epsilon \in \{-1,1\}^{C_j(r)}}
    p(r)
    \mathbb{I} \left(p(\epsilon_{C_j(r)} | \epsilon_{(M_{}'+1):(C_j(r)-1)}, r) > 0.5 + \alpha_1\right) 
   \prod_{M_{}'+1 \le i < C_j(r)} p(\epsilon_i | \epsilon_{(M_{}'+1):(i-1)}, r). 
\end{align*}
Note that in the latter equation, we slightly abuse notation and let $\epsilon$ denote a $C_j(r)$-length vector instead of a $p$-length vector. Next, observe that there are only $2$ possible values for $\epsilon_{C_j(r)} \in \{-1,1\}$. Therefore by the pigeonhole principle, there exists some \emph{fixed} $v \in \{-1,1\}$ such that
\begin{align}
    \alpha_2 / 2
    \le & 
    \sum_{r : p(r) > 0} \sum_{\epsilon \in \{-1,1\}^{C_j(r)-1}}  
    p(r)
    \mathbb{I} \left(p(v | \epsilon_{(M_{}'+1):(C_j(r)-1)}, r) > 0.5 + \alpha_1\right) 
   \prod_{M_{}'+1 \le i < C_j(r)} p(\epsilon_i | \epsilon_{(M_{}'+1):(i-1)}, r).
   \label{eq::grouprankcondhelper}
\end{align}
\underline{Step 2}: At this point, we will construct an event $A$ such that $\mathbb{P}(A)$ equals the right-hand side of (\ref{eq::grouprankcondhelper}) and furthermore such that the distribution of $\sorted(W_{G_j})$ is highly asymmetric conditional on $A$. To do this, recall that the definition of $C_j(r)$ guarantees that $\epsilon_{C_j(r)}$ corresponds to $\sorted(W_{G_j}))_1$. This plus the definition of $p(\epsilon_{C_j(r)} |\epsilon_{(M_{}'+1):(C_j(r)-1)}, r)$, from (\ref{eq::defprobepsi}), yields that
$$p(v | \epsilon_{(M_{}'+1):(C_j(r)-1)}, r) = \mathbb{P}\left( \sign(\sorted(W_{G_j}))_{1} = v \, \Bigg| \, \bigcap_{i=M_{}'+1}^{C_j(r)-1} \sign(\sorted(W_{G_{r_i}}))_{N_i(r)} = \epsilon_i \, ,\,  R = r \right). $$
Define $\Lambda_r$ to be the set of $\epsilon \in \{-1, 1\}^{C_j(r) - 1}$ such that  the above probability is greater than $0.5 + \alpha_1$. Formally,
$$\Lambda_r = \left \{\epsilon \in \{-1,1\}^{C_j(r) - 1} :    p(v | \epsilon_{(M_{}'+1):(C_j(r)-1)}, r) > 0.5 + \alpha_1\right \}. $$
Then define the event $A$ as below:
\begin{equation}\label{eq::defswaneventA}
A = \bigcup_{r : p(r) > 0} \left( \{R = r\} \cap \bigg \{ \sign(\sorted(W))_{1:(C_j(r)-1)} \in \Lambda_r \bigg \} \right).
\end{equation}
Note that $A$ is in the sigma-algebra generated by $[|\sorted(W_{G_j})|, W_{-G_j}]$ for two reasons. First, as noted in Theorem \ref{thm::avgequiproof}, $R$ is in the sigma-algebra generated by $\{|\sorted(W_{G_j})|\}_{j=1}^M$. Second, $v$ is fixed and $\sign(\sorted(W))_{1:C_j(r)-1}$ only depends on $R$ and $W_{-G_j}$, by the definition of $C_j(r)$. 

By the definition of $A$, we have that
\begin{align*}
    \mathbb{P}(A) = & 
    \sum_{r : p(r) > 0} p(r) \sum_{\epsilon \in \Lambda_r}  
    \prod_{M_{}'+1 \le i < C_j(r)} p(\epsilon_i | \epsilon_{(M_{}'+1):(i-1)}, r)
    \\ = & 
    \sum_{r : p(r) > 0} \sum_{\epsilon \in \{-1,1\}^{C_j(r)-1}}  
    p(r)
    \mathbb{I} \left(p(v | \epsilon_{(M_{}'+1):(C_j(r)-1)}, r) > 0.5 + \alpha_1\right) 
   \prod_{M_{}'+1 \le i < C_j(r)} p(\epsilon_i | \epsilon_{(M_{}'+1):(i-1)}, r) \\ 
   \ge &\,\,\, \alpha_2 / 2 \, \, \, \, \, \, \, \, \, \, \, \, \text{ by equation (\ref{eq::grouprankcondhelper}),}
\end{align*}
where the second step follows by definition of $\Lambda_r$, and in both steps, we slightly abuse notation so that $\epsilon$ is a $C_j(r)-1$ dimensional vector. Thus, the law of total probability plus the definition of $\Lambda_r$ implies that
$$\mathbb{P}\left(\sign(\sorted(W_{G_j}))_{1} = v \mid A \right) > 0.5 + \alpha_1$$
which implies that
$$d_{\tv}\left( \sorted(W_{G_j}), - \sorted(W_{G_j}) \mid A \right) > 2 \alpha_1.$$
However, this contradicts Corollary \ref{corr::condapprxreconst} since $2 \alpha_1 \alpha_2 / 2 > \frac{c_0 \sqrt{n} b}{m}$.
\end{proof}
\end{lemma}
\addtocounter{theorem}{-1}
\endgroup

\begingroup
\def\thetheorem{\ref{lem::binomial2geometric}}
\begin{lemma} Under the conditions and notation of Theorem \ref{thm::avgequiproof}, specifically regarding equation (\ref{eq::lastnontecheq}),
$$\sum_{k=C_0 M_{}'}^{\lfloor (1+q) p - 1 \rfloor} k \cdot \binom{\min(k, p) - M_{}'}{\lceil \frac{k+1}{1+q} \rceil - M_{}'} \cdot (0.5 + \alpha_1)^{(\min(k,p) - M_{}')/2 - \min(k,p)/\ell_0 } \le C_1 M_{}' + C_2$$
for sufficiently large $C_0$ and some universal constants $C_1, C_2$. Note we use the assumptions that $M_{}' \le \frac{p}{50}$, $\ell_0 > 200$, $\alpha_1 < 0.001$, and $q \le 0.1$.
\begin{proof}
 We first change variables by resetting $k$ to $k + C_0 M_{}'$.
\begin{align}
 = & 
    \sum_{k=0}^{\lfloor (1+q) p - 1 \rfloor - C_0 M_{}'} (k + C_0 M_{}')  
    \binom{\min(k + C_0 M_{}', p) - M_{}'}{\lceil \frac{k + C_0 M_{}' + 1}{1+q} \rceil - M_{}'} (0.5 + \alpha_1)^{(\min(k+C_0 M_{}',p)- M_{}')/2 - \min(k+C_0 M_{}',p)/\ell_0}
    \nonumber \\ 
    = & C_0 M_{}' \sum_{k=0}^{\lfloor (1+q) p - 1 \rfloor - C_0 M_{}'}
     \binom{\min(k + C_0 M_{}', p) - M_{}'}{\lceil \frac{k + C_0 M_{}' + 1}{1+q} \rceil - M_{}'}
    \cdot (0.5 + \alpha_1)^{(\min(k+C_0 M_{}',p)- M_{}')/2 - \min(k+C_0 M_{}',p)/\ell_0} \label{eq::udck1} \\
    & +  \sum_{k=0}^{\lfloor (1+q) p - 1 \rfloor - C_0 M_{}'} k \cdot 
     \binom{\min(k + C_0 M_{}', p) - M_{}'}{\lceil \frac{k + C_0 M_{}' + 1}{1+q} \rceil - M_{}'}
    \cdot (0.5 + \alpha_1)^{(\min(k+C_0 M_{}',p)- M_{}')/2 - \min(k+C_0 M_{}',p)/\ell_0} \label{eq::udck2}
\end{align}

We will show that (\ref{eq::udck1}) can be bounded by a linear function of $M_{}'$, and that (\ref{eq::udck2}) is bounded by a constant which does not grow with $M_{}'$ or $p$. Our initial task will be to analyze the components of these sums, with the eventual goal of comparing them to a geometric series. For both the binomial term and the exponential term, we will break analysis into two cases: first, when $k +C_0 M_{}' \le p$, and second, when $k + C_0 M_{}' > p$.

\underline{First}, we will analyze the exponential term. In the first case where $k + C_0 M_{}' \le p$, we have that
\begin{align*}
(0.5 + \alpha_1)^{(\min(k+C_0 M_{}',p)- M_{}')/2 - \min(k+C_0 M_{}',p)/\ell_0}
&= (0.5+\alpha_1)^{\left(\frac{1}{2} - \frac{1}{\ell_0} \right) (k+C_0 M_{}') - M_{}'/2} \\
& \le 2^{(\frac{1}{2} - \frac{1}{\ell_0}) \log_2(0.5 + \alpha_1) (k+C_0 M_{}') + M_{}'/2} \\
& \le 2^{-0.49 k + (-0.49 C_0 + 0.5) M_{}'},
\end{align*}
 where we obtain the $0.49$ constant using the fact that $\alpha_1 < 0.001$ and $\ell_0 > 200$. We can pick $C_0$ large enough such that this is less than
\begin{equation}\label{eq::avgequiexp1}
    \le 2^{-0.49 k - 0.48 C_0 M_{}'}.
\end{equation} 
In the second case, when $k + C_0 M_{}' > p$, we have that
\begin{equation}\label{eq::avgequiexp2}
    (0.5 + \alpha_1)^{(\min(k+C_0 M_{}',p)- M_{}')/2 - \min(k+C_0 M_{}',p)/\ell_0} \le 2^{-0.49 p - M_{}'/2} \le 2^{-0.48 p}
\end{equation} 
where the last equation follows from the assumption that $M_{}' \le \frac{p}{50}$.

\underline{Second}, we will analyze the binomial coefficient appearing in (\ref{eq::udck1}) and (\ref{eq::udck2}). We use the noncentral binomial bound below:
\begin{equation}\label{eq::noncentralbinbound}
\frac{2^{n H(d/n)}}{\sqrt{n}} \le \binom{n}{d} \le 2^{n H(d/n)}
\end{equation}
where $H(x) = - x \log_2(x) - (1 - x) \log_2(1-x)$ is the binary entropy function.
In the first case when $k + C_0 M_{}'  \le p$,
$$ 
 \binom{\min(k + C_0 M_{}', p) - M_{}'}{\lceil \frac{k + C_0 M_{}' + 1}{1+q} \rceil - M_{}'} \le
  2^{H\left( a_{} \right) (k + C_0 M_{}' - M_{}') },
$$
where we define
$$a_{} = \frac{\lceil \frac{k + C_0 M_{}' + 1}{1+q} \rceil - M_{}'}{k + (C_0 - 1)M_{}'} \ge \frac{\frac{k+(C_0 - 1)M_{}'}{1+q} - \frac{qM_{}'}{1+q}}{k + (C_0 - 1)M_{}'} \ge \frac{1}{1+q} - \frac{qM'}{(k + (C_0-1)M_{}')} \ge \frac{1}{1.1} - \frac{0.1}{C_0-1},$$
where in the last step on the right, we apply the fact that $q \le 0.1$. Note that many of the bounds above are quite loose. The main point, however, is that we always have that $a_{} > 0.5$ for $C_0 > 2$, and $H$ is continuous and decreasing on $[0.5,1]$. Note as $C_0$ gets larger, our lower bound on $a_{}$ approaches $1/(1.1)$. Thus, we can pick $C_0$ large enough to guarantee that $H(a_{}) \le H\left(\frac{1}{1.1} + \frac{0.1}{C_0 -1} \right) < 0.44$, where the number $0.44$ comes from the fact that $H(1/(1.1)) < 0.44$. All together, this implies that in the first case,
\begin{equation}\label{eq::avgequibin1}
      \binom{\min(k + C_0 M_{}', p) - M_{}'}{\lceil \frac{k + C_0 M_{}' + 1}{1+q} \rceil - M_{}'}
  < 2^{0.44 k + 0.44 C_0 M_{}'}.
\end{equation}
In the second case, when $k + C_0 M_{}' > p$, we will use the fact that $\binom{n}{d}$ is decreasing in $d$ when $n/2 \le d$. That assumption holds in this instance, as since $M_{}' \le p/50$,
$$\frac{\min(k + C_0 M_{}', p) - M_{}'}{2} \le \frac{p}{2} \le \left(\frac{1}{2} + \frac{1}{50} \right) p - M_{}' \le \left \lceil \frac{k+C_0 M_{}' + 1}{1+q} \right \rceil - M_{}' $$
where in the last equation, we use the fact that $q \le 0.1$, so $1/2 + 1/50 \le 1/1.1 \le \frac{1}{1+q}$. As a result, in the second case,
\begin{align}
\binom{\min(k + C_0 M_{}', p) - M_{}'}{\lceil \frac{k + C_0 M_{}' + 1}{1+q} \rceil - M_{}'}
&= 
    \binom{p-M_{}'}{\lceil \frac{k+C_0 M_{}'+1}{1+q} \rceil - M_{}'} \nonumber \\
&\le 
    \binom{p - M_{}'}{\lceil \frac{p}{1+q} \rceil - M_{}'} \nonumber \\
& \le 2^{H\left(\frac{\lceil p/(1+q) \rceil - M_{}'}{p - M_{}'} \right) (p - M_{}') } \label{eq::bincase2intermediate}.
\end{align} 
Analyzing the $H(\cdot)$ term in (\ref{eq::bincase2intermediate}), we use the assumptions that $M_{}' \le \frac{p}{50}$ and $q \le 0.1$ to find that
$$\frac{\lceil p/(1+q) \rceil - M_{}'}{p- M_{}'} \ge \frac{p/(1.1) - \frac{p}{50}}{p - \frac{p}{50}} = \frac{\frac{1}{1.1} - \frac{1}{50}}{\frac{49}{50}} \ge 0.907. $$
Note that $H(0.907) < 0.45$, so $H\left(\frac{\lceil p/(1+q) \rceil - M_{}'}{p- M_{}'}\right) < 0.45$. This implies that when $p < k + C_0 M_{}'$,
\begin{align}\label{eq::avgequibin2}
      \binom{\min(k + C_0 M_{}', p) - M_{}'}{\lceil \frac{k + C_0 M_{}' + 1}{1+q} \rceil - M_{}'}
  < & 2^{H\left(\frac{\lceil p/(1+q) \rceil - M_{}'}{p - M_{}'} \right) (p - M_{}') } \le  2^{0.45 p}.
\end{align}
Now, we will work with the sum (\ref{eq::udck1}). As a reminder, we need to show that this sum can be bounded by a universal constant not depending on $p$ or $M_{}'$. We split this sum into two parts which correspond to the cases where $k + C_0 M_{}' \le p$ and vice versa:
\begin{align}
    & \sum_{k=0}^{\lfloor (1+q) p - 1 \rfloor - C_0 M_{}'}
     \binom{\min(k + C_0 M_{}', p) - M_{}'}{\lceil \frac{k + C_0 M_{}' + 1}{1+q} \rceil - M_{}'}
    \cdot (0.5 + \alpha_1)^{(\min(k+C_0 M_{}',p)- M_{}')/2 - \min(k+C_0 M_{}',p)/\ell_0} \nonumber \\
\le & 
    \sum_{k=0}^{p - C_0 M_{}'} 2^{0.44 k + 0.44 C_0 M_{}'} 2^{-0.49 k - 0.48 C_0 M_{}'} + \sum_{k=p-C_0 M_{}' + 1}^{\lfloor (1+q) p - 1 \rfloor - C_0 M_{}'} 2^{0.45 p} 2^{-0.48 p} \nonumber
    \\
\le &
    2^{- 0.04 C_0 M_{}'} \sum_{k=0}^{p - C_0 M_{}'} 2^{-0.05 k} + (1+q) p \, 2^{-0.03 p}. \label{eq::udck1bound}
\end{align} 
The sum in the first term is bounded as $p \to \infty$, and all other terms are asymptotically decreasing in $p$ and $M_{}'$. This means that we can bound this by a constant which does not depend on $p$ or $M_{}'$.

Next, we will work with the sum (\ref{eq::udck2}) using almost exactly the same argument. In particular,
\begin{align}
    & \sum_{k=0}^{\lfloor (1+q) p - 1 \rfloor - C_0 M_{}'} k \cdot
     \binom{\min(k + C_0 M_{}', p) - M_{}'}{\lceil \frac{k + C_0 M_{}' + 1}{1+q} \rceil - M_{}'}
    \cdot (0.5 + \alpha_1)^{(\min(k+C_0 M_{}',p)- M_{}')/2 - \min(k+C_0 M_{}',p)/\ell_0} \nonumber \\
\le & 
    \sum_{k=0}^{p - C_0 M_{}'} k \cdot 2^{0.44 k + 0.44 C_0 M_{}'} 2^{-0.49 k - 0.48 C_0 M_{}'} + \sum_{k=p-C_0 M_{}' + 1}^{\lfloor (1+q) p - 1 \rfloor - C_0 M_{}'} k \cdot 2^{0.45 p} 2^{-0.48 p} \nonumber
    \\
\le &
    2^{- 0.04 C_0 M_{}'} \sum_{k=0}^{p - C_0 M_{}'} k 2^{-0.05 k} + (1+q)^2 p^2  2^{-0.03 p}. \label{eq::udck2bound}
\end{align}

Since the sum on the left converges and all other terms vanish as $M_{}', p \to \infty$, this equation can be uniformly bounded as well. Combining this with (\ref{eq::udck1}) and (\ref{eq::udck2}), there are constants $C_1$ and $C_2$ such that
 \begin{equation}
    \sum_{k=\lfloor (1+q) M_{}' - 1 \rfloor}^{\lfloor (1+q) p - 1 \rfloor} k \sum_{r : p(r) > 0} p(r) \sum_{\epsilon \in U_k^c(r)} p(\epsilon|r)
\le
    C_1 M_{}' + C_2
\end{equation}
holds uniformly under the conditions in the theorem.
\end{proof}
\end{lemma}
\addtocounter{theorem}{-1}
\endgroup

\subsection{Total variation bounds}

In this section, we will make some simple applications of the following theorem, which is reproduced from \cite{devroyetvbound2018}:

\begin{theorem}[From \cite{devroyetvbound2018}]\label{thm::devtvbound} Let $Z, Z'$ be $p$-dimensional multivariate normals with equal means but different covariance matrices $E_Z$ and $E_{Z'}$. Then 
$$\frac{1}{100} \le \frac{d_{\tv}(Z,Z')}{\min \{1, ||E_{Z'} E_Z^{-1} - I_p ||_F \}} \le \frac{3}{2}, $$
where $|| \cdot ||_F$ denotes the Frobenius norm. On the other hand, if $Z$ and $Z'$ are univariate Gaussians with means $\mu_1, \mu_2$ and equal variances $\sigma^2$, then
$$d_{\tv}(Z, Z') \le \frac{|\mu_1 - \mu_2|}{2 \sigma}. $$
\end{theorem}

\begin{lemma}\label{lem::tvbound} Let $X \sim \mathcal{N}(0, \Sigma)$ and let $Y \mid X \sim \mathcal{N}(X \beta, \sigma_0^2)$. Let $\beta' \in \mathbb{R}^p$ such that $\beta$ and $\beta'$ differ in at most one coordinate. Suppose without loss of generality that this is the first coordinate and $|\beta_1 - \beta_1'| = d$. Then let $Y' \mid X \sim \mathcal{N}(X \beta', \sigma_0^2)$. If $\bX, \by, \by'$ are the data matrix and response vectors for $n$ independent data points, then there is a universal constant $c_0$ such that
$$d_{\tv}([\bX, \by], [\bX, \by']) \le \frac{c_0 \sqrt{n} d}{\sigma_0}. $$

\begin{proof} First, consider a single observation $(X, Y, Y')$, i.e., the case where $n = 1$. Then by Theorem \ref{thm::devtvbound}, 
$$d_{\tv}([X,Y], [X,Y'] \mid X)= d_{\tv}(Y,Y' \mid X) \le \frac{|X_1 \beta_1 - X_1 \beta_1'|}{2 \sigma_0} = \frac{|X_1| |\beta_1 - \beta_1'|}{2 \sigma_0} = \frac{d}{2 \sigma_0} \cdot  |X_1|. $$
For any arbitrary random vectors $R_1, R_2, R_3$, we know $d_{\tv}\left(R_1, R_2 \right) \le \mathbb{E} \left[d_{\tv}(R_1, R_2 \mid R_3) \right]$. This plus the prior equation yields that
$$d_{\tv}([X, Y], [X, Y']) \le \frac{d}{2\sigma_0} \mathbb{E}[|X_1|] = \frac{d}{\sigma_0 \sqrt{2\pi}}, $$
since marginally $X_1 \sim \mathcal{N}(0,1)$, so $|X_1| \sim \chi_1$. Now consider the case where $n \ge 1$. Note that the i.i.d. rows $(X, Y)$ and $(X, Y')$ are both $p+1$ dimensional Gaussians with mean $0$. Let their covariance matrices be $\Gamma_1$ and $\Gamma_2$. Applying Theorem \ref{thm::devtvbound}, this implies
$$\min\{1, ||\Gamma_1 \Gamma_2^{-1} - I_{p+1}||_F \} \le \frac{C_0 d}{\sigma_0}, $$
where $C_0 \le \frac{100}{\sqrt{2\pi}}$. Note this bound on $C_0$ is fairly loose, as pointed out by \cite{devroyetvbound2018}, who did not optimize the constants in Theorem \ref{thm::devtvbound}.

Now, we can think of $(\bX, \by)$ and $(\bX, \by')$ as $n(p+1)$ dimensional multivariate Gaussians, with mean $0$ and block-diagonal covariance matrices  $\Gamma_1^{(n)}$ and $\Gamma_2^{(n)}$ (respectively), where
$$\Gamma_1^{(n)} = \begin{bmatrix} \Gamma_1 & 0 & \dots & 0 \\ 
                                   0 & \Gamma_1 & \dots & 0 \\
                                   \vdots & \vdots & \vdots & \vdots \\
                                   0 & 0 & \dots & \Gamma_1 \\ \end{bmatrix}  \text{ and } \Gamma_2^{(n)} = \begin{bmatrix} \Gamma_2 & 0 & \dots & 0 \\ 
                                   0 & \Gamma_2 & \dots & 0 \\
                                   \vdots & \vdots & \vdots & \vdots \\
                                   0 & 0 & \dots & \Gamma_2 \\ \end{bmatrix}.$$
As a result, we see that
$$||\Gamma_1^{(n)} \left(\Gamma_2^{(n)}\right)^{-1} - I_{n(p+1)} ||_F = \left|\left| \, \, 
\begin{bmatrix} \Gamma_1 & 0 & \dots & 0 \\ 
                       0 & \Gamma_1 & \dots & 0 \\
                       \vdots & \vdots & \vdots & \vdots \\
                       0 & 0 & \dots & \Gamma_1 \\ \end{bmatrix} 
\begin{bmatrix} \Gamma_2^{-1} & 0 & \dots & 0 \\ 
           0 & \Gamma_2^{-1} & \dots & 0 \\
           \vdots & \vdots & \vdots & \vdots \\
           0 & 0 & \dots & \Gamma_2^{-1} \\ \end{bmatrix} - I_{(p+1)n} \, \, \right| \right|_F $$
$$= \left|\left| \, \, 
\begin{bmatrix} \Gamma_1\Gamma_2^{-1} & 0 & \dots & 0 \\ 
           0 & \Gamma_1\Gamma_2^{-1} & \dots & 0 \\
           \vdots & \vdots & \vdots & \vdots \\
           0 & 0 & \dots & \Gamma_1\Gamma_2^{-1} \\ \end{bmatrix} - I_{(p+1)n} \, \, \right| \right|_F = \sqrt{n} ||\Gamma_1 \Gamma_2^{-1} - I_{p+1} ||_F \le \frac{C_0 \sqrt{n} d}{\sigma_0^2} .$$
           
Here we are using the fact that the inverse of a block-diagonal matrix is simply the inverse of the diagonals and applying the definition of the Frobenius norm. Combined with the second theorem from Devroye et. al, this tells us that for $n$ data points, 
$$d_{\tv}((\bX,\by), (\bX, \by')) \le  \frac{3 C_0 \sqrt{n} d}{2 \sigma_0}, $$
which proves the lemma if we reset $C_0$ to be $3/2$ times its original value.
\end{proof}
\end{lemma}

\subsection{Lemmas about random coefficients}\label{subsec::randomcoef}

Let $\mathcal{C}_{b,p} = [-b, b]^p$ be the $p$-dimensional cube with side-length $2b$ centered at $0$ and suppose $\beta$ is drawn uniformly from $\mathcal{C}_{b,p}$. Note this is equivalent to letting $|\beta_i| \iid \mathrm{Unif}(0, b)$, $\sign(\beta_i) \iid \pm 1$ with equal probability, with $\sign(\beta_i) \Perp |\beta_i|$. In this section, we analyze the distribution of the sizes of the buckets of coefficients defined in equation (\ref{eq::bucketdef}). As a quick reminder, we fix $m$ and let $D_k$ be the set of indices corresponding to block $k$. Then define
$$B_{j,k} = \left\{i \in D_k  : \frac{(j-1) \cdot b}{m} \le |\beta_i| \le \frac{j \cdot b}{m}  \right\}.$$
Furthermore, we let $B_{j,k}^+ = \{i \in B_{j,k} : \beta_i > 0\}$ and $B_{j,k}^- = \{i \in B_{j,k} : \beta_i < 0 \}$.
\begin{lemma}\label{lem::groupsizes} Given the prior definitions, fix integers $m, K > 0$ such that $m K < \frac{p}{4 \ell}$. Then for sufficiently large $m$,
$$\mathbb{P}\left( \min_{1 \le j \le m} \min_{1 \le k \le \ell}
|B_{j,k}| < K \right) \le \ell m \exp \left(- \frac{p}{5 \ell m} \right).$$
\begin{proof} Note that for any $j \in [m], k \in [\ell]$,
$$|B_{j,k}| \sim \text{Bin}\left(\frac{p}{\ell}, \frac{1}{m}\right). $$
Since $\frac{\ell K}{p} < \frac{1}{4m}$, we can use the Binomial bound in Theorem $1$ of \citet{binbound1989} which states that
$$\mathbb{P}\left(|B_{j,k}| \le K \right) \le \exp\left(- \frac{p}{\ell}  D\left(\frac{\ell K}{p} \bigg | \bigg | \frac{1}{m} \right) \right),$$
where $D(p_1||p_2)$ denotes the relative entropy between two Bernoulli random variables with probability $p_1$ and $p_2$. Define $x_1 =  \ell K / p$ and $x_2 = 1 / m$ and recall that we assume $x_1 < x_2 / 4$. When $m$ is sufficiently large, $x_2$ will approach zero, ensuring that $x_1$ does as well, and finally that $(x_2 - x_1)/(1-x_2)$ will  too. As a result, we can apply the approximation $\log(1 + (x_2 - x_1)/(1-x_2)) \ge c_0 (x_2 - x_1)/(1-x_2)$ for some $c_0 < 1$. Note that by the prior logic, for any fixed $c_0$, this approximation is valid for any sufficiently large $m$ under the theorem conditions, regardless of the values of $p$ or $K$. This approximation yields
\begin{align*}
D(x_1 || x_2) =& x_1 \log\left(\frac{x_1}{x_2}\right) + (1-x_2) \log\left(\frac{1-x_1}{1-x_2}\right) \\
=& x_1 \log\left(\frac{x_1}{x_2}\right) + (1-x_2) \log\left(1 + \frac{x_2 - x_1}{1 - x_2}\right) \\ 
\ge& x_1 \log\left(\frac{x_1}{x_2}\right) + c_0 (1-x_2) \frac{x_2 - x_1}{1-x_2} \\ 
=& x_1 \log\left(\frac{x_1}{x_2}\right) + c_0 x_2 - c_0 x_1 \\ 
\ge& x_1 \log\left(\frac{x_1}{x_2}\right) + \frac{3 c_0 x_2}{4}
\end{align*}
where the last step uses that $x_1 < \frac{x_2}{4}$. We will show that the final expression is greater than $x_2/5 = 1/(5m)$. To see this, note that for sufficiently large $m$, the above equation holds for $c_0$ large enough such that $3 c_0 /4 > (1/2 + 1/5)$. Then, we need only show that $x_2/2 > - x_1 \log\left(x_1/x_2\right)$. Note both quantities in the comparison are positive, so it suffices to show their ratio is greater than $1$. Using the fact that $x_2/x_1 > 4$, we see
$$\frac{x_2/2}{- x_1 \log(x_1/x_2)} = \frac{1}{2} \cdot \frac{x_2 / x_1}{\log(x_2/x_1)} > \frac{1}{2} \frac{4}{\log(4)} \ge 1,$$
where we use the fact that the function $\frac{x}{\log(x)}$ is increasing on $(e, \infty)$. This proves that $D(x_1 || x_2) > 1/(5m)$. Applying this to the initial binomial bound, we obtain 
$$\mathbb{P}\left(|B_{j,k}| \le K \right) \le \exp\left(- \frac{p}{5 \ell m} \right).$$
Therefore by the union bound, 
$$\mathbb{P}\left( \min_{1 \le j \le m} \min_{1 \le k \le \ell} |B_{j,k}| \le K \right) = \mathbb{P}\left( \bigcup_{1 \le j \le m} \bigcup_{1 \le k \le \ell} |B_{j,k}| \le K \right)\le \ell m \exp\left(- \frac{p}{5 \ell m} \right),$$
which completes the proof of the Lemma.
\end{proof}
\end{lemma}

\begin{lemma}\label{lem::groupsigns} In the previous setting, define
$$M_{}' = 2 \sum_{j=1}^m \sum_{k=1}^\ell \max\left(|B_{j,k}^+|, |B_{j,k}^-|\right) - p  = p - 2 \sum_{j=1}^m \sum_{k=1}^\ell \min(|B_{j,k}^+|, |B_{j,k}^-|).$$
Fix integers $m, K > 0$ such that $m K < \frac{p}{4 \ell}$. Then for sufficiently large $m$ and all $\delta > 0$,
$$\mathbb{P}(M_{}' >  2 \delta p) \le 2 \ell m \exp(-\delta^2 K) + \ell m \exp\left( - \frac{p}{5 \ell m} \right). $$
\begin{proof} Let $\mathcal{B}$ be the sigma-algebra generated by $\{|B_{j,k}|\}$. Observe for every $j,k$,
$$\max(|B_{j,k}^+|, |B_{j,k}^-|) \le \left(\frac{1}{2} + \delta\right) |B_{j,k}| + \frac{p}{\ell}  \mathbb{I}\left(\max(|B_{j,k}^+|, |B_{j,k}^-|) > \left(\frac{1}{2} + \delta\right) |B_{j,k}| \right).$$
Summing over $j$ and $k$, we find that
$$\sum_{j=1}^m \sum_{k=1}^\ell \max(|B_{j,k}^+|, |B_{j,k}^-|) \le \left(\frac{1}{2} + \delta\right) \sum_{j=1}^m \sum_{k=1}^\ell |B_{j,k}| + \frac{p}{\ell} \sum_{j=1}^m \sum_{k=1}^\ell \mathbb{I}\left(\max(|B_{j,k}^+|, |B_{j,k}^-|) > (1/2 + \delta) |B_{j,k}| \right).$$
Plugging this in yields
\begin{align}
    \mathbb{P} \left( M_{}' \ge 2 \delta p \, | \, \mathcal{B} \right) 
    = &
    \mathbb{P} \left(\sum_{j=1}^m \sum_{k=1}^\ell \max(|B_{j,k}^+|,|B_{j,k}^-|) > \left (\delta + \frac{1}{2}\right)  p \, | \, \mathcal{B} \right) 
    \nonumber \\
    \le &
    \mathbb{P} \left( \, \left(\delta + \frac{1}{2}\right) \sum_{j=1}^m \sum_{k=1}^\ell  |B_{j,k}| + \frac{p}{\ell} \sum_{j=1}^m \sum_{k=1}^\ell \mathbb{I}\left( \max(|B_{j,k}^+|, |B_{j,k}^-|) > \left(\delta + \frac{1}{2}\right) |B_{j,k}| \right) > \left (\delta + \frac{1}{2}\right)  p  \, | \, \mathcal{B} \right) 
    \nonumber
    \\
    = & \mathbb{P}\left( \sum_{j=1}^m \sum_{k=1}^\ell \mathbb{I}\left( \max(|B_{j,k}^+|, |B_{j,k}^-|) > \left(\delta + \frac{1}{2}\right) |B_{j,k}| \right) > 0 \, | \, \mathcal{B} \right),
    \label{eq:initTbound} 
    \end{align} 

where in the third line, we apply the fact that $\sum_{j,k} |B_{j,k}| = p$. At this point, we observe that
$$|B_{j,k}^+| \, \bigg| \, \mathcal{B} \sim \text{Bin}\left(|B_{j,k}|, \frac{1}{2}\right)  $$
and $|B_{j,k}^-|$ has the same distribution conditional on $\mathcal{B}$. Therefore we apply a union bound plus Hoeffding's inequality for binomials to conclude:
\begin{align}
     \mathbb{P}\left( \max(|B_{j,k}^+|, |B_{j,k}^-|) > \left(\delta + \frac{1}{2}\right) |B_{j,k}|  \, \big | \, \mathcal{B} \right)
     \le & \mathbb{P}\left (|B_{j,k}^+| > \left(\delta + \frac{1}{2}\right) |B_{j,k}|  \, \big |\,  \mathcal{B} \right ) + \mathbb{P}\left(|B_{j,k}^-| > \left(\delta + \frac{1}{2}\right) |B_{j,k}|  \, \big |\,  \mathcal{B}   \right) \nonumber \\
     = &  2 \mathbb{P}\left(|B_{j,k}^+| > \left(\delta + \frac{1}{2}\right) |B_{j,k}|  \, \big |\,  \mathcal{B} \right)  \nonumber \\
     \le & 2 \exp \left(- \delta^2 |B_{j,k}| \right) \, \, \text{ by Hoeffding's inequality } \nonumber \\
     \le & 2 \exp(-\delta^2 \min_{1 \le j \le m} \min_{1 \le k \le \ell} |B_{j,k}|). \label{eq:avgequisignhoeffding}
\end{align}

Applying a union bound to equations (\ref{eq:initTbound}) and  (\ref{eq:avgequisignhoeffding}), we obtain the conditional bound
$$\mathbb{P}_{\beta} \left( M_{}' \ge 2 \delta p \mid \mathcal{B} \right) \le 2 \ell m \exp(- \delta^2 \min_{1 \le j \le m} \min_{1 \le k \le \ell} |B_{j,k}| ).$$

This implies that unconditionally, for any $K$,
\begin{align}
        \mathbb{P}_{\beta} \left( M_{}' \ge 2 \delta p \right) 
    & \le 
        \mathbb{P}_{\beta}\left(M_{}' \ge 2 \delta
        p \, \big| \, \min_{1 \le j \le m} \min_{1 \le k \le \ell} |B_{j,k}| \ge K \right) + \mathbb{P}_{\beta}\left(\min_{1 \le j \le m} \min_{1 \le k \le \ell} |B_{j,k}| < K \right)
    \nonumber \\
    & \le
        2 \ell m \exp(-\delta^2 K) + \ell m \exp \left( - \frac{p}{5\ell m} \right),
    \nonumber 
\end{align}

where the last substitution combines the previous argument with Lemma \ref{lem::groupsizes}.

\end{proof}
\end{lemma}

\section{Theory of MRC knockoffs}\label{appendix::posresults}

In this section, we prove Proposition \ref{prop::mvrolsopt} and Theorem \ref{thm::olspower1}. We also provide a few extra simulations demonstrating the effect of reconstruction on power, as explained in Section \ref{sec::mvrknock}.

\subsection{Estimation error}

\begingroup
\def\theproposition{\ref{prop::mvrolsopt}} 
\begin{proposition}
Suppose $X \sim \mathcal{N}(0, \Sigma)$ for any $\Sigma$ and $Y\mid X \sim \mathcal{N}(X \beta, \sigma^2)$. Let $\betaext \in \mathbb{R}^{2p}$ be the concatenation of $\beta \in \mathbb{R}^p$ with $p$ zeros. Suppose $\hatbetaext \in \mathbb{R}^{2p}$ are OLS coefficients fit on $[\bX, \tilde{\bX}], \by$ and $n > 2p + 1$. Then
$$S_{\mathrm{MVR}} = \arg \min_S \mathbb{E}[||\hatbetaext-\betaext||_2^2].$$
\begin{proof} Fix $S$ such that $G_S$ is positive definite (otherwise we cannot fit OLS statistics). Since $\tilde{X} \Perp Y  \mid  X$, 
$$Y \mid [X, \tilde{X}] \sim \mathcal{N}([X, \tilde{X}] \betaext, \sigma^2). $$
Since $\hatbetaext$ is the OLS statistic on $[\bX, \tilde{\bX}]$ and $\by$,
\begin{equation}\label{eq:fxolsvar}
    \hatbetaext\mid [\bX, \tilde{\bX}] \sim \mathcal{N}\left(\betaext, \sigma^2 \left([\bX, \tilde{\bX}]^{\top} [\bX, \tilde{\bX}]\right)^{-1}\right).
\end{equation}
By construction, $[X, \tilde{X}] \sim \mathcal{N}(0, G_S)$, so $\left([\bX, \tilde{\bX}]^{\top} [\bX, \tilde{\bX}]\right) \sim \mathbb{W}_{2p}(n, G_S)$ where $\mathbb{W}_{2p}$ denotes the $2p$-dimensional Wishart distribution. When $n > 2p$, $\left([\bX, \tilde{\bX}]^{\top} [\bX, \tilde{\bX}]\right)^{-1} \sim \mathbb{W}_{2p}^{-1}(n, G_S)$ where $\mathbb{W}^{-1}_{2p}$ is the inverse-Wishart distribution. The law of iterated expectation implies that for any $j \in [2p]$,
$$\mathbb{E}\left[\left(\hatbetaext_j - \beta_j^{(\mathrm{ext})}\right)^2 \right] = \mathbb{E}\left[\sigma^2 \left( \left([\bX, \tilde{\bX}]^{\top} [\bX, \tilde{\bX}]\right)^{-1}\right)_{j,j}\right] = \frac{\sigma^2}{n - 2p - 1} (G_S^{-1})_{j,j}.$$
Summing over $j$, this yields
\begin{equation}\label{eq::l2intermsofmvr}
    \mathbb{E}[||\hatbetaext - \betaext||_2^2] = \frac{\sigma^2}{n - 2p - 1} \text{Tr}(G_S^{-1}) \propto L_{\mathrm{MVR}}(S).
\end{equation}
Since $S_{\mathrm{MVR}}$ minimizes $L_{\mathrm{MVR}}(S)$, it also minimizes $\mathbb{E}[||\hatbetaext - \betaext||_2^2]$.
\end{proof}
\end{proposition}
\addtocounter{proposition}{-1}
\endgroup

\subsection{Consistency}

Theorem \ref{thm::olspower1} tells us that when $X \sim \mathcal{N}(0, \Sigma)$ and $Y \mid X \sim \mathcal{N}(X \beta, \sigma^2)$, the power of OLS absolute coefficient difference statistics applied to MVR knockoffs converges to $1$ in low-dimensional settings. To prove this, we first prove that such feature statistics $W$ converge in squared $\ell_2$ norm to the linear coefficients $\beta$.

\begin{theorem}\label{thm::l2uniformconv} 
Suppose $X \sim \mathcal{N}(0, \Sigman)$, $Y\mid X \sim \mathcal{N}(X \betan, \sigma^2)$, and $\tilde{X}$ is generated using $S_{\mathrm{MVR}}$. Finally,  suppose $\hatbetaext \in \mathbb{R}^{2p}$ are OLS coefficients fit on $([\bX, \tilde{\bX}], \by)$, and let $W_j = |\hatbetaext_j| - |\hatbetaext_{j+p}|$ for $1 \le j \le p$.

Consider a sequence of covariance matrices $\Sigman$ such that the minimum eigenvalue of $\Sigman \in \mathbb{R}^{p \times p}$ is bounded uniformly above a fixed constant $\gamma \in \mathbb{R}^+$. Let $n, p \to \infty$ and assume that $p = o(n)$. Then for any $\epsilon > 0$,
$$\lim_{n \to \infty} \sup_{\betan \in \mathbb{R}^p} \mathbb{P}_{\betan}\left[ \big|\big|\, W - |\betan| \, \big|\big|_2^2 > \epsilon \right] = 0. $$
\begin{proof} Define $S^{(n)} = \gamma I_{p}$. For each $\Sigman$, let $G_{\mathrm{MVR}}^{(n)}$ denote the $G$-matrix formed from $\Sigman$ and the MVR solution, and let $G_{S}^{(n)}$ be the $G$-matrix from $\Sigman$ and $S^{(n)}$. Standard Schur-complement analysis yields that
$$\text{Tr}((G_{S}^{(n)})^{-1}) = \sum_{j=1}^{2p} \frac{1}{\lambda_j \left(G_{S}^{(n)}\right)} = \sum_{j=1}^p \frac{1}{\lambda_j \left(S^{(n)}\right)} + \sum_{j=1}^p \frac{1}{\lambda_j \left(2\Sigman - S^{(n)}\right)} \le \frac{2p}{\gamma} .$$
The last step follows because by definition all of the eigenvalues of $S^{(n)}$ are $\gamma$, and since $\lambda_j(\Sigman) \ge \gamma$ for each $j$, we must have that $\lambda_j(2 \Sigman - S^{(n)}) \ge \gamma$ as well. At this point, we note that by definition of the MVR loss, we have that
\begin{equation*}
    \text{Tr}\left(\left(G_{\mathrm{MVR}}^{(n)}\right)^{-1}\right) \le \text{Tr}\left(\left(G_S^{(n)}\right)^{-1}\right) \le  \frac{2p}{\gamma}. 
\end{equation*}
Using equation (\ref{eq::l2intermsofmvr}) from Proposition \ref{prop::mvrolsopt}, this implies that if we let $\betanext \in \mathbb{R}^{2p}$ be the concatenation of $\betan$ with $p$ zeros, 
$$ \mathbb{E}[||\hatbetaext - \betanext||_2^2] = \frac{\sigma^2}{n - 2p - 1} \text{Tr}\left(\left(G_{\mathrm{MVR}}^{(n)}\right)^{-1}\right) \le \frac{2p \sigma^2}{\gamma(n - 2p - 1)}.$$
This holds uniformly over $\betan$, implying
$$\lim_{n \to \infty} \sup_{\betan \in \mathbb{R}^p} \mathbb{E}[||\hatbetaext - \betanext||_2^2] \le \lim_{n \to \infty} \frac{2 p \sigma^2}{\gamma (n - 2p - 1)} = 0 $$
where the last step follows because $p = o(n)$. Next, observe that
\begin{align*}
\big|\big|\, W - |\betan| \, \big|\big|_2^2
&= \sum_{j=1}^p \left(|\hatbetaext_j| - |\hatbetaext_{j+p}| - |\betan_j|\right)^2 \\
&= \sum_{j=1}^p \left(|\hatbetaext_j|-|\betan_j|\right)^2 + \left(\hatbetaext_{j+p}\right)^2 - 2 \sum_{j=1}^p \left(|\hatbetaext_j| - |\betan_j|\right) \left|\hatbetaext_{j+p}\right|  \\
&\le \sum_{j=1}^p (\hatbetaext_j-\betan_j)^2 + (\hatbetaext_{j+p})^2 + 2 \sum_{j=1}^p \left|\hatbetaext_j - \betan_j\right| \left|\hatbetaext_{j+p}\right| \text{ by the reverse triangle inequality }  \\
&\le \left|\left|\hatbetaext - \betanext\right|\right|_2^2 + 2 \sqrt{\left|\left|\hatbetaext_{1:p} - \betan\right|\right|_2^2 \cdot \left|\left|\hatbetaext_{(p+1):2p}\right|\right|_2^2 } \text{ by the Cauchy---Schwartz inequality } \\
&\le 3 \left|\left|\hatbetaext - \betanext\right|\right|_2^2.
\end{align*}
This implies that $\big|\big|\, W - |\betan| \, \big|\big|_2^2$ must converge in expectation to zero uniformly over $\betan$. This implies that it also converges in probability to zero uniformly over $\betan$, which proves the theorem.
\end{proof}
\end{theorem}

\begingroup
\def\thetheorem{\ref{thm::olspower1}} 
\begin{theorem} Suppose $X \sim \mathcal{N}(0, \Sigman)$, $Y\mid X \sim \mathcal{N}(X \betan, \sigma^2)$, and $\tilde{X}$ is generated using $S_{\mathrm{MVR}}$. Suppose $\hatbetaext \in \mathbb{R}^{2p}$ are OLS coefficients fit on $([\bX, \tilde{\bX}], \by)$, and set $w([\bX, \tilde{\bX}],\by) = |\hatbetaext_{1:p}| - |\hatbetaext_{(p+1):2p}|$.

Let $n, p \to \infty$ such that $p = o(n)$ and consider a sequence of covariance matrices $\Sigman \in \mathbb{R}^{p \times p}$ such that the minimum eigenvalue of $\Sigman$ is bounded above a fixed constant $\gamma \in \mathbb{R}^+$. Suppose we sample a sequence of random $\betan$ as follows. Let all but a uniformly drawn subset of $\lceil s_0 p \rceil$ entries of $\betan$ equal zero, for a fixed constant $s_0 \in (0, 1]$, and then sample the remaining (non-null) entries of $\betan$ from a $\lceil s_0 p \rceil$-dimensional hypercube centered at $0$ with any fixed side-length. Then 
$$\power(w, \betan) \stackrel{p}{\to} 1. $$

\begin{proof} Fix $\epsilon > 0$. We will show that $\lim_{n \to \infty} \mathbb{P}(\power(w, \betan) \ge 1 - \epsilon) = 1 $. Note throughout this proof, we use $\mathbb{P}_{\betan}$ to denote a probability over the data for a fixed $\betan$, and we use $\mathbb{P}$ when the probability is over a random $\betan$. Fix $\epsilon_0 > 0$ and define $\bepsn = \{j : |\betan_j| > \epsilon_0 \}$, i.e., the set of all non-nulls with a coefficient values at least $\epsilon_0$ away from zero. Then note that
$$\big|\big|\, W - |\betan| \, \big|\big|_2^2 < \frac{\epsilon_0^2}{4} \implies \# \left\{ j \in \bepsn : W_j \le \frac{\epsilon_0}{2} \right \} = 0 \text{ and } \# \left \{j : W_j \le -\frac{\epsilon_0}{2} \right \} = 0 .$$
Intuitively, the former statement tells us that when $W$ approximates $|\betan|$ sufficiently closely in $\ell_2$ norm, all of the feature statistics corresponding to $\bepsn$ must be greater than $\epsilon_0/2$, and no feature statistics can be smaller than $- \epsilon_0/2$.

Note that by the definition of the knockoffs procedure, which ranks the $W$ statistics by absolute values, the events above imply that the $|\bepsn|$ largest feature statistics are all positive. When $|\bepsn| \ge \left \lceil \frac{1}{1+q} \right \rceil$, this implies that the knockoffs procedure will reject at least $|\bepsn|$ non-nulls. Let $\tau$ be the number of non-nulls rejected by the procedure. Then this means that for sufficiently large $|\bepsn|$, 
 $$\big|\big|\, W - |\betan| \, \big|\big|_2^2 < \frac{\epsilon_0^2}{4} \implies \tau \ge |\bepsn|.$$
 This statement plus Theorem \ref{thm::l2uniformconv} tells us that if $\cR^{(n)} = \left\{ \betan \in \mathbb{R}^p : |\bepsn| \ge \left \lceil \frac{1}{1+q} \right \rceil \right \}$, then
 $$ \lim_{n \to \infty} \sup_{\betan \in \cR^{(n)}} \mathbb{P}_{\betan}\left(\tau < |\bepsn| \right) = 0.$$
By the definition of power, this implies that for any $\epsilon_1 > 0$ and sufficiently large $p$, the following holds deterministically for any $\betan$ such that $|\bepsn| \ge \lceil \frac{1}{1+q} \rceil$:
$$\power(w, \betan) \ge (1 - \epsilon_1) \frac{|\bepsn|}{\lceil s_0 p \rceil}.$$
If $j$ is one of the $\lceil s_0 p \rceil$ non-null coordinates, then $|\betan_j| \iid \mathrm{Unif}(0, b)$. This implies $\mathbb{P}(|\betan_j| \ge \epsilon_0) = 1 - \frac{\epsilon_0}{b}$. Since each coordinate is independent, the law of large numbers implies $$\frac{|\bepsn|}{\lceil s_0 p \rceil} \stackrel{p}{\to} 1 - \frac{\epsilon_0}{b}.$$
Therefore, for any $\delta > 0$, there are sufficiently large $n$, $p$ such that 
$$\mathbb{P}\left(|\bepsn| > \left(1 - \frac{2 \epsilon_0}{b}\right) \lceil s_0 p \rceil \right) \ge 1 - \delta,$$
and note that as $p$ grows, this event guarantees that $\betan \in \cR^{(n)}$. All of the prior analysis implies that for any $\epsilon_0, \epsilon_1, \delta > 0$, there exist sufficiently large $n$, $p$ such that
$$\mathbb{P}\left(\power(w, \betan) > \left(1 - \epsilon_1\right) \left(1 - \frac{2\epsilon_0}{b}\right) \right) \ge 1 - \delta. $$
If we pick $\epsilon_0, \epsilon_1$ small enough such that $(1 - \epsilon_1) \left(1 - \frac{2\epsilon_0}{b} \right) \ge 1 - \epsilon$, this implies that
$$\lim_{n \to \infty} \mathbb{P}(\power(w, \betan) \ge 1 - \epsilon) = 1.$$
\end{proof}
\end{theorem}
\addtocounter{theorem}{-1}
\endgroup

\subsection{The importance of harshly penalizing high levels of reconstructability}\label{appendix::manywstats}

When sampling $\tilde{X}$, we will likely face trade-offs where to reduce the reconstructability of some feature $X_j$, we must increase the reconstructability of another feature $X_k$ in order maintain the pairwise exchangeability condition (\ref{eq::pairwiseexchange}). To navigate these trade-offs, in Section \ref{subsec::mvrdef}, we suggested a general principle of harshly penalizing high levels of reconstructability to ensure no feature $X_j$ is highly reconstructable from $X_{\text{-}j}, \tilde{X}$. In particular, we argued in Section \ref{subsec::mvrdef} that if a non-null $X_j$ is reconstructable using a knockoff $\tilde{X}_k$, then $\tilde{X}_k$ may be assigned $X_j$'s variable importance, causing the feature statistic $W_k$ to have a large magnitude but a negative sign.

As a concrete example, consider the case where $X \sim \mathcal{N}(0, \Sigma)$ for equicorrelated $\Sigma$ with $\rho \ge 0.5$, as discussed in Section \ref{sec::reconstruction}. Further suppose that $Y \mid X \sim \mathcal{N}(X \beta, 1)$ and we use SDP knockoffs, so $\Var(X_j | X_{\text{-}j}, \tilde{X}) = 0$ for all $j$. Fix $j,k\in [p]$ and suppose that $\beta_k \approx 1$ and $\beta_j \approx -1$. Even though SDP knockoffs are asymptotically powerless in this setting, the intuition in Section \ref{subsec::sdpopt} suggests that common feature statistics like lasso coefficients may correctly estimate the magnitude of $W_k$ and $W_j$, such that $|W_k|, |W_j| \to 1$. However, with approximately $50\%$ probability, the lasso will assign the feature importance of $X_k$ to $\tilde{X}_j$ and the feature importance of $X_j$ to $\tilde{X}_k$, so $(W_k, W_j) \stackrel{d}{\approx} (-W_j, -W_k)$, as in Corollary \ref{corr::condapprxreconst}.  This makes it difficult to discover \textit{any} non-null features, because the feature statistics often have large magnitudes and negative signs. 

\begin{figure}[h!]
    \centering
    \makebox[\textwidth]{\includegraphics[width=\textwidth]{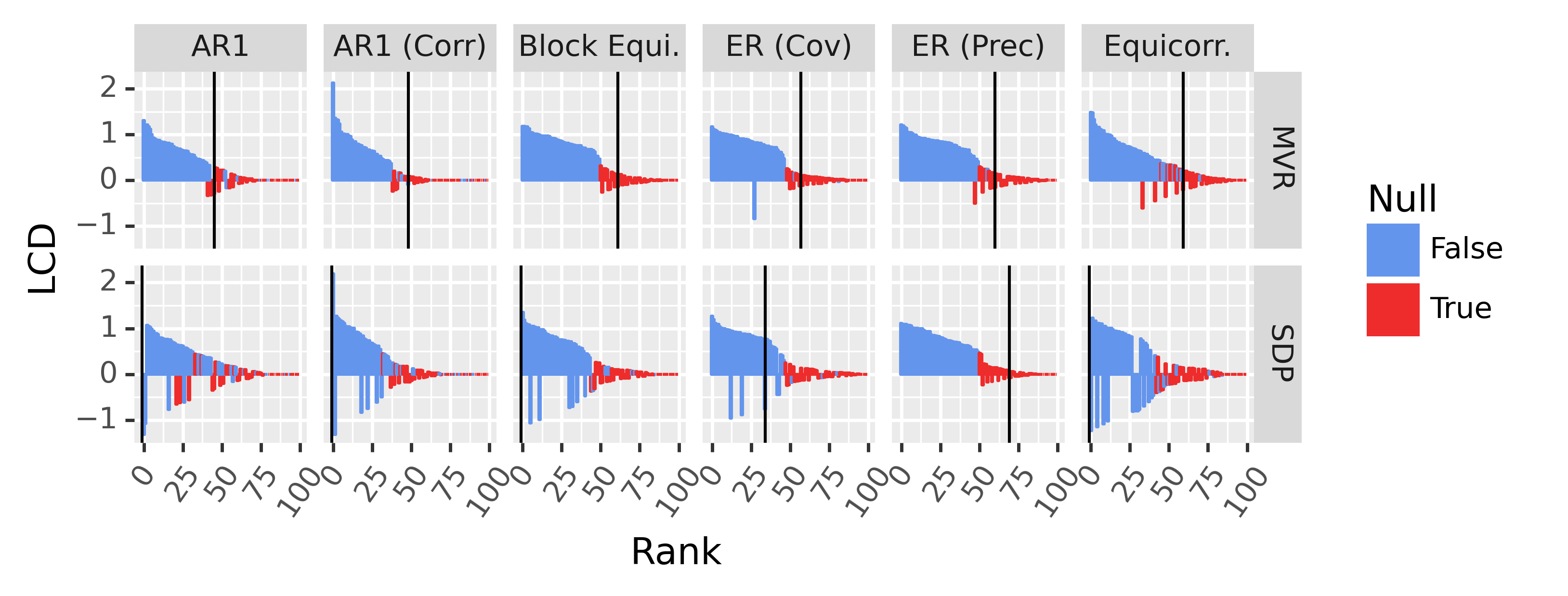}}
    \caption{We plot LCD statistics sorted in descending order of absolute value for several Gaussian linear models. The horizontal facets correspond to the design distributions specified above. Note $Y \mid X \sim \mathcal{N}(X \beta, 1)$ where $\beta$ has $50$ non-nulls with values $\pm 1$, with equal probability. The black lines denote the data-dependent thresholds.} 
    \label{fig::manywstats}
\end{figure}

Figure \ref{fig:wstats} in Section \ref{sec::mvrknock} demonstrates that for equicorrelated Gaussian random variables with a linear response, the reconstruction effect for SDP knockoffs causes many feature statistics to have large absolute values but negative signs. In Figure \ref{fig::manywstats}, we demonstrate that the same effect occurs for non-exchangeable Gaussian features. In particular, we compute feature statistics for all of the data-generating processes described in Section \ref{subsec::simgaussian}, with the following specific parameters. In all cases we set $p = 100$ and $n = 190$ with $Y \mid X \sim \mathcal{N}(X \beta, 1)$ where $\beta$ has $50$ non-nulls with values $\pm 1$. We set $\rho = 0.6$ for the equicorrelated and block-equicorrelated designs, $a = 3$ for the  AR1 data-generating processes, and the sparsity parameter equal to $0.2$ for the ErdosRenyi designs.

\section{Convexity and computation for MRC knockoffs}\label{appendix::computation}

\subsection{Proof of convexity}\label{appendix::convexity}

In this section, we prove that the MVR formulation in the Gaussian case can be reduced to a simple semidefinite program. Note \cite{multiknock2018} have previously shown a similar result for the \smaxent formulation.

\begin{lemma}\label{lem::mvrconvex} Consider the MVR optimization problem $\min_S \text{Tr}(G_S^{-1})$ such that $0 \preccurlyeq S \preccurlyeq 2 \Sigma $ with $S$ diagonal. Let $R_1, R_2 \in \mathbb{R}^{p \times p}$ be slack variables and, as usual, $S$ is a diagonal matrix. Then the MVR optimization problem is equivalent to the following SDP. 
\begin{align*}
 \text{\emph{minimize }} & \text{Tr}(R_1) + \text{Tr}(R_2) \\ 
    \text{\emph{subject to }} & B(R_1, R_2, S) \equiv \begin{bmatrix} 
    R_1 & 0 & I_p & 0 \\
    0 & R_2 & 0 & I_p \\
    I_p & 0 & S & 0 \\ 
    0 & I_p & 0 & 2 \Sigma - S \\
    \end{bmatrix} \succcurlyeq 0.
\end{align*}
\begin{proof} Recall that $L_{\mathrm{MVR}}(S) \propto \text{Tr}(G_S^{-1}) = \text{Tr}((2 \Sigma - S)^{-1}) + \text{Tr}(S^{-1}) $. Schur complement analysis yields
$$B(R_1, R_2, S) \succcurlyeq 0 \text{ if and only if } \begin{bmatrix} R_1 & 0 \\ 0 & R_2 \\ \end{bmatrix} - \begin{bmatrix} S & 0 \\ 0 & 2 \Sigma - S \end{bmatrix}^{-1} \succcurlyeq 0  $$
which in turn implies that for any feasible solution,
$$\text{Tr}(R_1) + \text{Tr}(R_2) \ge \text{Tr}(S^{-1}) + \text{Tr}((2 \Sigma - S)^{-1}) .$$
Additionally, for a fixed $S$, one can always set $R_1 = S^{-1}$ and $R_2 = (2 \Sigma - S)^{-1}$ and achieve $\text{Tr}(R_1) + \text{Tr}(R_2) = \text{Tr}(S^{-1}) + \text{Tr}((2 \Sigma - S)^{-1})$. Therefore, minimizing $\text{Tr}(R_1) + \text{Tr}(R_2)$ is equivalent to minimizing $\text{Tr}(S^{-1}) + \text{Tr}(2 \Sigma - S)^{-1}) \propto L_{\mathrm{MVR}}(S)$. 
\end{proof}
\end{lemma}

Unfortunately, generic solvers for this problem may be as slow as $O(p^6)$ \citep{kypsdp}. In the next two sections, we develop a much faster algorithm to compute $S_{\mathrm{MVR}}$ and $S_{\mmaxent}$.  

\subsection{Computing MVR knockoffs}\label{appendix::computemvr}

In this section, we introduce Algorithm \ref{alg::mvrstable}, which computes $S_{\mathrm{MVR}}$ in $O(n_{\mathrm{iter}} p^3)$. This algorithm is inspired by \cite{fanok2020} and uses their overall strategy, although the technical details differ. 

The key idea behind the algorithm is as follows. Fix $j \in [p]$ and a diagonal matrix $S \succcurlyeq 0$ such that $D \equiv 2 \Sigma - S \succcurlyeq 0$. Furthermore, let $M$ be the matrix of all zeros except $M_{j,j} = 1$. We first use Schur complements to decompose
\begin{equation}\label{eq::mvrcoorddesc1}
L_{\mathrm{MVR}}(S + \delta_j M) = \text{Tr}(2 \Sigma - S - \delta_j M)^{-1} + \text{Tr}(S + \delta_j M)^{-1}.
\end{equation}
We seek to find $\delta_j$ which minimizes this quantity. Since $S + \delta_j M$ is diagonal, $\text{Tr}(S + \delta_j M)^{-1} = 1 / (S_{j,j} + \delta_j) +  \sum_{k \ne j} \frac{1}{S_{k,k}}$, and only the term $1 / (S_{j,j} + \delta_j)$ depends on $\delta_j$. Then, we apply the Sherman--Morrison rank-one inversion formula to note
\begin{equation}\label{eq::mvrcoorddesc2}
(D - \delta M)^{-1} = D^{-1} - \frac{- \delta_j D^{-1} M D^{-1}}{1 - \delta_j D^{-1}_{j,j}},
\end{equation}
where $D$ is constant with respect to $\delta_j$. Let $c_n = - \text{Tr}(D^{-1} M D^{-1})$ and let $c_d = D^{-1}_{j,j}$. Then, equations (\ref{eq::mvrcoorddesc1}) and (\ref{eq::mvrcoorddesc2}) imply that
\begin{equation}\label{eq::mvrcoorddesc3}
L_{\mathrm{MVR}}(S + \delta_j M) = \frac{1}{S_{j,j} + \delta_j} - \frac{\delta_j c_n}{1 - \delta_j c_d} + c
\end{equation} 
for some constant $c$. Taking the derivative of equation (\ref{eq::mvrcoorddesc3}) with respect to $\delta_j$ yields the quadratic optimality condition (\ref{eq::mvrquadraticoptimalitycond}), which can be solved in constant time:
\begin{equation}\label{eq::mvrquadraticoptimalitycond}
(- c_n - c_d^2) \delta_j^2 + 2 (- c_n S_{j,j} + c_d) \delta_j + (- c_n S_{j,j}^2 - 1) = 0. 
\end{equation} 
Note equation (\ref{eq::mvrquadraticoptimalitycond}) and the constraint from (\ref{eq::mvrcoorddesc3}) that $- S_{j,j} < \delta_j < \frac{1}{c_d}$ yield a unique solution for $\delta_j$. To compute $c_n$ and $c_d$ efficiently, we could initially compute $D^{-1}$ and then update $D^{-1}$ at each step using rank-$1$ updates, which has time complexity $O(p^2)$. Unfortunately, as \cite{fanok2020} observed, the rank-$1$ updates to $D^{-1}$ are numerically unstable. Following their approach, we detail an alternative which is equally efficient but maintains a Cholesky decomposition of $D$ instead of  maintaining $D^{-1}$. 

Indeed, given a Cholesky decomposition of $D = LL^{\top}$, we can compute $c_n$ and $c_d$ in $O(p^2)$. To see how, let $e_j$ be the $j$th basis vector. Using the fact that $\text{Tr}(u v^{\top}) = u^{\top} v$ for any two vectors $u, v \in \mathbb{R}^p$, we can then represent
\begin{equation}\label{eq::computecncdeq1}
    c_n = - (D^{-1} e_j)^{\top} (D^{-1} e_j) \text{ and } c_d = e_j^{\top} D^{-1} e_j.
\end{equation}
To compute $c_n$, note if we let $v_n = D^{-1} e_j$, then $c_n = -||v_n||_2^2$. However, we can solve for $v_n$ by solving the system $LL^{\top} v_n = e_j$. Since $L$ is triangular, this can be solved in $O(p^2)$ using forward-backward substitution. Second, we observe that $e_j^{\top} D^{-1} e_j = e_j^{\top} (LL^{\top})^{-1} e_j = e_j^{\top} (L^{\top})^{-1} L^{-1} e_j = (L^{-1} e_j)^{\top} (L^{-1} e_j)$. Therefore, if we let $v_d = L^{-1} e_j$, then $c_d = ||v_d||_2^2$. Since $L$ is triangular, we can find $v_d$ as the solution to $L v_d = e_j$ in $O(p^2)$.

This motivates the following algorithm, which maintains a running copy of $L L^{\top}$. After updating the value of $S_{j,j}$, we perform a rank one update of $L$, which can be done in $O(p^2)$. The result is Algorithm \ref{alg::mvrstable}, which is numerically stable and runs in $O(n_{\mathrm{iter}} p^3)$. 

\begin{algorithm}[h!]
\caption{Stable Coordinate Descent for MVR Knockoffs}\label{alg::mvrstable}
\begin{algorithmic}[1]
        \State Initialize $S = \lambda_{\mathrm{min}}(\Sigma) \cdot I_p$
        \State Solve for $L$ such that $L L^{\top} = 2 \Sigma - S$
        \For {$\ell=1,2,\dots,n_{\mathrm{iter}}$}
            \For {$j=1,2,\dots,p$}
        		\State Compute $c_n$ and $c_d$ using $L$ as detailed in (\ref{eq::computecncdeq1})
    			\State Solve $\delta_j^*$ as the solution to the optimality condition (\ref{eq::mvrquadraticoptimalitycond})
    			\State Set $S_{j,j} = S_{j,j} + \delta_j^*$
    			\State Compute the rank-$1$ update to $L$ for $\delta_j^*$
    		\EndFor
		\EndFor
	\end{algorithmic} 
\end{algorithm}

\subsection{Coordinate descent for \smaxent knockoffs}\label{appendix::computemaxent} 

In this section, we introduce a coordinate descent algorithm to efficiently compute the $S$-matrix for \smaxent knockoffs. Note that this algorithm is similar to the barrier formulation for the SDP posed in \cite{fanok2020}, with only minor changes to account for a different optimization function.

In particular, \cite{fanok2020} observed that if we let $D = 2 \Sigma - S$, we can write
\begin{equation}\label{eq::fanokdeteq}
\text{log} \, \text{det} \left( D \right) = \log (D_{j,j} - D_{-j,j}^{\top} D_{-j,-j}^{-1} D_{-j,j}) + \log \det \left( D_{-j,-j} \right),
\end{equation}
where only the term on the left depends on $S_{j,j}$. We can therefore write
\begin{equation}\label{eq::separatedmaxent}
\log \det (G_S) = \log \det(2 \Sigma - S) + \log \det (S) = \log (S_{j,j}) + \log (2 \Sigma_{j,j} - S_{j,j} - D_{-j,j}^{\top} D_{-j,-j}^{-1} D_{-j,j}) + c,
\end{equation}
where $c$ is a constant not depending on $S_{j,j}$. It also may be helpful to note that $D_{\text{-}j,j}$ and $D_{\text{-}j,\text{-}j}$ do not depend on $S_{j,j}$. Taking the derivative of equation (\ref{eq::separatedmaxent}) with respect to $S_{j,j}$, we find that we should update
\begin{equation}\label{eq::maxentoptcondition}
S_{j,j} = \frac{2 \Sigma_{j,j} - D_{-j,j}^{\top} D_{-j,-j}^{-1} D_{-j,j}}{2}.
\end{equation}
To efficiently calculate $c_m = D_{-j,j}^{\top} D_{-j,-j}^{-1} D_{-j,j}$, \cite{fanok2020} observed that $c_m = ||v_m||_2^2$ where $v_m$ is the solution to the equation $L v_m = u$ for $u \in \mathbb{R}^p$ where $u_i = 2 \Sigma_{i,j}$ if $i \ne j$ and $u_j = 0$ otherwise. Since $L$ is triangular and we can update the Cholesky decomposition $2 \Sigma - S = L L^{\top}$ using rank-$1$ updates, computing $c_m$ takes only $O(p^2)$. This yields Algorithm \ref{alg::maxent}.

\begin{algorithm}[h!]
\caption{Stable Coordinate Descent for \smaxent Knockoffs}\label{alg::maxent}
\begin{algorithmic}[1]
        \State Initialize $S = \lambda_{\mathrm{min}}(\Sigma) \cdot I_p$
        \State Solve for $L$ such that $L L^{\top} = 2 \Sigma - S$
        \For {$\ell=1,2,\dots,n_{\mathrm{iter}}$}
            \For {$j=1,2,\dots,p$}
        		\State Compute $c_m$ using $L$
    			\State Solve $S_{j,j}$ as the solution to the optimality condition (\ref{eq::maxentoptcondition})
    			\State Compute the rank-$1$ update to $L$ corresponding to the new value of $S_{j,j}^*$
    		\EndFor
		\EndFor
	\end{algorithmic} 
\end{algorithm}

\subsection{Solutions for equicorrelated Gaussian designs}\label{appendix::equimvreqmmi}

\begin{figure}
    \centering
    \makebox[\textwidth]{\includegraphics[width=\textwidth]{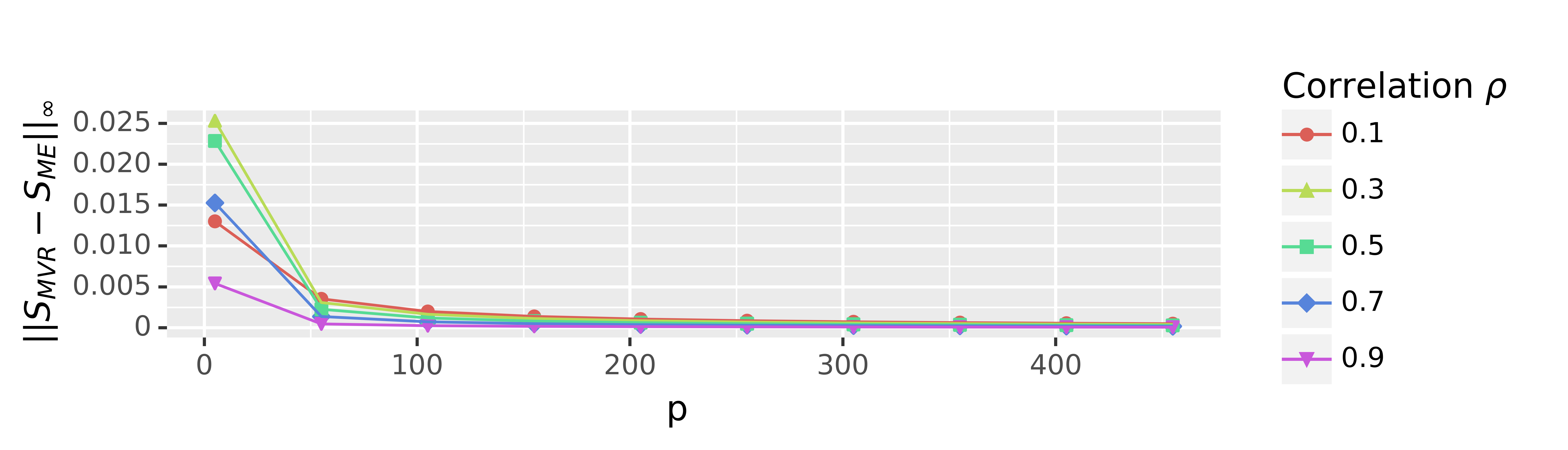}}
    \caption{MVR and \smaxent solutions for exchangeable features: we plot $||S_{\mathrm{MVR}} - S_{\mmaxent}||_{\infty}$ for equicorrelated $\Sigma$. We vary the correlation $\rho$ and the dimension $p$ and observe that the norm converges towards zero quite quickly.
    } 
    \label{fig::equimvreqmmi}
\end{figure}

In this section, we prove that when $\Sigma$ is equicorrelated, the same solution, $(1 - \rho) I_p$, is asymptotically optimal for both $L_{\mathrm{MVR}}$ and $L_{\mmaxent}$. Our result is asymptotic in $p$, but Figure \ref{fig::equimvreqmmi} demonstrates via simulations that $||S_{\mathrm{MVR}} - S_{\mmaxent}||_{\infty}$ quickly converges to $0$ for finite $p$. As we mentioned in Section \ref{subsec::maxentdef}, this means that we should expect \smaxent knockoffs to achieve approximately the same optimality guarantees as MVR knockoffs for Gaussian equicorrelated features.

\begin{theorem}\label{thm::mvreqmmi} Let $X \sim \mathcal{N}(0, \Sigma)$ for equicorrelated $\Sigma$ with correlation $\rho$. Then $S^* = (1 - \rho) I_p$ is asymptotically optimal for both the MVR and \smaxent optimization problems. In particular,
$$\lim_{p \to \infty} L_{\mathrm{MVR}}(S_{\mathrm{MVR}}) - L_{\mathrm{MVR}}(S^*) = \lim_{p \to \infty} L_{\mmaxent}(S_\mmaxent) - L_{\mmaxent}(S^*) = 0.$$
This result is nontrivial in the sense that the limit of each individual objective as $p \to \infty$ is $\infty$.
\begin{proof} We begin by showing that $S^*$ is asymptotically optimal for the \smaxent problem. Recall that the eigenvalues of $G_{S^*}$ are those of $S^*$ and $2 \Sigma - S^*$. Since $\Sigma$ is a rank-one update to the identity (see Lemma \ref{lem::equicorrsoln}), we can check that $G_{S^*}$ has two eigenvalues: $1-\rho$ with multiplicity $2p-1$ and $\lambda_{\mathrm{max}}(G_{S^*}) = 2 p \rho + 1 - \rho$ with multiplicity $1$. This implies $\log \det(G_{S^*}^{-1}) = - (2p-1) \log \left(1-\rho\right) - \log\left(\lambda_{\mathrm{max}}(G_{S^*}\right)$.

The primal \smaxent problem is to maximize $\log \det G_S$ subject to $G_S \succcurlyeq 0$, or equivalently to minimize $\log \det G_S^{-1}$ subject to the same constraint. The dual of this problem (see \cite{boydmaxdet1998}) is written below for a variable $R \in \mathbb{R}^{2p \times 2p}$:
\begin{align}
\text{ maximize } & \mathcal{O}_{\mmaxent}(R) \equiv \log \det\left(R\right) - \text{Tr}\left(G_0 R \right) + 2p \label{eq::dualmmi} \\
\text{ such that } & R \succcurlyeq 0 \\
                   & R_{j,j+p} = R_{j+p,j} = 0  \, \,\forall \, j \, \in [p], \nonumber
\end{align}
where $G_0$ is the $G_S$ matrix which corresponds to setting $S = 0 \cdot I_p$. To show $S^*$ is asymptotically optimal for the \smaxent problem, it suffices to show that we can find a sequence of $R^*$ such that the dual gap between $R^*$ and $S^*$ vanishes, i.e., $\lim_{p \to \infty} \log \det G_{S^*}^{-1} - \mathcal{O}_{\mmaxent}(R^*) = 0$. In particular, let $c_d = \frac{1}{1-\rho}$ and $c_r = \frac{1}{2p-2} \cdot \left(\frac{1}{\lambda_{\mathrm{max}}(G_{S^*})} - \frac{1}{1-\rho} \right)$. Then set
$$D = c_r  \mathbf{1} \mathbf{1}^{\top} + (c_d - c_r) I_p \text{ and } R^* = \begin{bmatrix} D & D - c_d I_p \\ D - c_d I_p & D \end{bmatrix}$$ 
Intuitively, $R^*$ is the $2p \times 2p$ matrix which takes the value $c_d$ on its diagonal, $0$ on the diagonals of its off-diagonal blocks, and $c_r$ everywhere else. We first observe that
$$\text{Tr}(G_0 R^*) = \sum_{j=1}^{2p} \sum_{k=1}^{2p} R^*_{jk} (G_0)_{jk} = 2p \cdot c_d + (4p^2 - 4p) \rho c_r.  $$
The first term in the last equality corresponds to the $2p$ diagonal elements of $G_0$ (which equal $1$) and those of $R^*$ (which equal $c_d)$. The second term corresponds to the other $4p^2 - 4p$ nonzero elements of $R^*$, which equal $c_r$, and the corresponding elements of $G_0$, which equal $\rho$. Simplifying yields that
\begin{equation*}
\text{Tr}(G_0 R^*) = \frac{2p}{1-\rho} + 2p \rho  \left(\frac{1}{2p \rho + 1 - \rho} - \frac{1}{1-\rho}\right) = 2p - \frac{2p \rho}{2 p \rho + 1 - \rho} .
\end{equation*}
This implies that $\lim_{p \to \infty} - \text{Tr}(G_0 R^*) + 2p = 1$. As a result, it suffices to show that $\lim_{p \to \infty} \log \det(G_{S^*}^{-1}) - \log \det(R^*) = - 1$. Note the eigenvalues of $R^*$ are those of $2D - c_d I_p$ and those of $c_d I_p$. Applying rank one theory, this yields eigenvalues of $c_d = \frac{1}{1-\rho}$ with multiplicity $p$, $(2p-2) c_r + \frac{1}{1-\rho} = \lambda_{\mathrm{max}}(G_{S^*})$ with multiplicity $1$, and $c_d - 2 c_r$ with multiplicity $p-1$. Note that the first $p+1$ of these eigenvalues agree with $p+1$ of the eigenvalues of $G_{S^*}$, as previous calculated. This yields
\begin{align}
\log \det(G_{S^*}^{-1}) - \log \det(R^*) 
    &= 
(p-1) \left[ \log\left(\frac{1}{1-\rho}\right) - \log \left( \frac{1}{1-\rho} - \frac{1}{p-1} \left(\frac{1}{2 p \rho + 1 - \rho} - \frac{1}{1-\rho} \right) \right) \right] \nonumber \\
    &= (p-1) \left[ \log\left(\frac{1}{1-\rho}\right) - \log\left(\frac{1}{1-\rho} \left(1 + \frac{1}{p-1} - \frac{1-\rho}{(p-1)(2 p \rho + 1 - \rho)} \right) \right) \right] \nonumber \\
    &= - (p-1) \log \left(1 + \frac{1}{p-1} - \frac{1-\rho}{(p-1)(2 \rho p + 1 - \rho)} \right). \label{eq::mmisolnudck}
\end{align}
To show (\ref{eq::mmisolnudck}) converges to $-1$, it suffices to show the following Lemma, where we take $x = p-1$.
\begin{lemma} Let $a,b \in \mathbb{R}$ for $a \ne 0$. Then
$$\lim_{x \to \infty} x \log\left(1 + \frac{1}{x} + \frac{1}{ax^2 + bx} \right) = 1 $$
\begin{proof} By L'Hopital's rule, 
\begin{align}
    \lim_{x \to \infty} \frac{\log\left(1 + 1/x + 1/(ax^2 + bx) \right)}{1/x}
    &= 
    \lim_{x \to \infty} \frac{-x^2}{1 + 1/x + 1/(ax^2+bx)} \cdot \left(-\frac{1}{x^2} - \frac{2ax +b}{(ax^2 + bx)^2}\right) \nonumber \\
    &= 
    \lim_{x \to \infty} \frac{1}{1 + 1/x + 1/(ax^2 + bx)} \cdot \left(1 + \frac{x^2 (2ax + b)}{(ax^2 + bx)^2} \right) \label{eq::mmisolnudckpf}
\end{align}
and (\ref{eq::mmisolnudckpf}) converges to $1$, as the left term in the product converges to $1$, and $\frac{x^2 (2ax + b)}{(ax^2 + bx)^2} \to 0$ since the numerator has degree $3$ but the denominator has degree $4$.
\end{proof}
\end{lemma}

Next, we show the same asymptotic optimality of $S^*$ for the MVR problem. The formulation of the MVR problem in Lemma \ref{lem::mvrconvex} is a standard SDP, which admits the following dual. Let $R_1, R_{21}, R_{22}, R_{23}, R_{24}, R_{41}, R_{42}, R_{44} \in \mathbb{R}^{p \times p}$. Then the dual is
\begin{align}
\text{ maximize } & \mathcal{O}_{\mathrm{MVR}}(R) \equiv - 2\text{Tr}(R_{21}) - 2\text{Tr}(R_{24}) - 2\text{Tr}(\Sigma R_{44}) \label{eq::mvrdual} \\
\text{ such that } & (R_{41})_{j,j} = (R_{44})_{j,j} \, \, \forall \, j \in [p]
\nonumber \\
& R \equiv \begin{bmatrix} I_p & R_1 & R_{21} & R_{22} \\ R_1 & I_p & R_{23} & R_{24} \\ R_{21} & R_{23} & R_{41} & R_{42} \\ R_{22} & R_{24} & R_{42} & R_{44}  \end{bmatrix}  \succcurlyeq 0. \nonumber
\end{align}
As before, we will find a sequence of dual feasible $R^*$ such that $\lim_{p \to \infty} L_{\mathrm{MVR}}(S^*) - \mathcal{O}_{\mathrm{MVR}}(R^*) = 0$. In particular, we will pick $R^*$ of the form 
\begin{equation}\label{eq::rstarform}
R^* = \begin{bmatrix} I_p & 0 & b I_p & 0 \\ 0 & I_p & 0 & a I_p -\frac{a}{p} \mathbf{1} \mathbf{1}^{\top} \\ b I_p & 0 & \left(c - \frac{c}{p}\right) I_p & 0 \\ 0 & a I_p - \frac{a}{p} \mathbf{1} \mathbf{1}^{\top} & 0 & c I_p - \frac{c}{p} \mathbf{1} \mathbf{1}^{\top}  \end{bmatrix}
\end{equation}
for the constants $a, b, c$, defined below. Note we define $a, b, c$ in terms of each other and in terms of the function $f(p) = (\sqrt{2p(2p-1)} - \sqrt{2p(2p-3) +2})^2$ for convenience of notation. For now, it is best to ignore $f(p)$---intuitively, $f(p)$ scales the constants $a,b,c$ by a number that converges to $1$ from above, which will help ensure that $R^*$ is feasible. Now, let
$$c = f(p) \left( \frac{2p-1}{2p} \right) \left(\frac{1}{1-\rho}\right)^2 \left( \frac{p}{p-1} \right),$$
$$b = - \sqrt{c - \frac{c}{p}},$$
$$a = - \frac{p}{p-1} \left(\frac{2p-1}{p(1-\rho)} + b\right).$$
For convenience, we will write $R^* = \begin{bmatrix} I_{2p} & R_2^* \\ R_2^* & R_4^* \end{bmatrix}$ where $R^*_2, R^*_4$ can also be represented as block matrices as in (\ref{eq::mvrdual}). Note this construction guarantees that the diagonals of $R^*_{41}$ and $R^*_{44}$ match. Therefore, we only need to show that $R^* \succcurlyeq 0$ to show it is dual feasible. Using the block formulation of $R^*$, it suffices to show that (i) $R^*_4 \succcurlyeq 0$ and (ii) $R^*_4 - R^*_2 (R^*_2)^{\top} \succcurlyeq 0$. To do this, we will repeatedly use the fact that for a matrix of the form $D = d_1 I_p - d_2 \mathbf{1}_p \mathbf{1}_p^{\top}$ for $d_1,d_2 > 0$,  $D \succcurlyeq 0$ if $d_2  \le d_1 / p$. (This is a simple consequence of Sherman--Morrison rank one theory.)

To show (i), note $R^*_4 = \begin{bmatrix} R^*_{41} & 0 \\ 0 & R^*_{44} \end{bmatrix}$, so we need only show that $R^*_{41} \succcurlyeq 0$ and $R^*_{44} \succcurlyeq 0$. $R^*_{41} = \left(c - \frac{c}{p}\right) I_p \succcurlyeq 0$ follows because $c - \frac{c}{p} > 0$. Furthermore, $R^*_{44} = c I_p - \frac{c}{p} \mathbf{1} \mathbf{1}^{\top} \succcurlyeq 0$ by the previous consequence of Sherman--Morrison theory. To show (ii), note
$$R^*_4 - R^*_2 (R^*_2)^{\top} = \begin{bmatrix} (c - \frac{c}{p} - b^2) I_p & 0 \\ 0 & c I_p - \frac{c}{p} \mathbf{1} \mathbf{1}^{\top} - \left(a I_p - \frac{a}{p} \mathbf{1} \mathbf{1}^{\top}\right) \left(a I_p - \frac{a}{p} \mathbf{1} \mathbf{1}^{\top}\right)^{\top} \end{bmatrix}.$$
The top-left block is positive-semidefinite since $c - \frac{c}{p} - b^2 \ge 0$ holds with equality by the definition of $b$. Simplifying the bottom-right block, we want to show
$$(c - a^2) I_p + \left(-\frac{c}{p}  + \frac{2a^2}{p} - p \cdot \frac{a^2}{p^2}\right) \mathbf{1} \mathbf{1}^{\top} = (c - a^2) I_p - \left(\frac{c - a^2}{p} \right) \mathbf{1} \mathbf{1}^{\top} \succcurlyeq 0, $$
which holds with equality based on the Sherman--Morrison theory. Lastly, we need to show that the coefficient on $I_p$ is nonnegative, i.e., $c - a^2 \ge 0$. To see this, we plug in the definitions:
\begin{align*}
    c - a^2
=&
    f(p) \left( \frac{2p-1}{2p} \right) \left( \frac{1}{1-\rho} \right)^2 \left( \frac{p}{p-1}\right)
    -
    \left( \frac{p}{p-1}\right)^2 \left(\frac{2p-1}{p(1-\rho)} - \sqrt{f(p) \left(\frac{2p-1}{2p}\right) \left( \frac{1}{1-\rho} \right)^2 } \right)^2 \\
=& 
    \left(\frac{1}{1-\rho} \right)^2 \left(\frac{p}{p-1} \right) 
    \left[
        f(p) \left( \frac{2p-1}{2p} \right)
        - \left( \frac{p}{p-1} \right) \left(\frac{2p-1}{p} - \sqrt{f(p) \left(\frac{2p-1}{2p}\right)} \right)^2 
    \right] \\
=& 
    \left(\frac{1}{1-\rho} \right)^2 \left(\frac{p}{p-1} \right) 
    \left[
        - \left(\frac{f(p)}{p-1} \right) \left( \frac{2p-1}{2p} \right)
        - \left( \frac{p}{p-1} \right) \left(\frac{2p-1}{p}\right)^2 
        + 2 \left( \frac{p}{p-1} \right) \left(\frac{2p-1}{p}\right) \sqrt{f(p) \left(\frac{2p-1}{2p}\right)}
    \right],
\end{align*}
where in the last line, we expand the squared term in the brackets and then use the fact that $f(p)\left(\frac{2p-1}{2p} \right) - \left( \frac{p}{p-1}\right) \left(\sqrt{f(p) \left(\frac{2p-1}{2p} \right)} \right)^2 = - \left(\frac{f(p)}{p-1} \right) \left( \frac{2p-1}{2p} \right)$. Continuing, we can factor $\frac{1}{p-1}$ and $\frac{2p-1}{p}$ out of the terms in the brackets:
\begin{align*}
    c-a^2
=& 
    \frac{2p-1}{(1-\rho)^2 (p-1)^2}
    \left[
        - \frac{f(p)}{2}
        - (2p-1)
        + 2 p \sqrt{f(p) \left(\frac{2p-1}{2p}\right)}
    \right] \\
=& 
    \frac{2p-1}{(1-\rho)^2 (p-1)^2}
    \left[
        - \frac{f(p)}{2}
        + \sqrt{f(p)} \sqrt{(2p)(2p-1)}
        - (2p-1)
    \right].
\end{align*}
At this point, we will show that the term in the brackets equals zero. This holds because that expression is a quadratic function of $\sqrt{f(p)}$, with roots at
$$\frac{-\sqrt{2p(2p-1)} \pm \sqrt{2p(2p-1) - 4 \cdot - \frac{1}{2} \cdot -(2p-1)} }{2 \cdot - \frac{1}{2}} = \sqrt{2p(2p-1)} \pm \sqrt{2p(2p-3) + 2}.$$
Since by definition, $\sqrt{f(p)} = \sqrt{2p(2p-1)} - \sqrt{2p(2p-3) + 2}$, we have that $c - a^2 \ge 0$ with equality.

Now that we know $R^*$ is dual feasible, we need to show the difference between $\mathcal{O}_{\mathrm{MVR}}(R^*)$ and $L_{\mathrm{MVR}}(S^*)$ vanishes asymptotically. It may be helpful to note at this point that by construction, $c - \frac{c}{p} = f(p) \cdot \frac{2p-1}{2p(1-\rho)^2}$. Now, we can easily check that $$\text{Tr}(\Sigma R^*_{44}) = \sum_{j,k \in [p]} \Sigma_{j,k} (R^*_{44})_{j,k} = p \left(c - \frac{c}{p}\right) -  (p^2 - p) \rho \frac{c}{p} = f(p) \left[ \frac{2p-1}{2 (1-\rho)^2} - \rho \left(\frac{2p-1}{2(1-\rho)^2} \right)\right] = f(p) \cdot \frac{2p-1}{2(1-\rho)}, $$
where as usual, in the second equality, the first term corresponds to the $p$ diagonal elements of $\Sigma$ and $R^*_{44}$, and the second term corresponds to the $p^2 - p$ off-diagonal elements. Finally, we note that by the definition of $a$, we have that $a - \frac{a}{p} + b = - \frac{2p-1}{p(1-\rho)}$. This yields that
$$\text{Tr}(R^*_{21}) + \text{Tr}(R^*_{24}) = p \left(b + a - \frac{a}{p}\right) = - \frac{2p-1}{1-\rho}.$$
Therefore $$\mathcal{O}_{\mathrm{MVR}}(R^*) = - 2 \text{Tr}(R^*_{21}) - 2\text{Tr}(R^*_{24}) - 2\text{Tr}(\Sigma R^*_{44}) = 2 \cdot \frac{2p-1}{1-\rho} - f(p) \cdot \frac{2p-1}{1-\rho} = \frac{2p-1}{1-\rho} + (1 - f(p)) \frac{2p-1}{1-\rho}.$$
Intuitively, since $f(p)$ converges to $1$ extremely quickly, $O_{\mathrm{MVR}}(R^*) \approx \frac{2p-1}{1-\rho}$. Recall from previous computations of the eigenvalues of $G_{S^*}$ that $L_{\mathrm{MVR}}(S^*) = \frac{2p-1}{1-\rho} + \frac{1}{2 p \rho + 1 - \rho}$.
As a result, 
$$L_{\mathrm{MVR}}(S^*) - O_{\mathrm{MVR}}(S^*) = (1-f(p)) \frac{2p-1}{1-\rho} + \frac{1}{2 \rho p + 1 - \rho}, $$
where the last term clearly vanishes as $p \to \infty$. Therefore, to show $L_{\mathrm{MVR}}(S^*) - O_{\mathrm{MVR}}(S^*)$ vanishes, it suffices to show that $(1-f(p)) \cdot p$ vanishes. In particular, we can apply the definition of $f(p)$ and do some algebra to find that
\begin{align*}
    (1-f(p)) p 
&= 
    p - p\left(\sqrt{2p(2p-1)} - \sqrt{2p(2p-3) + 2} \right)^2 \\
&= 
    p - p \left((2p)(2p-1) - 2 \sqrt{(2p)(2p-1)(2p(2p-3) + 2)} + 2p(2p-3) + 2 \right) \\
&= 
    p - p \left(8p^2 - 8p + 2 - 2 \sqrt{4 p (p-1) (2p-1)^2} \right) \\
&= 
    p - p \left(8p^2 - 8p + 2 - 4 (2p-1) \sqrt{p(p-1)}\right) \\
&= 
    p \left(1 - 8p^2 + 8p - 2 + 8p \sqrt{p(p-1)} - 4 \sqrt{p(p-1)}  \right) \\
&= 
    -p \left(8p^2 - 8p + 1 - 8p \sqrt{p(p-1)} + 4\sqrt{p(p-1)} \right).
\end{align*}
At this point, we multiply and divide by $8p^2 - 8p + 1 + 8p \sqrt{p(p-1)} - 4 \sqrt{p(p-1)}$ to rationalize the expression:
\begin{align*}
&= 
    \frac{-p\left((8p^2 - 8p + 1)^2 - (8p \sqrt{p(p-1)} - 4 \sqrt{p(p-1)})^2 \right)}{8p^2 - 8p + 1 + 8p \sqrt{p(p-1)} - 4 \sqrt{p(p-1)}} \\
&= \frac{-p}{8p^2 - 8p + 1 + 8p \sqrt{p(p-1)} - 4 \sqrt{p(p-1)}}.
\end{align*}
where in the second step, we use the fact that $(8p^2 - 8p + 1)^2 - (8p \sqrt{p(p-1)} - 4 \sqrt{p(p-1)})^2 = 1$. Since the denominator of this fraction has higher degree than the numerator, this proves that $\lim_{p \to \infty} (1-f(p))p = 0$. By the previous analysis, this proves that the dual gap vanishes asymptotically, so $S^*$ is asymptotically optimal for both the MVR and \smaxent losses.
\end{proof}
\end{theorem}

\subsection{Speedups for structured covariance matrices}\label{subsec::blockdiagapprx}

While our coordinate descent algorithms are substantially faster than generic semidefinite program solvers, they are still prohibitively expensive when $p$ is very large. One way out of this is to follow the approach of \cite{fanok2020} and approximate $\Sigma$ using a rank-$k$ factor model, in which case both algorithms will run in $O(n_{\mathrm{iter}} p k^2)$ (see \cite{fanok2020} for details). Alternatively, following the approach of \cite{mxknockoffs2018}, we note below that the MVR and \smaxent optimization problems can be efficiently parallelized when $\Sigma$ is approximated as a block-diagonal matrix.

\begin{lemma}\label{lem::amrc}  Suppose $\Sigma = \mathrm{blockdiag}(\Sigma_1, \dots, \Sigma_{\ell})$. Let $S_j^*$ be the result of the Gaussian MVR optimization problem for $\Sigma_j$. Then $S_{\mathrm{MVR}} = \mathrm{blockdiag}(S_1^*, \dots, S_{\ell}^*)$. The same holds for the \smaxent problem.
\begin{proof} Showing this result requires showing that the constraints and the objective function in both problems are separable across the blocks. To deal with the constraints, observe that $S = \mathrm{blockdiag}(S_1, \dots, S_l) \succcurlyeq 0$ if and only if $S_j \succcurlyeq 0$ for all $1 \le j \le \ell$,and similarly $2 \Sigma - S \succcurlyeq 0$ if and only if $2 \Sigma_j - S_j \succcurlyeq 0$ for all $1 \le j \le \ell$. For the MVR loss, note that the inverse of a block-diagonal matrix is the block-diagonal matrix of its blocks' inverses. Therefore
$$L_{\mathrm{MVR}}(S) \propto \text{Tr}((2\Sigma - S)^{-1}) + \text{Tr}(S^{-1}) = \sum_{j=1}^\ell \text{Tr}(2\Sigma_j - S_j)^{-1} + \text{Tr}(S_j^{-1}). $$
To see this result for \smaxent knockoffs, note that $$\log \det \left(G_S \right) = \sum_{j=1}^\ell \log \det \left(\begin{bmatrix} \Sigma_j & \Sigma_j - S_j \\ \Sigma_j - S_j & \Sigma_j \end{bmatrix} \right).$$
For both problems, the constraints and objective function are separable, which completes the proof.
\end{proof}
\end{lemma}

This motivates the AMVR (resp. AME) construction:

\textbf{Step 1.} Approximate $\Sigma$ as a block-diagonal matrix $\Sigma_{\mathrm{approx}}$ and find $S_{\mathrm{approx}}$ as the solution to the MVR (resp. \smaxent) problem for $\Sigma_{\mathrm{approx}}$.

\textbf{Step 2.} Run a grid search over $[0,1]$ to find $\gamma = \arg \max_{\gamma \in [0,1]} L_{\mathrm{MVR}}(\gamma \cdot S_{\mathrm{approx}}) \text{ s.t. } 2 \Sigma \succcurlyeq \gamma \cdot S_{\mathrm{approx}} $ For AME, replace $L_{\mathrm{MVR}}$ with $L_{\mmaxent}$.

Finally, we return $S^* = \gamma \cdot \hat S$.

\subsection{Runtime Simulations}\label{appendix::runtime}

The state-of-the-art algorithms to compute SDP knockoffs have the same or slower computational complexities than MVR and ME knockoffs \citep{fanok2020}. However, such analysis hides constant factors that may have substantial effects in practice. In this section, we plot the average computation time for MVR, ME, and SDP Gaussian knockoffs, including settings where we use block-diagonal approximations and factor approximations of $\Sigma$ to speed up computation, as in Section \ref{subsec::blockdiagapprx}. To compute SDP knockoffs, we use two algorithms: one which directly solves an SDP \citep{mxknockoffs2018} and one which uses coordinate descent \citep{fanok2020}.

\begin{figure}[h]
    \centering
    \makebox[\textwidth]{\includegraphics[width=\textwidth]{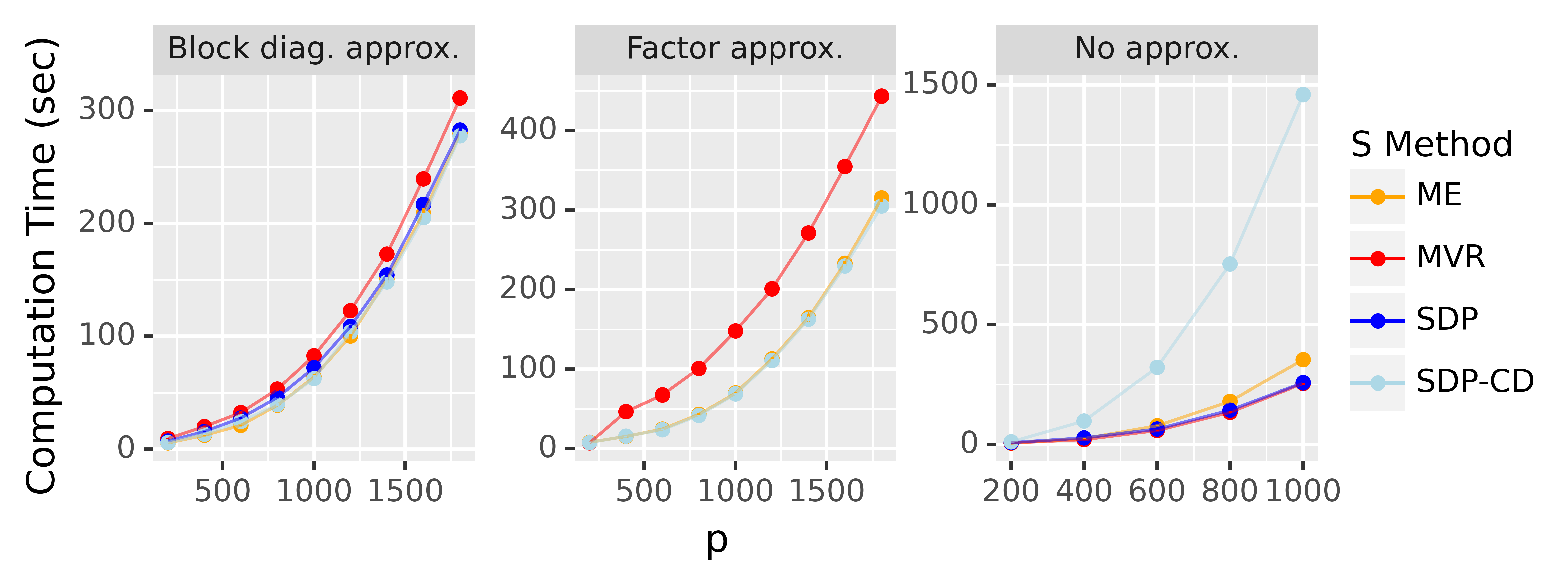}}
    \caption{Runtimes for MVR, ME, and SDP Gaussian knockoffs. We generate $\Sigma$ as in the ``AR1" setting from Section \ref{sec::sim}, with $p$ varied between $200$ and $2000$. In the left and middle panels, we apply factor and block-diagonal approximations of $\Sigma$ with $25$ factors and a maximum block size of $100$, respectively. In the right panel, we generate exact MVR, ME, and SDP knockoffs for $p$ between $200$ and $1000$. The ``SDP" and ``SDP-CD" curves correspond to computing SDP knockoffs by directly solving a SDP or using coordinate descent, respectively. Note that the ``SDP" method cannot take advantage of factor approximations, so it does not appear in the middle panel.}
    \label{fig::runtimes}
\end{figure}

Figure \ref{fig::runtimes} shows that the runtimes for each of these algorithms are very similar for the ``AR1" covariance matrix from Section \ref{sec::sim}, with the exception of the coordinate descent SDP method, which takes longer to converge in the ``no approximation" setting. Other than this, for each approximation strategy, all average runtimes are roughly within a factor of two of each other, and many runtimes are indistinguishable. Of course, these results should be taken with a grain of salt, since the precise runtime of each algorithm depends on implementation-specific details.  Despite this, Figure \ref{fig::runtimes} shows that generating MVR and ME knockoffs should not take substantially longer than generating SDP knockoffs.

\section{\smaxent knockoffs for discrete features}\label{appendix::discretemmi}

In this section, we describe a formulation for exact \smaxent knockoffs for discrete features, as discussed in Sections \ref{sec::mvrknock} and \ref{sec::discussion}. Although this formulation is too computationally expensive to be practical, it may provide a starting point for future work, and furthermore it allows us to prove Lemma \ref{lem::mminondegen}. For simplicity, we will consider the binary case where $[X, \tilde{X}] \in \{0,1\}^{2p}$, but this analysis generalizes naturally as long as each feature has a finite support.

To begin with, let $v_k = \mathbb{P}\left([X, \tilde{X}] = \text{binary}(k) \right)$ where $\text{binary}(k) \in \{0,1\}^{2p}$ is the binary representation of $k \in \{0, \dots, 2^{2p}-1\}$.  To define a valid joint knockoff distribution for $[X, \tilde{X}]$, we need the variables $v_k$ to preserve the marginal distribution of $X$ and to satisfy the pairwise exchangeability condition (\ref{eq::pairwiseexchange}). First, we show how we can ensure that constraint on the marginal distribution of $X$ using linear constraints. For each configuration $\epsilon \in \{0,1\}^{p}$, this corresponds to ensuring
\begin{equation}\label{eq::discretemmimarginal}
    \sum_{k : \text{binary}(k)_{1:p} = \epsilon} v_k = \mathbb{P}(X = \epsilon).
\end{equation}
There are $2^p$ values for $\epsilon$ and therefore $2^p$ such linear constraints. Next, to define notation, we let $\text{binary}(k)_{\swap(j)}$ be equal to $\text{binary}(k)$ except with the digits in locations $j$ and $j+p$ swapped. The pairwise exchangeability condition (\ref{eq::pairwiseexchange}) requires that for any $k, k' \in \{0, \dots, 2^{2p}-1\}$,
\begin{equation}\label{eq::discretemmipairwise}
v_k = v_{k'} \text{ if } \exists \, j \in [p] \, \text{ s.t. }  \text{binary}(k)_{\swap(j)} = \text{binary}(k').
\end{equation}
Constraint (\ref{eq::discretemmipairwise}) can be viewed as the linear constraint $v_k - v_{k'} = 0$. Alternatively, it may be more efficient to consolidate the variables $v_k$ and $v_{k'}$ into a single variable,\footnote{It can be shown that doing this for each pairwise exchangeability constraint leaves $3^p$ optimization variables overall.} but to ease notation, we will not do so.

Let $v = (v_1, \dots, v_{2^{2p}})$ be the vector of optimization variables. Additionally, let $b \in \mathbb{R}^{2^p}$ be the vector of probabilities of $X = \text{binary}(k)$ for $k \in \{0, \dots, 2^{p}-1\}$. Let $m$ be the number of distinct pairwise exchangeability constraints. Then for a suitable $A \in \mathbb{R}^{2^{2p} \times 2^p}$ corresponding to (\ref{eq::discretemmimarginal}) and a matrix $B \in \mathbb{R}^{2^{2p} \times m}$ corresponding to the pairwise exchangeability constraints, we can solve
\begin{align}
\text{ maximize } & - \sum_{k=0}^{2^{2p}} v_k \log(v_k) \label{eq::discretemmi} \\ 
\text{ s.t. }   & A v = b \nonumber \\
                & B v = 0 \nonumber \\ 
                & 0 \le v_k \le 1 \, \, \, \, \, \, \, \forall \, k \in \{1, \dots, 2^{2p}\} \nonumber
\end{align}
to find the distribution for $[X, \tilde{X}]$ which has the maximal total entropy while maintaining the marginal distribution of $X$ and the pairwise exchangeability condition. Note that in (\ref{eq::discretemmi}) and throughout, we use the convention that $0 \cdot \log(0) = \lim_{x \to 0^+} x \log(x) = 0$.

This optimization problem is convex and has been well-studied, especially since $A$ is sparse \citep{persson1986, boyd2004}. Unfortunately, the number of variables and constraints grows exponentially in $p$, making this formulation intractable. One way to resolve this may be to use special structure, such as conditional independence properties of the distribution of $X$, to simplify the constraints. Furthermore, one might settle for minimizing mutual information over a slightly restricted class of knockoff distributions in order to reduce the number of optimization variables. A last approach might be to carefully analyze concentration of measure results, which indicate that for some constrained maximum entropy problems, almost all feasible solutions approximately maximize the entropy objective \citep{grunwald2011}. We leave more concrete analysis to future work.

This formulation allows us to prove Lemma \ref{lem::mminondegen}. We begin with a technical lemma regarding the finite case. In general, we use the convention that $0 \cdot \log(0) = \lim_{x \to 0} x \log(x) = 0$.

\begin{lemma}\label{lem::mmimaxsupport} Let $X$ have finite support $\mathcal{X}$. Suppose there exists a valid distribution $[X, \tilde{X}]$ with support $A \subset \mathcal{X} \times \mathcal{X}$. Then when $\tilde{X}$ are \smaxent knockoffs, the support of $[X, \tilde{X}]$ contains $A$.
\begin{proof}
As discussed above, we can represent the \smaxent problem as follows. Let $p(x, \tilde{x})$ represent the probability mass function of $[X, \tilde{X}]$. Then we want to solve
\begin{align}
\text{ maximize } & H(p) \equiv -\sum_{x, \tilde{x} \in \mathcal{X} \times \mathcal{X}} p(x, \tilde{x}) \log \left(p(x, \tilde{x}) \right) \nonumber \\ 
\text{ s.t. } & \, \, \sum_{\tilde{x} \in \mathcal{X}} p(x, \tilde{x}) = p(x)  \, \, \, \, \, \forall \, x \in \mathcal{X} \label{eq::preservemarg} \\
              & p([x, \tilde{x}]_{\swap(j)}) = p([x, \tilde{x}]) \, \, \, \, \, \, \, \,  \forall \, j \in [p], \forall \, [x, \tilde{x}] \in \mathcal{X} \times \mathcal{X}. \label{eq::preserveswap}
\end{align}
Suppose a feasible solution $p_0$ has support $A$. Suppose for the sake of contradiction that the optimal solution $p^*$ has support $A^*$ where $A \not \subset A^*$. Note that for all $\alpha \in [0,1]$, $\alpha p^* + (1 - \alpha) p_0$ is a feasible solution to the problem as well, as it obeys (\ref{eq::preservemarg}) and (\ref{eq::preserveswap}). We use this property twice. First, define $p' = 0.5 p_0 + 0.5 p^*$, which is a feasible solution that has support $A \cup A^*$. We will use $p'$ to construct a contradiction. Second, since all convex combinations $\alpha p^* + (1-\alpha) p'$ are feasible as well, we note
\begin{align}
\frac{\partial}{\partial \alpha} H(\alpha p^* + (1 - \alpha) p') =&  - \sum_{(x, \tilde{x}) \in A \cup A^*} (p^*(x, \tilde{x})- p'(x, \tilde{x})) \log\left(\alpha p^*(x, \tilde{x}) + (1 - \alpha) p'(x, \tilde{x}) \right)\, \, \label{eq::Hderiv} \\ 
=&  - \sum_{(x, \tilde{x}) \in A \setminus A^*} -p'(x, \tilde{x}) \log((1-\alpha)p'(x, \tilde{x})) \label{eq::mminondegentopsum} \\ 
& - \sum_{(x, \tilde{x}) \in A^*} (p^*(x, \tilde{x})- p'(x, \tilde{x})) \log\left(\alpha p^*(x, \tilde{x}) + (1 - \alpha) p'(x, \tilde{x}) \right) \label{eq::mminondegenbottomsum},
\end{align}
where (\ref{eq::mminondegentopsum}) follows because we assume $p^*(x, \tilde{x}) = 0$ for all $(x, \tilde{x}) \not \in A$.

We will show that $\lim_{\alpha \to 1} \frac{\partial}{\partial \alpha} H(\alpha p^* + (1 - \alpha) p') = - \infty$. To show this, first observe that the top term (\ref{eq::mminondegentopsum}) approaches $- \infty$ as $\alpha \to 1$. This follows because for $(x, \tilde{x}) \in A^* \setminus A$, $p'(x, \tilde{x}) > 0$ by assumption, and for each $(x, \tilde{x}) \in A \setminus A^*$, $\lim_{\alpha \to 1}\log((1-\alpha) p'(x, \tilde{x})) = - \infty$.  Second, note that the limit of the bottom term (\ref{eq::mminondegenbottomsum}) as $\alpha \to 1$ is $- \sum_{(x, \tilde{x}) \in A^*} (p^*(x,\tilde{x}) - p'(x, \tilde{x})) \log(p^*(x. \tilde{x}))$, which is finite because it is a finite sum of finite elements. Thus, $\lim_{\alpha \to 1} \frac{\partial}{\partial \alpha} H(\alpha p^* + (1 - \alpha) p') = - \infty$.

This means that for all $\alpha$ sufficiently close to $1$, $\frac{\partial}{\partial\alpha} H(\alpha p^* + (1-\alpha) p')$ is negative. Since $H$ is continuous, this implies that for some $\alpha^* \in [0, 1)$, $H(\alpha^* p^* + (1 - \alpha^*) p') > H(p^*)$. This is a contradiction, since we assumed $p^*$ is the optimal solution. Therefore $A \subset A^*$.
\end{proof}
\end{lemma}

\begingroup
\def\thetheorem{\ref{lem::mminondegen}} 
\begin{lemma} Let $X$ have finite support $\mathcal{X}$. For any $j \in [p]$ and any $x, \tilde{x} \in \mathcal{X} \times \mathcal{X}$, if there exists a valid distribution $[X, \tilde{X}]$ such that $\Var(X_j | X_{\text{-}j}=x_{\text{-}j}, \tilde{X}=\tilde{x}) > 0$, then \smaxent knockoffs $\tilde{X}$ satisfy this property as well.
\begin{proof} Without loss of generality let $j = 1$. Suppose $p_0$ is a valid joint feature-knockoff distribution such that under $p_0$,  $\Var(X_1|X_{-1}=x_{-1}, \tilde{X}=\tilde{x}) > 0$. This implies that there exist at least two values $x_1^{(1)}, x_1^{(2)}$ such that $p_0((x_1^{(1)}, x_{-1}, \tilde{x})), p_0((x_1^{(2)}, x_{-1}, \tilde{x})) > 0$. By Lemma \ref{lem::mmimaxsupport}, if $p^*$ is the joint PMF of the features and \smaxent knockoffs, then $p^*(x_1^{(1)}, x_{-1}, \tilde{x}), p^*(x_1^{(2)}, x_{-1}, \tilde{x}) > 0$ as well. This implies that for the \smaxent knockoffs $\tilde{X}$, $\Var(X_j | X_{\text{-}j}=x_{\text{-}j}, \tilde{X}=\tilde{x}) > 0$.
\end{proof}
\end{lemma}
\addtocounter{theorem}{-1}
\endgroup

\section{Further simulation results}\label{appendix::sims}

\subsection{Further simulations for equicorrelated designs}\label{appendix::equicorrelated}

\begin{figure}[h!]
    \centering
    \includegraphics{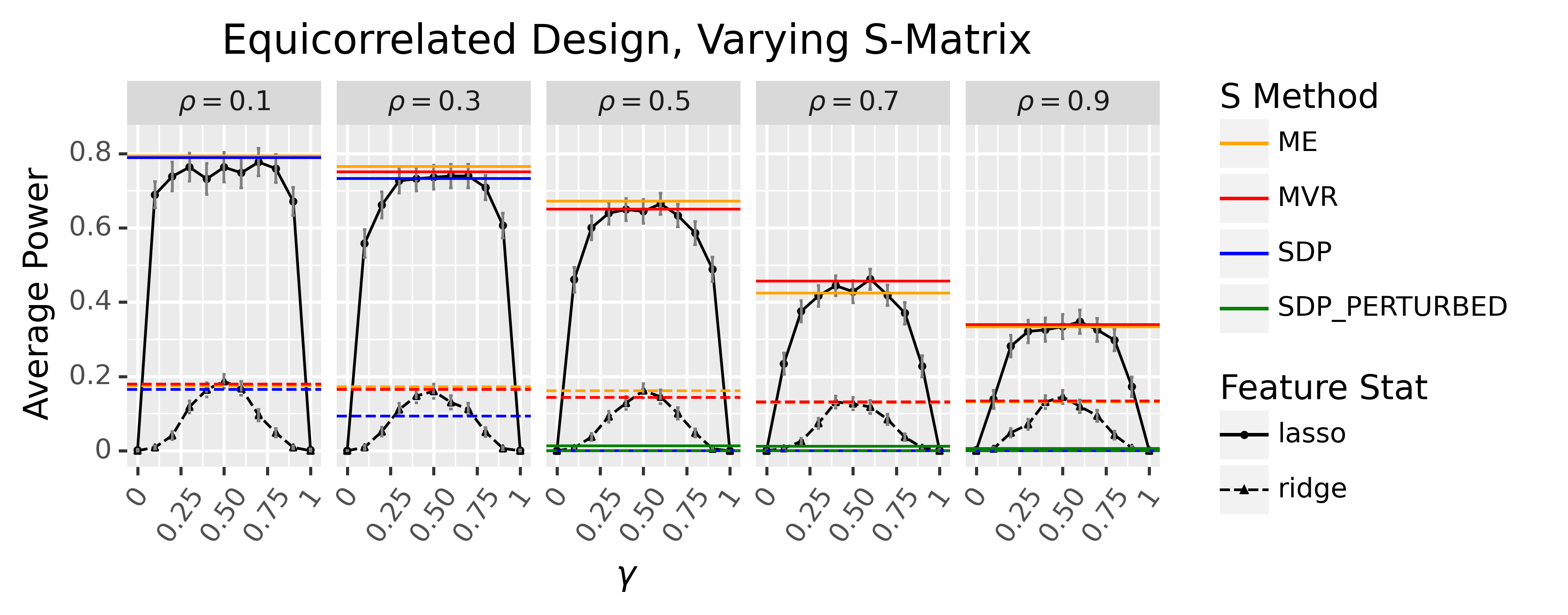}
    \caption{Empirical powers when $X$ is an exchangeable Gaussian design and $Y\mid X \sim \mathcal{N}(X \beta, 1)$. The number of non-nulls is $50$ with coefficients sampled independently from $\mathrm{Unif}\left([-\delta,-\delta/2] \cup [\delta/2, \delta]\right)$. We let $\delta = 2$ for $\rho = 0.9$ and $\delta = 1$ otherwise. We use lasso and ridge coefficient differences and compare the performance of different $S$-matrices, as defined in equation (\ref{eq:gammadef}). Here, $n = 250$ and $p = 500$. All horizontal lines have standard errors less than $1.8\%$.}
    \label{fig:equiplot2}
\end{figure}

In this section, we present Figure \ref{fig:equiplot2}, which demonstrates that $S_{\mathrm{MVR}}$ and $S_{\mmaxent}$ are approximately optimal over all $S$-matrices of the form $\gamma \cdot 2 \lambda_{\mathrm{min}}(\Sigma) I_p$ when $X$ is equicorrelated. Note that the ``maximum" $S$-matrix corresponding to $\gamma = 1$ is not always equal to $S_{\mathrm{SDP}}$, because the SDP will never set any value of $\mathrm{diag}(S)$ to greater than $1$. Indeed, Lemma \ref{lem::equicorrsoln} tells us $S_{\mathrm{SDP}}$ corresponds to $\gamma = \frac{1}{2 - 2\rho}$ when $\rho < 0.5$ and $\gamma = 1$ when $\rho \ge 0.5$.

\subsection{Comparison to conditional independence knockoffs}\label{appendix::cicomp}

In this section, we compare the MRC framework to the CI knockoffs introduced by \cite{ciknockoff2019} and further analyzed by \cite{ke2020}. We make two observations about the CI knockoff framework. 

\begin{figure}[h!]
    \centering
    \includegraphics{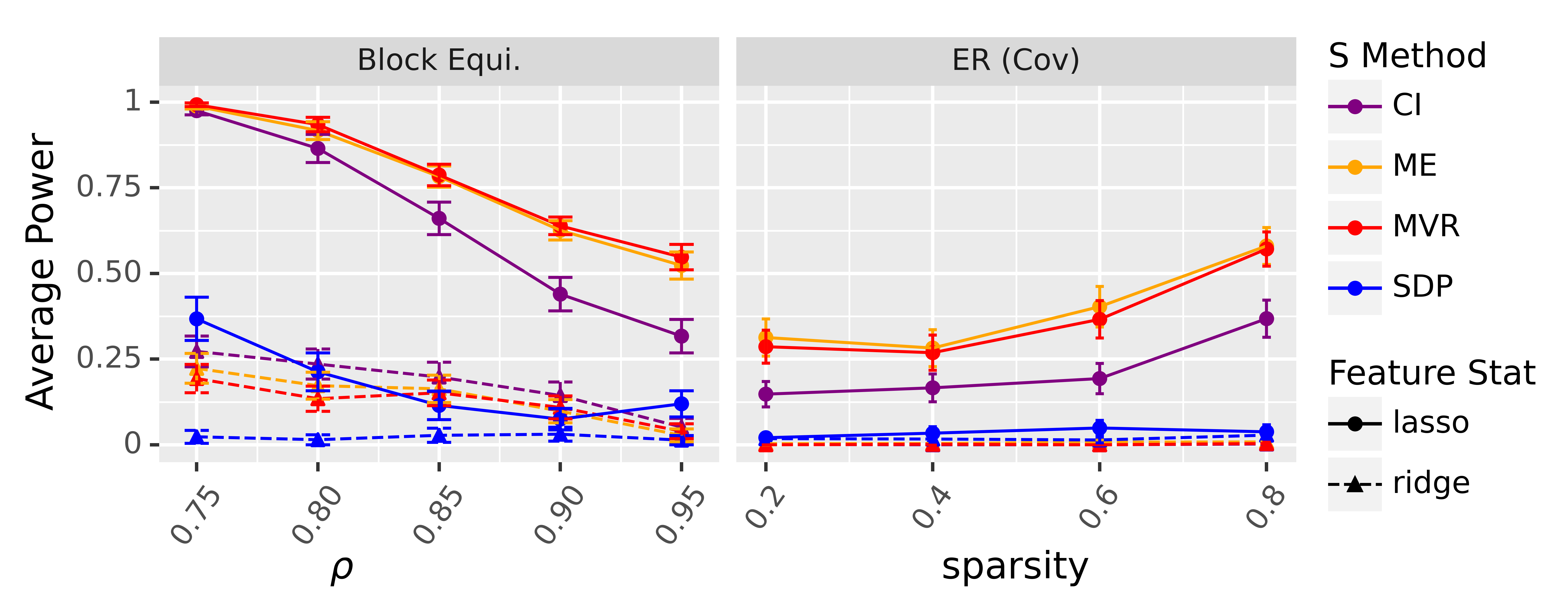}
    \caption{We let $X \sim \mathcal{N}(0, \Sigma)$ and $Y \mid X \sim \mathcal{N}(X \beta, 1)$. In both cases, $p = 500$, the number of non-nulls is $50$. On the left panel, the non-nulls are sampled as independent symmetric random signs, and the covariance matrix is block-equicorrelated with block-size $\ell=2$ and correlation $\rho$. We set $n=333$ and we control the FDR at level $q = 0.05$. On the right panel, the non-nulls are sampled independently from $\mathrm{Unif}\left([1, -1/2] \cup [1/2, 1]\right)$ we use an ErdosRenyi covariance matrix (see Appendix \ref{appendix::simgaussiandescription}), and we vary the sparsity of the covariance matrix between $20\%$ and $80\%$, with $n=375$ and $q=0.1$.} 
    \label{fig::cicomp}
\end{figure}

First, even when CI knockoffs are well-defined, the conditional independence condition $X_j \Perp \tilde{X}_j \mid X_{\text{-}j}$ does not tell us whether $X_j$ is reconstructable from the joint information in $X_{\text{-}j}, \tilde{X}$. To illustrate this, suppose $X \sim \mathcal{N}(0, \Sigma)$ where $\Sigma$ is block-equicorrelated with a block-size of $\ell=2$ and correlation $\rho \ge 0.5$, which is one of the correlation structures analyzed by \cite{ke2020}. In this case, as $\rho$ approaches $1$, the conditional independence approach yields almost exactly the same answer as the MAC-minimizing approach. In particular, $S_{\mathrm{SDP}} = (2 - 2\rho) I_p$, $S_{\mathrm{CI}} = (1-\rho^2) I_p$, and $ S_{\mathrm{MVR}} \approx S_{\mmaxent} \approx (1-\rho) I_p$, so if we set $\rho = 0.9$ as an example, then $S_{\mathrm{SDP}} = 0.2 \cdot I_p$, $S_{\mathrm{CI}} = 0.19 \cdot I_p$, and $S_{\mathrm{MVR}} \approx S_{\mmaxent} \approx 0.1 \cdot I_p$. As a result, CI knockoffs become less powerful than MRC knockoffs as $\rho$ increases, as demonstrated by the left panel of Figure \ref{fig::cicomp}.

\begin{figure}[h!]
    \centering
    \includegraphics{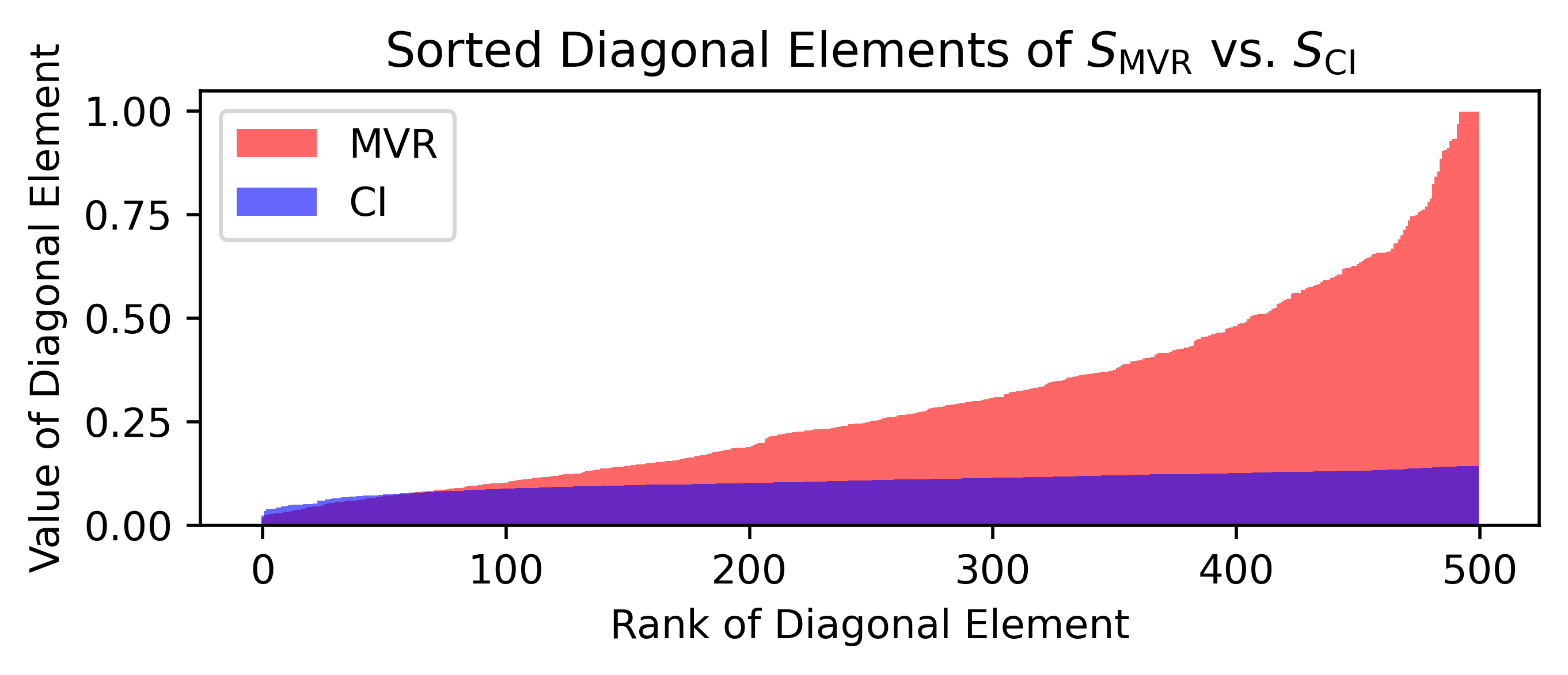}
    \caption{We let $p = 500$ and sample one ErdosRenyi covariance matrix (see Appendix  \ref{appendix::simgaussiandescription}) which is $99\%$ sparse and compute $S_{\mathrm{MVR}}$ and $S_{\mathrm{CI}}$ using the binary search method from \cite{ke2020}. We plot the sorted diagonal elements of both methods to illustrate that $S_{\mathrm{CI}}$ has very small diagonal elements compared to $S_{\mathrm{MVR}}$, explaining the power loss in Figure \ref{fig::cicomp}.} 
    \label{fig::splot_cicomp}
\end{figure}

Second, as noted by \cite{ciknockoff2019}, CI knockoffs are only well-defined for a fairly restrictive class of Gaussian designs. Recently, \cite{ke2020} proposed the following extension to the case where $X \sim \mathcal{N}(0, \Sigma)$ for general $\Sigma$. They suggest computing $S_{\mathrm{CI}} = (\mathrm{diag}(\Sigma^{-1}))^{-1}$ naively and then performing a binary search to find the maximum $\gamma \in [0,1]$ such that $\gamma \cdot S_{\mathrm{CI}}$ satisfies $0 \preccurlyeq \gamma \cdot S_{\mathrm{CI}} \preccurlyeq 2 \Sigma$. Unfortunately, the $\gamma$ produced by this binary search may be quite small, limiting the power of these generalized CI knockoffs. For example, consider an ErdosRenyi covariance matrix where $99\%$ of the entries of $\Sigma$ equal zero. In this example, as illustrated by Figure \ref{fig::splot_cicomp}, the binary search sets most of the diagonal elements of $\gamma \cdot S_{\mathrm{CI}}$ to approximately equal $0.1$, whereas $S_{\mathrm{MVR}}$ chooses a variety of values ranging between $0$ and $1$. As shown in the right panel of Figure \ref{fig::cicomp}, this substantially increases the power of the MRC methods relative to CI knockoffs.

\subsection{Examples from the literature}\label{appendix:replications}

In this section, we discuss two examples in the literature where SDP knockoffs fail to have any power. First, \cite{xing2019} ran simulations in the setting where $X$ is Gaussian and equicorrelated, and $Y \mid X \sim \mathcal{N}(X\beta, 1)$. They let $n = 1000$, $p = 300$, and vary  $\rho$ between $0$ and $0.8$. The linear coefficients $\beta$ have $60$ non-nulls sampled independently from $\mathcal{N}(0, 20 / \sqrt{n})$. They found that the performance of model-X knockoffs fell to $0$ when $\rho \ge 0.5$, including when they estimated the covariance matrix using the data. They implemented a ``model-X-fix" method, which involves projecting and slightly perturbing the design matrix. However, the power of the model-X-fix method still drops off substantially as $\rho$ grows larger.

\begin{figure}[h!]
    \centering
    \makebox[\textwidth]{\includegraphics[width=\textwidth]{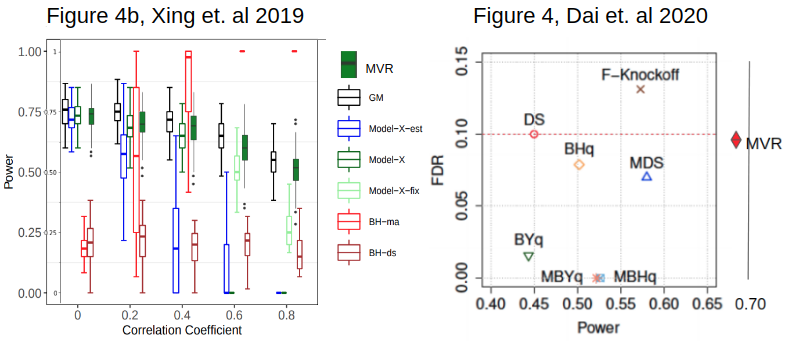}}
    \caption{Left: The replication of Figure $4b$ from \cite{xing2019} with the addition of MVR knockoffs in solid dark green. Right: The replication of Figure $4$ from \cite{dai2020} with the addition of MVR knockoffs as a solid red diamond. The standard errors for the MVR dot are less than $0.005$ for power and $0.0025$ for FDR.
    } 
    \label{fig::replications}
\end{figure}

We rerun these experiments using MVR knockoffs instead of SDP knockoffs. The MVR knockoffs outperform both the SDP knockoffs as well as the ``model-X-fix" method by large margins, as much as $60$ and $25$ percentage points, respectively. Although the Gaussian mirror method slightly outperforms knockoffs in this setting, the performance gap is fairly small (on the order of $2-5\%$, depending on the value of $\rho$). These results are depicted in the left panel Figure \ref{fig::replications}.

Second, \cite{dai2020} ran simulations with a Gaussian equicorrelated design with a constant pairwise correlation of $\rho \in [0,0.8]$. They set the number of data points $n = 500$ and vary the dimensionality $p \in [500, 2000]$. For the response, they let $Y\mid X \sim \mathcal{N}(X \beta, 1)$ where $\beta$ has $50$ non-nulls with a signal size of $\mathcal{N}(0, 10\sqrt{\log(p)/n})$. They use lasso coefficient differences as feature statistics. In the high-dimensional case where $p = 2000$, the SDP MX-knockoffs have almost zero power. In the low-dimensional case where $p = 500$, MX-knockoffs have very low but nonzero power.\footnote{This result may seem to contradict Theorem \ref{thm::avgequi}, which states that SDP knockoffs should have zero power asymptotically since the feature-knockoff covariance matrix should have rank $p+1$. However, in practice, many statistical packages attempt to prevent \textit{exact} low rank structure, which likely accounts for the low but nonzero power of SDP knockoffs.} In both cases, MX-knockoffs have the lowest power compared to every competitor. 

\begin{figure}[h]
    \centering
    \makebox[\textwidth]{\includegraphics[width=\textwidth]{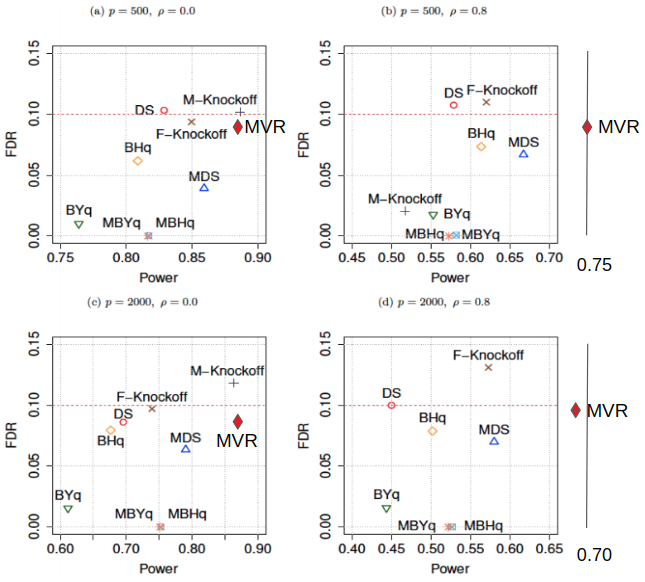}}
    \caption{The replication of Figure $4$ from \cite{dai2020} with the addition of MVR knockoffs as solid red diamonds.
    All standard errors are less than $0.005$ for power and $0.0025$ for FDR.} 
    \label{fig:dai2020}
\end{figure}

We rerun these experiments in exactly the same setting, except we generate MVR knockoffs instead of SDP knockoffs. As a sanity check, we note that the power of MVR knockoffs agrees with the original paper when $\rho = 0$, since in this simple case, $S_{\mathrm{MVR}} = S_{\mathrm{SDP}} = I_p$. On the other hand, when $\rho = 0.8$, we see that MVR knockoffs outperform every competitor by a wide margin---they are so powerful they do not even fit on the initial charts from the paper. These results are presented in the right panel of Figure \ref{fig::replications} for the high-dimensional case where $\rho = 0.8$. We present the full results for various $\rho$ and $p$ in Figure \ref{fig:dai2020}.

\subsection{Simulation details for sections \ref{subsec::simgaussian}, \ref{subsec::simrobustness}, \ref{subsec::simfx}}\label{appendix::simgaussiandescription}

In this section, we provide additional details on the covariance matrices $\Sigma$ in Sections \ref{subsec::simgaussian} through \ref{subsec::simfx} as well as the sparse nonlinear models in Figure \ref{fig::nongaussianlinearpower}. The covariance matrices are defined below. In all cases, we rescale each covariance matrix $\Sigma$ such that $\Sigma_{jj} = 1$ for all $j \in [p]$ after construction.

\begin{itemize}
    \item \underline{Equi.} or \underline{Equicorrelated} covariance matrices refer to $\Sigma$ where $\Sigma_{ij} = \rho$ for $i \ne j$ and $1$ otherwise. Throughout, unless specified otherwise, we will set $\rho = 0.5$.
    \item \underline{Block equi.} or \underline{block equicorrelated} covariance matrices refer to $\Sigma$ where $\Sigma$ is block-diagonal with $p/\ell$ $\ell \times \ell$ equicorrelated blocks with correlation $\rho$ inside the blocks. We set $\ell = 5$ and $\rho = 0.5$ unless specified otherwise. In this setting, signals are clustered according to the blocks, meaning that all features in a block are either null or non-null. This simulation setting, including the clustering of the signals, follows \cite{daibarber2016}.
    \item \underline{ER} (ErdosRenyi) matrices are constructed as follows. We generate a random upper-triangular matrix $V$ where for $i > j$,
    $$V_{ij} \iid \pm \delta_{ij} \cdot U_{ij}   $$
    where $\delta_{ij} \iid \text{Bern}(0.2)$ and $U_{ij} \iid \mathrm{Unif}(0.1, 1)$. In the \underline{ER (Cov)} setting, we set
    $$\Sigma = (V + V^{\top}) + (0.1 + \lambda_{\mathrm{min}}(V^{\top} + V)) \cdot I_p $$
    In the \underline{ER (Prec)} setting, we set the precision matrix $\Sigma^{-1}$ equal to the same quantity. In both cases, we rescale $\Sigma$ to be a correlation matrix. This simulation set-up roughly follows \cite{nodewiseknock}.
    \item Finally, we generate \underline{AR1} covariance matrices where we can represent $X_1 \sim \mathcal{N}(0,1)$, and for $j \in \{2, \dots, p\}$,
    $$X_j = \rho_j X_{j-1} + \sqrt{1 - \rho_j^2} Z_j $$
    for $Z_j \iid \mathcal{N}(0,1)$. Throughout, we sample $\rho_j \iid \text{Beta}(3,1)$ to simulate a challenging setting where features are highly correlated. After sampling the correlations, if $\lambda_{\mathrm{min}}(\Sigma) < 0.001$, we add $(0.001 - \lambda_{\mathrm{min}}(\Sigma)) I_p$ to increase the eigenvalues above a numerical tolerance and then rescale to a correlation matrix. Since non-nulls are often clustered together in genetic studies, in the \underline{AR1 (Corr)} setting, we partially capture this idea by letting the non-nulls lie along a single continuous block of features.
\end{itemize}

Note in Section \ref{subsec::simrobustness}, we use the package sklearn for implementations of the graphical lasso and Ledoit--Wolf covariance estimation methods \citep{sklearn2011}.

For the nonlinear simulations in Figure \ref{fig:gaussianrfplot}, we run simulations involving the five following conditional means:
\begin{itemize}
    \item \underline{cos}: $\mu(X) = \text{cos}(X) \beta$, where $\cos$ denotes the elementwise cosine operation.
    \item \underline{cubic}: $\mu(X) = (X^3)\beta - X \beta$ where $X^3$ denotes the elementwise cubing operation.
    \item \underline{pairint}: $\mu(X) = \sum_{i,j \in [p]} \beta_{i,j} X_i X_j$. 
    \item \underline{quadratic}: $\mu(X) = (X^2) \beta$.
    \item \underline{trunclinear}: $\mu(X) = \sum_{j=1}^p \sign(\beta_i) \mathbb{I}\left(X_j \beta_j > 0 \right)$.
\end{itemize}
In all cases, we sample nonzero coefficients from $\pm5$ with equal probability. In all settings except the pairint setting, we choose $30$ non-nulls uniformly at random. In the pairint setting, we select $30$ non-null features uniformly at random and sequentially group them into disjoint pairs, from left to right. For example, if features $1,13,52,61$ were the first four non-nulls, then features $1$ and $13$ would have an interaction and features $52$ and $61$ would have an interaction. This means each feature may participate in at most one pairwise interaction.

\subsection{Further experiments in the AR1 setting}\label{appendix::ar1corr}

\begin{figure}[h!]
    \centering
    \makebox[\textwidth]{\includegraphics[width=\textwidth]{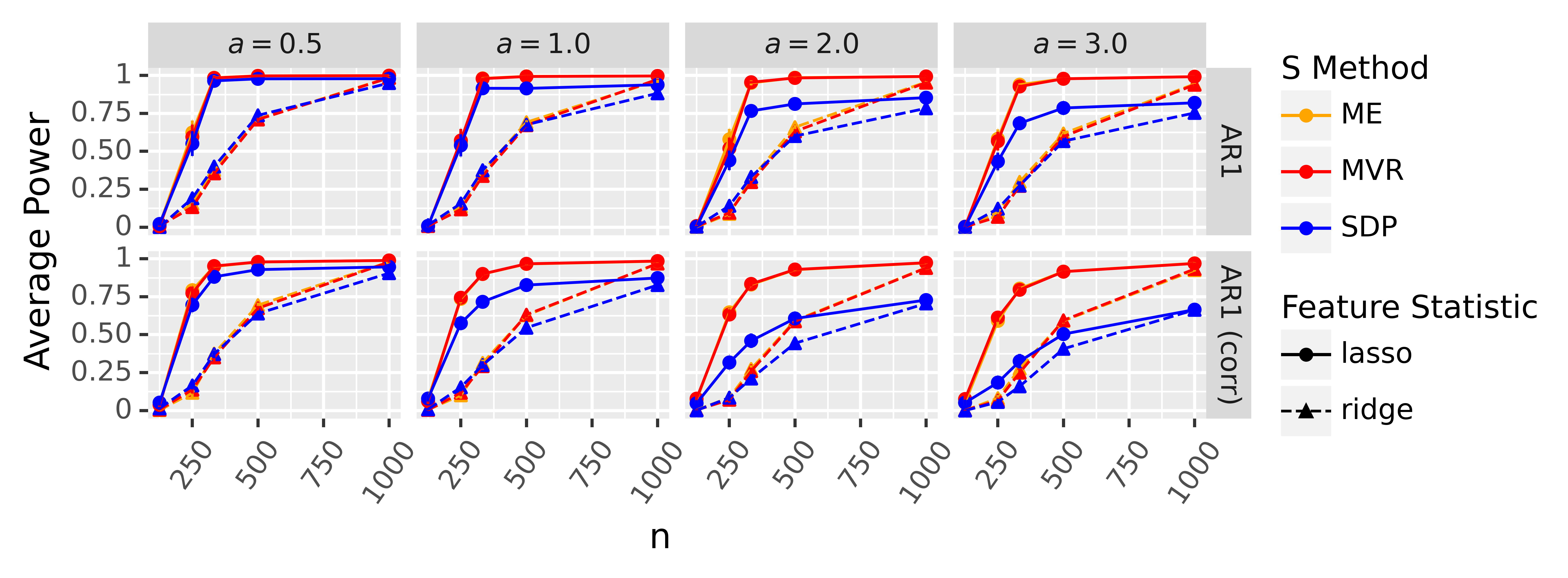}}
    \caption{Power for Gaussian AR1 designs with varied correlations. We sample $X \sim \mathcal{N}(0, \Sigma)$ for the AR1 designs as defined in Appendix \ref{appendix::simgaussiandescription} where $\text{Cor}(X_j, X_{j+1})$ is sampled independently from $\text{Beta}(a, 1)$ for $a \in \{0.5, 1, 2, 3\}$. We let $Y \sim \mathcal{N}(X \beta, 1)$ with $p = 500$ and $50$ non-nulls, where the non-null coefficients are sampled independently from $ \mathrm{Unif}([-2, -1] \cup [1, 2])$.  We use cross-validated lasso and ridge coefficient differences as feature statistics.}
    \label{fig::ar1corrplota}
\end{figure}

\begin{figure}[h!]
    \centering
    \makebox[\textwidth]{\includegraphics[width=\textwidth]{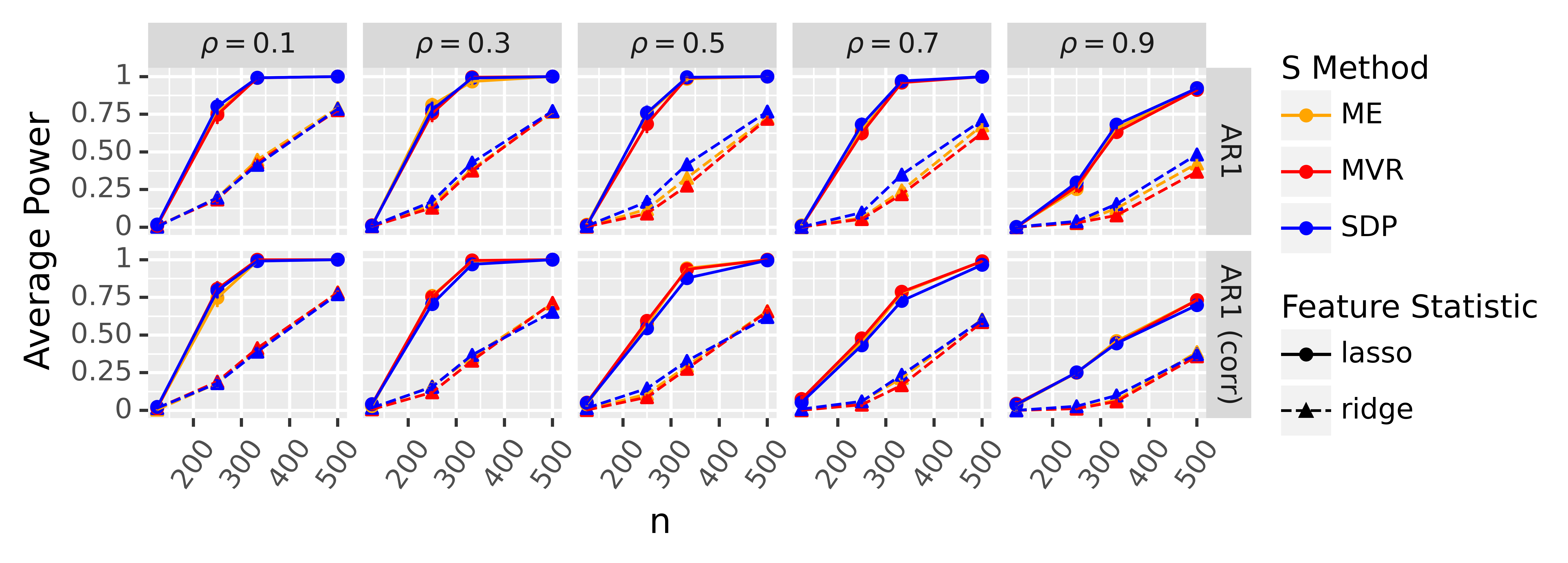}}
    \caption{The same setting as Figure \ref{fig::ar1corrplota}, except we set $\text{Cor}(X_j, X_{j+1}) = \rho$ for a constant $\rho$, and we sample the non-null coefficients independently from $\mathrm{Unif}\left([-1, -0.5], [0.5, 1]\right)$. }
    \label{fig::ar1corrplotrho}
\end{figure}

In this section, we vary the correlation in the AR1 setting detailed in Appendix \ref{appendix::simgaussiandescription}. Previously, we sampled $\text{Cor}(X_j, X_{j+1}) \iid \text{Beta}(a,b)$ for $a = 3$ and $b = 1$. In Figure \ref{fig::ar1corrplota}, we present the results for when we vary $a \in \{0.5, 1, 2, 3\}$. As expected, we see that the MRC methods outperform the SDP by higher margins when the correlation is higher. We also present results when the correlation $\text{Cor}(X_j, X_{j+1}) = \rho$ for a constant $\rho \in (0,1)$. Interestingly, in this case, MRC knockoffs do not outperform their SDP counterparts---as shown in Figure \ref{fig::ar1corrplotrho}, MRC knockoffs seem to very slightly outperform SDP knockoffs when the non-nulls are clustered together, but they seem to slightly under-perform SDP knockoffs when the non-nulls are not clustered together. We conjecture that the key reconstructability condition (\ref{eq::genreconstructioncond}) will hold approximately for some $J \subset [p]$ with high probability when we randomly sample $\Sigma$, but when $\Sigma$ has special structure such as constant between-feature correlations, the condition (\ref{eq::genreconstructioncond}) may not hold.

\begin{figure}[h!]
    \centering
    \makebox[\textwidth]{\includegraphics[width=\textwidth]{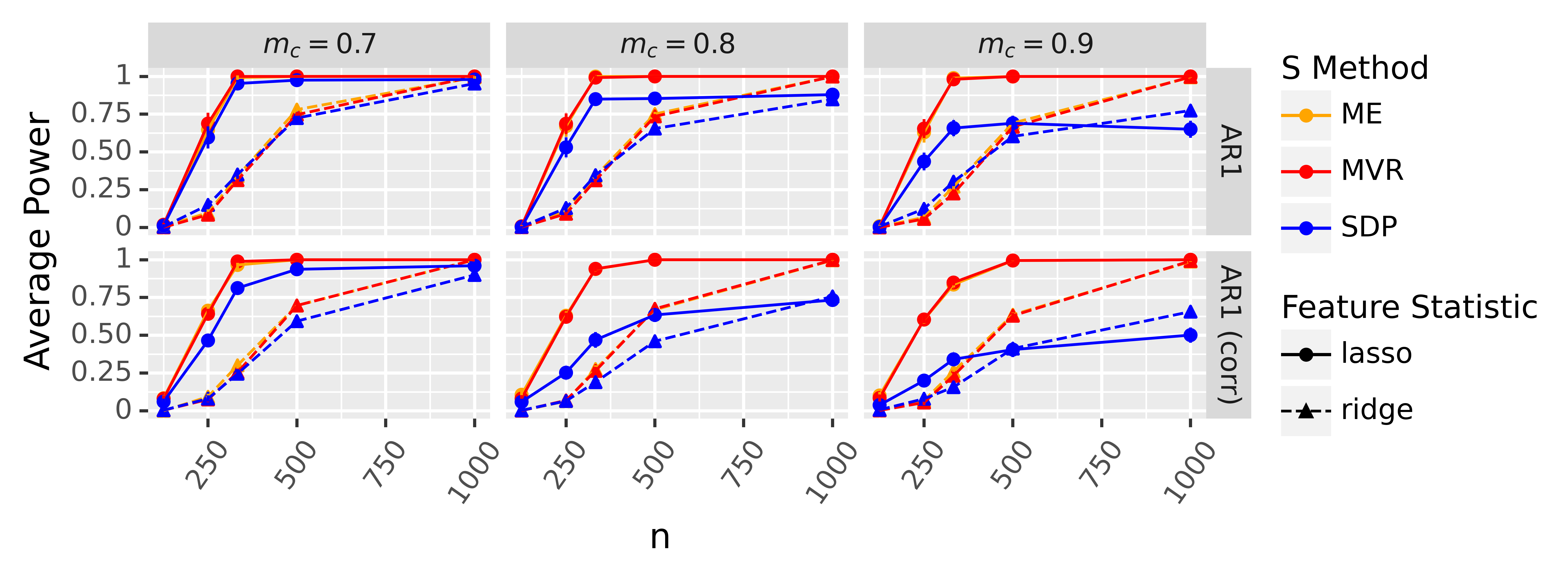}}
    \caption{The same setting as Figure \ref{fig::ar1corrplota}, except we sample $\text{Cor}(X_j, X_{j+1}) \iid \min(m_c, \mathrm{Beta}(3,1))$ for a ``maximum correlation" value $m_c$.}
    \label{fig::ar1maxcorr}
\end{figure}

Of course, when sampling $\text{Cor}(X_j, X_{j+1}) \iid \mathrm{Beta}(3,1)$, many correlations will be extremely close to $1$. One might wonder what the effect of these extremely large correlations is. To analyze this, in Figure \ref{fig::ar1maxcorr}, we sample pairwise correlations $\text{Cor}(X_j, X_{j+1}) \iid \min(m_c, \mathrm{Beta}(3,1))$, for $m_c \in \{0.7, 0.8, 0.9\}$. The results show that even when $m_c = 0.7$, lasso-based MRC knockoffs can outperform their SDP counterparts by as much as $20$ percentage points. Thus, the advantage of MRC knockoffs persists even without extremely high correlations.

\subsection{Further experiments for nonlinear responses}\label{appendix::nonlinearstats}

In this section, we present further results where $Y \mid X$ is nonlinear. In particular, we simulate the power of DeepPINK and LCD feature statistics.

\begin{figure}[h!]
    \centering
    \makebox[\textwidth]{\includegraphics[width=\textwidth]{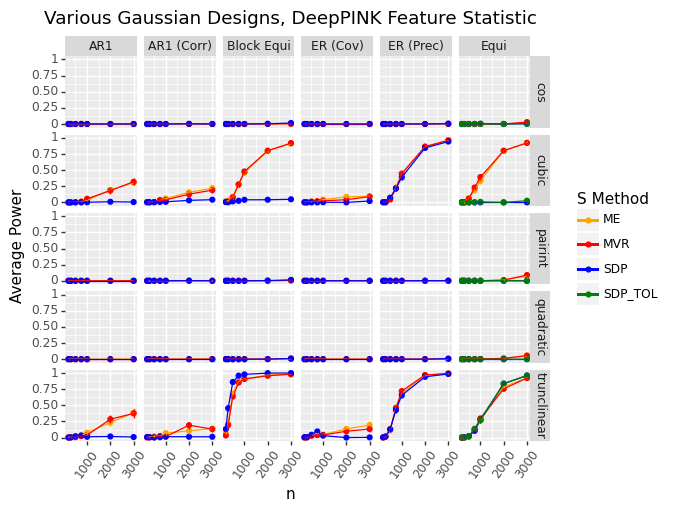}}
    \caption{We let $X \sim \mathcal{N}(0, \Sigma)$ for various $\Sigma$ and $Y\mid X \sim \mathcal{N}(\mu(X), 1)$. The precise definitions of the covariance matrices and conditional responses are presented  in Appendix \ref{appendix::simgaussiandescription}. We let $p = 200$ with $30$ non-nulls. We use DeepPINK feature statistics (see \cite{deeppink2018}).}
    \label{fig::gaussiandeeppinkplot}
\end{figure}

For the DeepPINK statistics, in Figure \ref{fig::gaussiandeeppinkplot}, we again see that MRC knockoffs tend to outperform their SDP counterparts, although the DeepPINK statistic has low power for all methods for some conditional means. We also observe that occasionally, adding \textit{more data} to the DeepPINK feature statistic seems to worsen performance. Although this is strange, this phenomenon has been observed fairly consistently for some deep models in the machine learning literature \citep{doubledescent2019}. Note that our DeepPINK implementation is very similar to but not identical to that of \cite{deeppink2018}---for example, we use a different set of hyper-parameters (such as batchsize during training) than the original paper. Please see our code for more details.

\begin{figure}[h!]
    \centering
    \makebox[\textwidth]{\includegraphics[width=\textwidth]{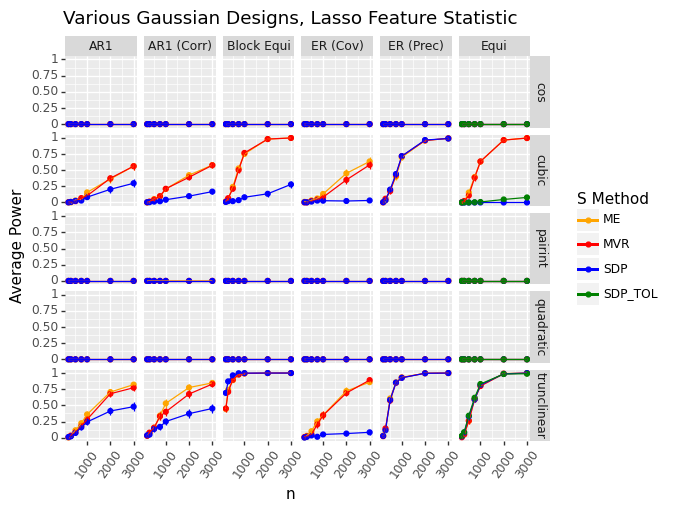}}
    \caption{We let $X \sim \mathcal{N}(0, \Sigma)$ for various $\Sigma$ and $Y\mid X \sim \mathcal{N}(\mu(X), 1)$. The precise definitions of the covariance matrices and conditional responses are presented  in Appendix \ref{appendix::simgaussiandescription}. We let $p = 200$ with $30$ non-nulls. We use lasso coefficient differences as our feature statistics.}
    \label{fig::nonlinlasso}
\end{figure}

For the LCD statistics in Figure \ref{fig::nonlinlasso}, we often see large gains for MRC knockoffs over SDP knockoffs for the trunclinear and cubic responses. For the other conditional means, which are highly nonlinear, the LCD coefficient statistics unsurprisingly have zero power for all methods.

\subsection{Details for non-Gaussian simulations in Section \ref{subsec::simnongaussian}}\label{appendix::metrodetails}

\begin{figure}[h!]
    \centering
    \makebox[\textwidth]{\includegraphics[width=\textwidth]{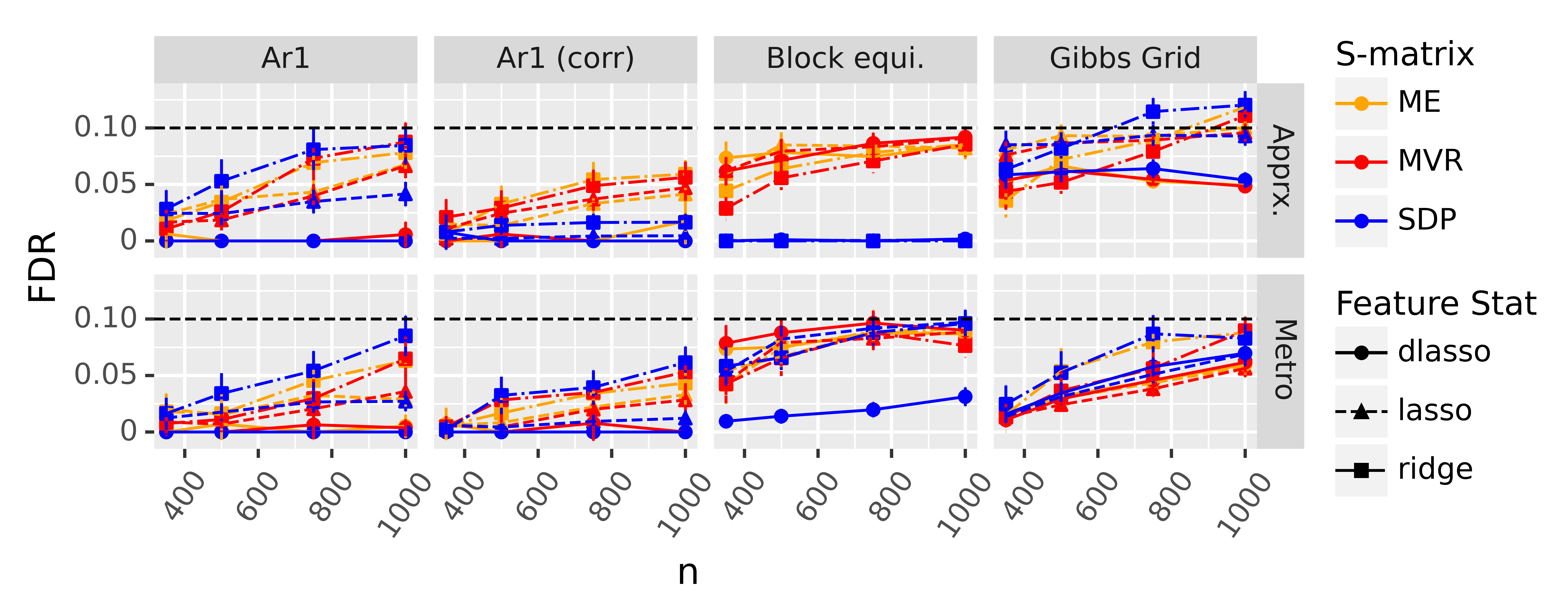}}
    \caption{The corresponding FDR plot for Figure \ref{fig::nongaussianlinearpower}.}
    \label{fig::nongaussianlinearfdr}
\end{figure}

In this section, we present precise details about the three design distributions we simulate in Section \ref{subsec::simnongaussian}. We also include Figure \ref{fig::nongaussianlinearfdr}, which is the corresponding FDR plot for Figure \ref{fig::nongaussianlinearpower}. 

\underline{Block-Equicorrelated $t$}: First, as an analogy to the block-equicorrelated Gaussian case, we partition $X_1, \dots, X_p$ into blocks of size $5$. We let $X_J \sim t_{\nu}\left(0, \frac{\nu-2}{\nu} \Sigma_J\right)$ where $\Sigma_J \in \mathbb{R}^{5 \times 5}$ is equicorrelated with correlation $\rho = 0.5$ and $\nu = 3$. 

\underline{Heavy-Tailed Markov Chain}: Second, we consider the $t$-tailed Markov chain discussed in \cite{metro2019}. In particular, assume $R_j \iid t_{\nu}$, and let $X_1 = \sqrt{\frac{\nu-2}{\nu}} R_1$ and $X_{j+1} = \rho_j X_j + \sqrt{1 - \rho_j^2} \sqrt{\frac{\nu - 2}{\nu}} R_{j+1}$. In our simulations, we set  $\nu = 3$. Similar to the AR1 setting in Section \ref{subsec::simgaussian}, we sample the values of $\rho_j \iid \text{Beta}(3,1)$. As in Appendix \ref{appendix::simgaussiandescription}, we refer to the default case with the name ``AR1", and we also refer to a case where the non-nulls are clustered together as the ``AR1 (Corr)" case.

\underline{Gibbs Grid Model}: Lastly, we consider a $d \times d$ discrete grid, where the density $\Phi$ of $X$ factors over a set of cliques $\mathcal{C}$ which follow a grid structure similar to an Ising model \citep{ising1925}. In particular,
$$\Phi(x) = \prod_{\{j_1, j_2\} \in \mathcal{C}} \psi_{j_1, j_2}(x_{j_1}, x_{j_2}) \text{ for } x \in \mathbb{R}^p. $$
We set $\psi_{\{j_1, j_2\}}(x_{j_1},x_{j_2}) = \exp\left(\beta_{\{j_1, j_2\}} |x_{j_1} - x_{j_2}| \right)$ and we let $\beta_{{j_1, j_2}} = \pm 1$ with equal probability. We allow each feature $X_j$ to take one of $20$ evenly spaced values between $-2.375$ and $2.375$ and we choose our covariance-guided proposals with probability proportional to the relevant Gaussian density on each of these values. We employ a Gibbs sampler to sample design matrices $X$. Furthermore, since this grid has a complicated conditional dependence structure, we use two additional tricks introduced in \cite{metro2019} to increase the efficiency of the Metro sampler.

First, as outlined in \cite{metro2019}, efficient Metropolized knockoff (Metro) sampling requires that the estimated covariance $\hat \Sigma$ used to generate the proposals $X^*$ satisfies certain conditional independence properties. Namely, if the distribution of $X$ is represented as an undirected graphical model (UGM), for any features $j_1$ and $j_2$ which are not connected in the UGM, we must have $\hat \Sigma^{-1}_{j_1, j_2} = 0$. However, unlike in our experiments with $t$-distributions, this property does not even hold when using the true covariance matrix $\Sigma$. As a result, we use the glasso package from the R programming language to estimate $\hat \Sigma$ for the Gibbs grid data to ensure that $\hat \Sigma^{-1}$ has the necessary sparsity pattern for efficient sampling. Note that since $\hat \Sigma$ is a ``compatible proposal" in that it allows for efficient knockoff generation and the probability of our discrete proposals depend only on $\hat \Sigma$, our discrete proposals will be compatible as well. We use the same estimated covariance matrix for each of our replications.

Second, the time complexity of sampling from the Gibbs grid model is prohibitively complex. In particular, for a grid of width $d_1 \times d_2$, the computational complexity of sampling knockoffs for the grid runs in $O(d_1 d_2 2^{\min(d_1, d_2)})$. \cite{metro2019} suggested a ``divide-and-conquer" approach to reduce complexity, where we set the knockoffs for several rows of the grid to be deterministically equal to the features. Following this approach, we set every $5$th row of knockoffs deterministically equal to the features. To increase power for these points, however, we repeat this for both rows and columns on two different translations of the dividing rows, leading to four different divide and conquer mechanisms, which we apply to one quarter of the rows of the data each.

Lastly, we will explain how we generate the proposals $X^*$ for the discrete Gibbs grid model. Suppose $X \sim [0, \Sigma]$, and fix $j \in [p]$. Also fix the $G_S$ matrix, which we will refer to as $G$ for notational convenience. Recall that in the continuous case, when generating proposal $X_j^*$, we sample $X_j^*$ from a univariate Gaussian with mean $\mu_j(X, X_{1:(j-1)}^*)$ and variance $\sigma_j^2$, where
$$\mu_j(X, X_{1:(j-1)}^*) = G_{1:(p+j-1),p+j}^{\top} G_{1:(p+j-1),1:(p+j-1)}^{-1} [X, X^*_{1:(j-1)}]$$ and
$$\sigma_j^2 = G_{p+j, p+j} - G_{1:(p+j-1),p+j}^{\top} G_{1:(p+j-1),1:(p+j-1)}^{-1} G_{1:(p+j-1),p+j}.$$ In the discrete case, if $X_j$ can take one of $K$ values $v_1, \dots, v_K$, we sample $X_j^*$ such that $\mathbb{P}\left(X_j^* = v_k \mid X, X_{1:j}^*\right) \propto \varphi(v_k; \mu_j(X, X_{1:(j-1)}^*), \sigma_j^2)$ where $\varphi(\cdot \, ; \mu, \sigma^2)$ is the PDF of a univariate Gaussian with mean $\mu$ and variance $\sigma^2$.

\section{Overview of \package}\label{appendix::knockpy}

In this section, we briefly list some of the features of \package. See \url{https://github.com/amspector100/knockpy} for more detailed tutorials and documentation.

\textbf{S-Matrix computation for Gaussian knockoffs}:
\package \ offers efficient algorithms to compute a wide variety of $S$-matrices, including a fast SDP solver and implementations of Algorithm \ref{alg::mvrstable} and Algorithm \ref{alg::maxent}, which respectively compute $S_{\mathrm{MVR}}$ and $S_{\mmaxent}$.

\textbf{Metropolized knockoff sampler}: \package \ includes a fully general Metropolized knockoff sampler which can generate exactly valid covariance-guided Metropolized knockoffs for any unnormalized density $\Phi$. The Metropolized sampler can also take further advantage of conditional independence structure of $X$ to speed up queries to $\Phi$, yielding an $O(p)$ improvement over the only previous implementation we know of (which did not use covariance-guided proposals).

\textbf{Feature statistics}: \package\ offers a whole suite of built-in feature statistics, including cross-validated lasso, ridge, and group-lasso coefficients, lasso-path statistics, the DeepPINK statistic \citep{deeppink2018} and random forest statistics with swap and swap integral importances \citep{knockoffsmass2018}.

\textbf{Modularity for development}: \package \ is built to be modular, such that researchers and analysts can easily layer functionalities on top of it. For example, in under three lines of code, \package\ can wrap any python class capable of predicting $Y$ given $X$ and generate feature importances via the swap and swap integral importance procedures from \citep{knockoffsmass2018}. Each of the major classes of \package\ offer explicit ways for users to mix and match \package's features with their own code.

\textbf{Miscellaneous features}: \package\ also supports a host of other features from the knockoffs literature, including fixed-X knockoffs \citep{fxknock}, some early support for group knockoffs \citep{daibarber2016} and knockoff ``recycling" \citep{splitting2019}.

\end{document}